\documentclass[iop, numberedappendix, apj,twocolappendix]{emulateapj}
\usepackage{epsfig}
\usepackage{graphicx}
\usepackage{xspace}
\usepackage{amsfonts}
\usepackage{amsmath}
\usepackage{mathrsfs}
\usepackage{amssymb}
\usepackage{rotating}
\usepackage{tabularx}
\usepackage{adjustbox}
\usepackage{threeparttable}
\usepackage{booktabs}
\usepackage[usenames, dvipsnames]{color}
\usepackage[breaklinks,colorlinks,citecolor=blue,linkcolor=blue]{hyperref}
\usepackage[all]{hypcap}
\usepackage{natbib}
\usepackage{etoolbox}

\makeatletter

\patchcmd{\NAT@citex}
  {\@citea\NAT@hyper@{%
     \NAT@nmfmt{\NAT@nm}%
     \hyper@natlinkbreak{\NAT@aysep\NAT@spacechar}{\@citeb\@extra@b@citeb}%
     \NAT@date}}
  {\@citea\NAT@nmfmt{\NAT@nm}%
   \NAT@aysep\NAT@spacechar\NAT@hyper@{\NAT@date}}{}{}

\patchcmd{\NAT@citex}
  {\@citea\NAT@hyper@{%
     \NAT@nmfmt{\NAT@nm}%
     \hyper@natlinkbreak{\NAT@spacechar\NAT@@open\if*#1*\else#1\NAT@spacechar\fi}%
       {\@citeb\@extra@b@citeb}%
     \NAT@date}}
  {\@citea\NAT@nmfmt{\NAT@nm}%
   \NAT@spacechar\NAT@@open\if*#1*\else#1\NAT@spacechar\fi\NAT@hyper@{\NAT@date}}
  {}{}

\usepackage{color}
\usepackage{ulem}
\usepackage{xspace}
\usepackage{cancel}

\newcommand{\lya} {Ly$\alpha$\xspace}
\newcommand{\hi}{${\rm HI}$ }
\newcommand{\nh}{$n_{\rm H}$ }

\newcommand{\cMpc}{\, {\rm cMpc}}
\newcommand{\ckpc}{\, {\rm ckpc}}
\newcommand{\Mpc}{\, {\rm Mpc}}
\newcommand{\kpc}{\, {\rm kpc}}
\newcommand{\K}{\, {\rm K}}
\begin{document}

 \title{A Fundamental Test for Galaxy Formation Models: Matching the Lyman-$\alpha$ Absorption Profiles of Galactic Halos Over Three Decades in Distance} 
 
\author{Daniele Sorini\altaffilmark{1, 2}
			Jos\'e O$\tilde{\textrm {n}}$orbe\altaffilmark{1},
		   Joseph F. Hennawi\altaffilmark{1, 3}, 
		   Zarija Luki\'c\altaffilmark{4}}
		   %% DS Accent added on Jose
\altaffiltext{}{E-mail: sorini@mpia-hd.mpg.de}		   
\altaffiltext{1}{Max-Planck-Institut f\"ur Astronomie, K\"onigstuhl 17, D-69117 Heidelberg, Germany}
\altaffiltext{2}{Fellow of the International Max Planck Research School for Astronomy and
Cosmic Physics at the University of Heidelberg (IMPRS-HD)}
\altaffiltext{3}{Department of Physics, Broida Hall, University of California, Santa Barbara, CA 93106-9530, USA}
\altaffiltext{4}{Lawrence Berkeley National Laboratory, CA, USA}
%\slugcomment{Draft Version}
\shorttitle{Testing Galaxy Formation Models Using \lya Absorption Profiles of Galactic Halos}
\shortauthors{Sorini et al.}

\begin{abstract}
  
Galaxy formation depends critically on the physical state of gas in
the circumgalactic medium (CGM) and its interface with the
intergalactic medium (IGM), determined by the complex interplay
between inflow from the IGM and outflows from supernovae and/or AGN
feedback. The average Lyman-$\alpha$ absorption profile around galactic halos
represents a powerful tool to probe their gaseous environments. We compare
predictions from Illustris and Nyx hydrodynamical simulations with the
observed absorption around foreground quasars, damped Lyman-$\alpha$ systems,
and Lyman-break galaxies. We show how large-scale BOSS and
small-scale quasar pair measurements can be combined to precisely
constrain the absorption profile over three decades in transverse
distance $20 \kpc \lesssim b \lesssim 20\Mpc$. Far from galaxies
$\gtrsim 2 \Mpc$, the simulations converge to the same profile 
and provide a reasonable match to the observations. This asymptotic agreement arises
because the $\Lambda$CDM model successfully describes the ambient IGM,
and represents a critical advantage of studying the mean absorption
profile.  However, significant differences between the 
simulations, and between simulations and observations are present on scales
$20\,\kpc \lesssim b \lesssim 2\Mpc$, illustrating the challenges of
accurately modeling and resolving galaxy formation physics.
It is noteworthy that these differences are observed as far out as
$\sim 2\Mpc$, indicating that the `sphere-of-influence' of galaxies
could extend to approximately $\sim 7$ times the halo
virial radius. 
Current observations are very
precise on these scales and can thus strongly
discriminate between different galaxy formation models. We
demonstrate that the Lyman-$\alpha$ absorption profile is primarily sensitive to
the underlying temperature-density relationship of diffuse gas around
galaxies, and argue that it thus provides a
fundamental test of galaxy formation models.

\end{abstract}


%________________________________________________________________
\maketitle

\section{Introduction}
\label{sec:intro}

Several fundamental processes for the buildup of galaxies, like accretion, gas flows and feedback, determine the physical
state of the circumgalactic medium \citep[CGM; for a brief review, see][]{CGM_review}. The CGM lies at the interface between galactic halos and diffuse baryons in the intergalactic medium (IGM), which traces the large-scale structure of the
Universe, and can be precisely described by the $\Lambda$CDM cosmological model (see \citealt{Meiksin_review} and \citealt{McQuinn_2016} for recent reviews). As such, 
understanding the physical properties of the CGM and IGM around galaxies
is crucial for advancing our understanding
of galaxy formation and evolution in a cosmological context.

Diffuse gas around galaxies and quasars (henceforth QSOs) can be probed via absorption line measurements
in the spectra of background QSOs at
small transverse separations from the foreground object. 
For example, the spectra of 15 highly luminous QSOs in the Keck Baryonic Structure Survey (KBSS) revealed that the CGM of foreground star-forming galaxies in the redshift range $2\lesssim z \lesssim 3$ presents an excess of neutral hydrogen (HI) absorption with respect to the IGM \citep[][and references therein]{Steidel_2010, Rakic_2012, Rudie_2012_CGM, Rudie_2013}. 
\cite{Turner_2014} confirmed these results and also observed an increased optical depth for metal lines 
in the vicinity of the galaxies. Other observations probed the \lya transmission
up to galactocentric distances of $\sim 10 \, h^{-1} \Mpc$, thus giving insight into the physics of both the CGM and IGM \citep{Adelberger_2003, Adelberger_2005, Crighton_2011}. Through a systematic study of high-optical-depth \lya absorbers in the IGM within the redshift range $2.6\lesssim z \lesssim 3.3$, the MAMMOTH (MApping the Most Massive Overdensity Through Hydrogen) project \citep{MAMMOTH_1, MAMMOTH_2} led to the discovery of overdensities in the cosmic web on scales $\sim 10-20 \Mpc$, enabling constraints on structure formation models.

The Quasars Probing Quasars (QPQ) project uncovered a large sample of
projected QSO pairs with small impact parameters \citep{Hennawi_2004, 
Hennawi_2006_binary_paper, Hennawi_2010}, which enabled the first
studies of the CGM of quasars \citep{Hennawi_2006,QPQ2} at $z\sim 2-3$
(see also \citealt{Bowen_2006, Farina_2011, Farina_2013,Johnson_2013,Farina_2014,Johnson_2015,Johnson_2015_environment,Johnson_2016} for similar
work at lower $z$). 
Using an
enlarged statistical sample, \cite{Prochaska_2013} observed an excess
of \lya absorption within $1 \Mpc$ from the foreground QSOs (see also
\citealt{Prochaska_2013_QPQ5}). This enhanced absorption due to $\rm
HI$, as well as metals \citep{QPQ7,Lau_2016},
implies the presence of a substantial reservoir
of cool ($T\sim 10^4 \, \rm K$) metal-enriched gas around quasars \citep{Prochaska_2013_QPQ5}.

The QPQ observations were extended to larger scales (impact parameters $1-80 \, h^{-1} \cMpc$) by \cite{Font-Ribera_2013}, who used background sightlines from the Baryonic Oscillation Spectroscopic Survey (BOSS; \citealt{Ahn_2012}) to measure the \lya forest - QSO cross-correlation function. This statistic is equivalent to the average \lya forest transmission profile around QSOs measured in the QPQ project. With the same technique, \cite{Font-Ribera_2012b} characterized the average \lya absorption out to $60 \, h^{-1} \cMpc$ from Damped \lya Absorbers (DLAs) at $z\sim 2-3.5$. 
More recently, \cite{Rubin_2015} extended this measurement to small scales using close QSO pairs, whereby
one sightline is used to identify the DLA, and the neighboring one probes
\lya and metal line absorptions at a small impact parameter. These observations
provide evidence for excess \lya absorption within $200 \kpc$ from DLAs.

There is thus now a large amount of data characterizing the strength of HI \lya absorption in the CGM and IGM surrounding star-forming galaxies, QSOs, and DLAs (see also \citealt{Fu_2016,Fu_2017} for recent progress
characterizing the CGM of sub-millimeter galaxies.). 
The strength of HI absorption around galaxies is determined by the abundance and physical state (density, temperature, cloud size) of cool ($T \lesssim 10^5\,{\rm K}$) gas.
While the inflow of cool material from the IGM to the CGM is predicted ab initio by cosmological hydrodynamic simulations, it is ultimately the interplay between inflows and complex feedback processes that determines its physical state.
However feedback processes are still poorly understood, and nearly all simulations
model them via `sub-grid' implementations, the details of which vary from case to case and code to code \citep[][for a review]{Somerville_review}. These feedback prescriptions can be constrained by comparing the predictions of simulations with measurements of absorption features in the CGM and IGM. 

Most past numerical work treating the CGM has focused primarily on
the covering factor of optically thick absorbers around galaxies and
QSOs at $z\sim 2-3$. Broadly speaking,
current simulations \citep{Ceverino_2012, Dekel_2013, Shen_2013,Meiksin_2015,
  Suresh_2015,Meiksin_2017} are able to reproduce the
\cite{Rudie_2012_CGM} measurements of this covering factor around
galaxies, but struggle \citep{Fumagalli_2014,
  Faucher-Giguere_2015} to explain the high covering factor of
optically thick gas observed around QSOs
\citep{Prochaska_2013}. \cite{Faucher-Giguere_2016} 
reproduced this high covering factor with the FIRE zoom-in simulations
\citep{Hopkins_2014}, and argued that high-resolution was crucial to
achieving this match, whereas \cite{Rahmati_2015} managed to match
these observations with the much lower-resolution EAGLE suite of
cosmological hydrodynamic simulations \citep{Crain_2015, Schaye_2015}.
In light of this debate about resolution, 
and the fact that both star-formation and AGN
feedback are implemented in EAGLE, while only the former is included
in FIRE, it appears that reproducing absorption line observations
around QSOs remains an important open question.

Another well-studied and related statistic is the column density distribution
function (CDDF) of \hi absorbers at $z \sim 2-3$.
Good agreement with observations
\citep{Kim_2002, Peroux_2005, Zwaan_2005, O_Meara_2007, Noterdaeme_2009,
  Prochaska_2009b, Prochaska_2010, Noterdaeme_2012,
  Kim_2013, Rudie_2013} was found in the OWLS \citep{Schaye_2010} and
``Sherwood'' \citep{Bolton_2017} suites of cosmological simulations
\citep{Altay_2011, Altay_2012, Rahmati_2013, Rahmati_2013b,
  Rahmati_2015} and in hydrodynamic simulations based on the
\textsc{Arepo} code \citep{Bird_2013, Bird_DLA}, as well as with other
codes \citep{Fumagalli_2011,McQuinn_2011_CDDF}. Finally, at lower redshift $z<0.2$, 
\citep[][see also \citealt{Stinson_2012}]{Gutcke_2017} 
find that the NIHAO
\citep{Wang_2015} suite of zoom-in simulations does a good job or
reproducing the observed distribution of \hi column densities around galaxies
\citep{Prochaska_2011, Tumlinson_2013}.

Perhaps the simplest statistics characterizing the CGM and IGM around
galaxies is the average \lya transmission profile. Several numerical
studies \citep{Kollmeier_2003, Kollmeier_2006, Rakic_2012, Rakic_2013,
  Meiksin_2014, Meiksin_2015, Meiksin_2017, Turner_2017} have compared
large-volume hydrodynamical simulations equipped with different
feedback implementations, to observations of the \lya flux decrement
around LBGs \citep{Adelberger_2003, Adelberger_2005, 
Steidel_2010,Crighton_2011, Rakic_2012, Turner_2014}, 
and QSOs \citep{Prochaska_2013}
In general the observations are well reproduced by these simulations, with the exception of
measurements within the virial radius of the respective foreground objects.
While \cite{Meiksin_2017} argued that stronger stellar feedback would improve agreement there
for LBGs and QSOs, \cite{Turner_2017} find that the \lya optical depth in the CGM of LBGs is weakly
dependent on the stellar feedback model.

In this paper, we build upon this body of work and compare recent \lya
absorption measurements around LBGs, QSOs, and DLAs to both the
state-of-the-art ``Illustris'' cosmological hydrodynamical simulation
\citep{Vogelsberger_2014N, Vogelsberger_2014, Genel_2014, Illustris_public,
  Sijaki_2015} and a large-box run of the ``Nyx''
hydrodynamics code \citep{Almgren_2013, Lukic_2015}.  We show how
large-scale measurements from BOSS \citep{Font-Ribera_2012b,
  Font-Ribera_2013} can be combined with small-scale observations
using QSO pairs \citep{Prochaska_2013, Rubin_2015} to precisely constrain the
average \lya transmission profiles over a large dynamic range in
impact parameter ($20 \kpc \lesssim b \lesssim 20\Mpc$). The transmission
profiles around galactic halos predicted by both simulations converge
to the same answer on large scales $\gtrsim 10\Mpc$, where they also
agree well with the measurements -- a manifestation of the success of the $\Lambda$CDM model in describing
the IGM far from galaxies.
However, discrepancies between the data and the simulations are observed
on scales comparable to the virial radius, resulting from the complexities of galaxy formation.
Support for this conclusion comes from the
fact that the simulations also disagree on these scales,
reflecting the
impact of star-formation and AGN feedback present in Illustris and absent in Nyx.
Surprisingly, we find that these differences persist out to $\sim 5-7$ virial radii
(or $\sim 2 \Mpc$) indicating that current measurements already
have the power to discriminate between different feedback models. 
We show how the underlying temperature-density relationship of gas in the CGM and CGM-IGM
interface determines the \lya transmission profile. 

This manuscript is organized as follows. In \S~\ref{sec:sims}, we describe the details of the simulations adopted. In \S~\ref{sec:modeling}, we explain how we constructed the sample of simulated spectra of \lya absorption. We compare the predictions of the simulations with observations in \S~\ref{sec:results}. In  \S~\ref{sec:discussion}, we discuss the physical implications of our results and compare them with other findings in the literature in \S~\ref{sec:previous_work}. We summarize our results and discuss the perspectives of this work in \S~\ref{sec:conclusions}. Throughout the paper, distances are expressed in physical units (e.g., $\kpc$ and $\Mpc$). When using co-moving units, we add a ``c'' in front of the unit of measure (e.g $\ckpc$ and $\cMpc$).

\section{Simulations}
\label{sec:sims}

To study the \lya absorption in the CGM, we used a large-box run of the Nyx hydrodynamic code, and the publicly available Illustris cosmological hydrodynamic simulation. In this section, we briefly describe the main characteristics of the two simulations.

\subsection{Nyx}
\label{sec:Nyx}

Nyx \citep{Almgren_2013, Lukic_2015} is a hydrodynamic code, treating dark matter (DM) as self-gravitating Lagrangian particles, and modeling baryons as an ideal gas on a uniform Cartesian grid. Shock waves are accurately captured by solving the Eulerian equations of gas dynamics through a second-order-accurate piecewise parabolic method. The adaptive mesh refinement (AMR) option provided by Nyx is switched off in the present work, since we are interested in simulating the \lya absorption in the entire box, and not only within the most overdense regions (i.e. halos). More details about the numerical methods implemented in Nyx and relevant scaling tests can be found in \cite{Almgren_2013} and \cite{Lukic_2015}. 

The physical processes necessary to model the \lya forest are included in the code. The gas is considered to have a primordial composition, with abundances $X_{\rm p}=0.76$ and $Y_{\rm p}=0.24$, respectively. Furthermore, the inverse-Compton cooling off the microwave background is included, and the net loss of the thermal energy resulting from atomic collisional processes is accounted for. The values of the recombination, collisional ionization, dielectric recombination rates, as well as cooling rates, which are used in this work, are given in \cite{Lukic_2015}. The ionizing ultraviolet background (UVB) is given by \cite{Haardt_Madau_2012}. Self-shielding modeled following \citealt{Rahmati_2013} (see \S~\ref{sec:sim_absorption} for further discussion). Nyx does not incorporate any star formation or feedback prescription.

Since there is no star-formation in Nyx and hence no star particles,
cells at the center of a dark matter halo can reach very high 
densities. To avoid potential artefacts arising from these rare high density
cells, we threshold the density at $\delta_{\rm th}=1000$ when computing the \lya
optical depth \footnote{We also considered other values of $\delta_{\rm th}$, ranging from 200 to 3000. By visual inspection, we verified that simulating the \lya absorption spectra with $\delta_{\rm th}=1000$ gives physically reasonable results.}. 

All simulations are initialized at redshift $z_{\rm ini} = 200$, making sure that nonlinear evolution is not compromised \citep[for a detailed discussion on this issue see, e.g.,][]{Onorbe_2014}. The cosmological model assumed is the $\Lambda$CDM model with parameters consistent with Planck data \citep{Planck}: $\Omega_{\mathrm{m}}=0.3$, $\Omega_{\Lambda}=1-\Omega_{\mathrm{m}}=0.7$, $\Omega_{\mathrm{b}}=0.047$, $h=0.685$, $\sigma_8=0.8$, $n_s=0.965$.

In this work, we consider a run with a volume of $(146 \cMpc)^3$, $4096^3$ gas cells and as may DM particles. This run represents a state-of-the-art hydrodynamic cosmological simulation. The box size is comparable with the largest hydrodynamic cosmological simulations in the literature \citep[e.g.][]{Vogelsberger_2014, Crain_2017}. The resolution of $35.6 \, \ckpc$ for baryons grants a precision at percent level in the probability density function (PDF) of the \lya forest flux and within $\sim5\%$ in the 1D \lya flux power spectrum up to $k\sim 0.1 \, \rm s \, km^{-1}$ \citep{Lukic_2015}. For the comparison with the observations considered in this manuscript, we use the snapshots at redshift $z=2.4$ and $z=3$. To test the impact of the redshift of foreground objects on the predictions given by Nyx, we use also the snapshot at redshift $z=2$ (see the Appendix \ref{sec:redshift}).

\subsection{Illustris}
\label{sec:Illustris}

Illustris \citep{Vogelsberger_2014, Vogelsberger_2014N, Genel_2014, Sijaki_2015} is a hydrodynamic simulation based on the \textsc{Arepo} code \citep{Springel_2010}. Dark matter is treated with a Lagrangian approach, while baryons are described as an ideal gas on a moving mesh constructed with a Voronoi tessellation. A Tree-PM scheme \citep{Xu_1995} is used to compute gravitational forces; a particle-mesh method calculates long-range forces, while a hierarchical algorithm \citep{Barnes_1986} is used to determine short-range forces. Gas dynamics is implemented by solving the viscosity-free Euler equations.

Several fundamental astrophysical processes for galaxy formation are included: primordial and metal-line cooling, a sub-resolution model of the interstellar medium, stochastic star formation above a density threshold of $0.13 \: \rm cm^{-3}$, gas recycling and chemical enrichment. Kinetic stellar feedback is implemented through supernovae-driven winds, determined by the velocity dispersion of the host halo \citep{Vogelsberger_2013}. Illustris also includes supermassive black hole seeding, accretion and merging. The AGN feedback implementation follows the dual modeling described in
\cite{Sijacki_2007}. For high black hole accretion rates, the radiated energy is thermally coupled to the surrounding gas (``quasar-mode'' AGN feedback). For low accretion rates, a mechanical ``radio-mode'' AGN feedback injects hot gas bubbles in the halo atmosphere. 
The free parameters governing the feedback prescriptions were constrained matching the overall star formation efficiency predicted by smaller scale simulations \citep{Vogelsberger_2013} to observed data \citep{Guo_2011,Moster_2013,Behroozi_2013}.

Photo-ionization and heating are implemented with the UVB by \cite{Faucher-Giguere_2009}. Self-shielding and ionization from nearby AGN are taken into account when calculating the heating and cooling of the simulated gas cells. 
Consistent with what we are doing with the Nyx simulation, when computing the \lya optical depth from Illustris, we also include collisional ionization and self-shielding, following \citealt{Rahmati_2013} (see \S~\ref{sec:sim_absorption} for further discussion). 

The simulation is initialized at $z_{\rm ini}=127$ \cite[see][for details]{Vogelsberger_2014N}. The cosmological model assumed is the $\Lambda$CDM model with parameters consistent with the 9-year data release of
WMAP \citep{Hinshaw_2013}: $\Omega_{\mathrm{m}}=0.2726$, $\Omega_{\Lambda}=1-\Omega_{\mathrm{m}}=0.7274$,
$\Omega_{\mathrm{b}}=0.0456$, $h=0.704$, $\sigma_8=0.809$, $n_s=0.963$. We utilize the snapshots at redshift $z=2.44$ and $z=3.01$ for the comparison with the observations considered in this work. To test the impact of the redshift of foreground objects on the predictions given by Illustris, we use also the snapshot at redshift $z=2$ (see the Appendix \ref{sec:redshift}).Unless otherwise stated, we refer to the ``Illustris-1'' run, which is the highest-resolution one. The volume of the simulation is $(106.5 \cMpc)^3$; with $1820^3$ DM particles, and as many gas Voronoi cells, the mean interparticle separation is $58.5 \ckpc$. 
The mass resolution is $6.3\times10^6\: M_{\odot}$ and $1.3\times10^6\: M_{\odot}$ for DM and gas, respectively.

\section{Modeling}
\label{sec:modeling}

\begin{figure*}[]
  \centering
   \includegraphics[width=\textwidth]{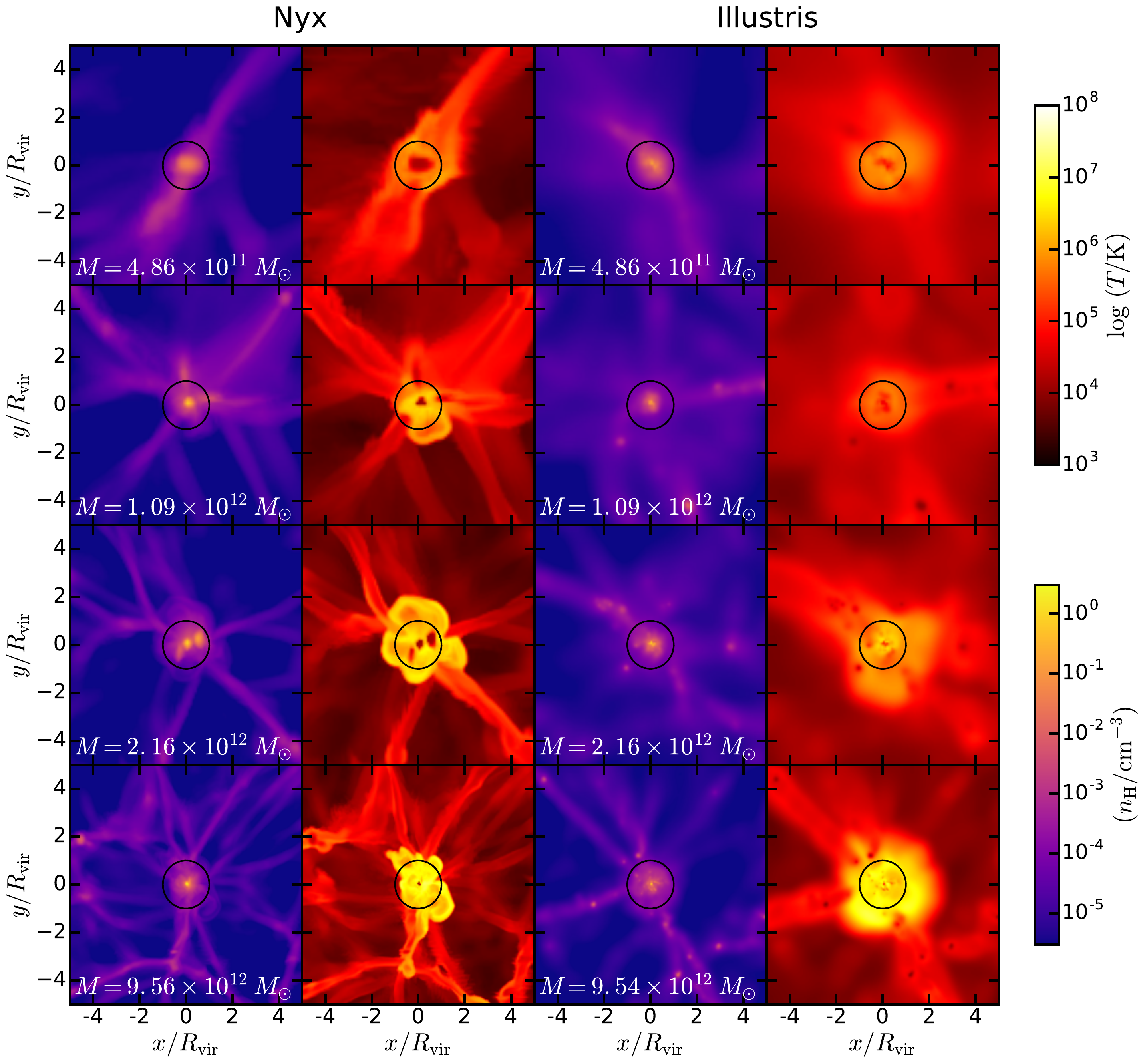}
  \caption{\textit{Top panels}: Temperature and hydrogen density slices around a sample of 4 halos selected from Nyx and Illustris at $z=2.4$ and $z=2.44$, respectively. The first and second columns from the left show the hydrogen density and temperature around Nyx halos, respectively, while the third and fourth columns the same quantities for Illustris halos. The slices are one pixel thick, are centered around the halos, and span an area of $5\times5$ virial radii. The virial radius is marked with a black circle. The mass of each halo is written in the hydrogen density panel. In every row, the halos taken from Nyx and Illustris are chosen to have the same mass (within 0.2\%), but are different halos, i.e. evolved with different codes from different initial conditions. For the most massive halos, Illustris generally presents more extended bubbles of hot gas with respect to Nyx.} \label{fig:slices}
\end{figure*}

In this section, we explain how we create mock absorption spectra (skewers) to reproduce the observations. Specifically, we consider measurements of the \lya absorption in the CGM of QSOs, DLAs and LBGs measured from the spectra of background QSOs
at different impact parameters from the foreground objects. 

Reproducing the observations from simulations requires three steps. First, we need to select a sample of halos representing the foreground objects. Then, we extract a sample of lines of sight at different impact parameters from each halo, and finally we compute mock \lya absorption spectra along such skewers. 

\subsection{Selection of Halos}
\label{sec:sel_halos}

To begin with, one needs to identify massive DM halos in Nyx and Illustris. Then, one has to select samples of halos hosting the foreground QSOs, LBGs and DLAs. For Illustris, we use the publicly available catalogs of halos, which are identified with a friends-of-friends algorithm \citep{Davis_1985} with a linking length equal to 0.2 times the mean particle separation. 
In Nyx, halos are identified by finding the topological connected components with density above 138 times the mean density, which gives very similar results to the particle-based FOF algorithm with a linking length equal to 0.168 times the mean particle separation (Luki\'c et al. in prep.).\footnote{The fact that the Nyx halo finder corresponds to a FOF linking length different than the one adopted by the Illustris FOF algorithm does not impact our analysis (see the Appendix \ref{app:halo_masses} for further discussion).}

At face value, the halo masses in the simulations are not guaranteed to match the mass of real halos. Furthermore, the halo masses in the Nyx and Illustris catalogs may not be consistent, being determined with different halo-finding algorithms and mass definitions. To ensure a physically meaningful comparison, we calibrate the halo masses of the two simulations with the same observations of QSO and LBG clustering. Specifically, for each simulation, we determine the minimum halo mass $M_{\rm min}$ above which the 3D auto-correlation function of halos matches the same quantity estimated for QSOs and LBGs by \cite{White_2012} and \cite{Bielby_2011}, respectively. For the QSO sample, we obtain $M_{\rm min}=10^{12.4} \, M_{\odot}$ and $M_{\rm min}=10^{12.5} \, M_{\odot}$ at $z\approx2.4$ in Illustris and Nyx, respectively. These values are in broad agreement with the similar analysis by \cite{White_2012}. 

We assume that all halos above the threshold $M_{\rm min}$ are populated with QSOs. We adopt this simple model because the observational constraints on the halo mass of QSO hosts at the redshift of interest cannot currently break degeneracies among more refined halo occupation distribution (HOD) models. In fact, a mass-independent HOD over the halo mass range considered in our work appears to be consistent with the uncertainties on the parameters of such models \citep[see, e.g.,][]{Rodriguez-Torres_2017}.

For the LBG sample, the value of $M_{\rm min}$ at  $z\approx2.4$  is $10^{11.6} \, M_{\odot}$ and $10^{11.7} \, M_{\odot}$ in Illustris and Nyx, respectively. These values are consistent with the typical mass of LBG-hosting halos in the KBSS, $\sim10^{12}\, M_{\odot}$ \citep{Adelberger_2005a, Conroy_2008, Trainor_2012, Rakic_2013, Turner_2014}.

Regarding the measurements of the \lya absorption around DLAs, we considered the same sample of halos selected for the LBGs as the DLA host halos. This choice is justified by the fact that the typical mass of DLAs estimated by \citet{Font-Ribera_2012b}  ($\sim 10^{12} \, M_{\odot}$) is the same as the one measured for LBGs.
 This estimate was obtained measuring the \lya-DLA cross-correlation from
 a large sample of BOSS quasar spectra and fitting it with the prediction of the $\Lambda$CDM model within linear theory. From the amplitude of the cross-correlation function one can derive the bias factor of DLAs, which in turn provides an estimate of the typical mass of a DLA-hosting halo.

For all our catalogs we make the simplifying assumption that the 
LBG, QSO, or DLA lies at the center of the DM halo. This assumption 
is justified for the LBGs and QSOs, since it is well known that halo 
model fits to their clustering indicate that the satellite fraction is very 
low \citep{Allen_2005,White_2008,Conroy_2013, Barone-Nugent_2014}. For 
DLAs there are no halo models of clustering, but this is a fair assumption 
given that they occupy halos of the same mass as the LBG hosts. 

We summarize all derived values of $M_{\rm min}$ in Table \ref{tab:sim_params}. Further details on the procedure to determine $M_{\rm min}$ can be found in the Appendix \ref{app:halo_masses}. 

\begin{table}
	\begin{center}
	\caption{Parameters used in simulations to reproduce observations.}\label{tab:sim_params}
		\begin{threeparttable}
		\begin{tabular}{lcccc} 
			\hline
			Observations \tnote{a} & \multicolumn{2}{c}{$z$ \tnote{b}} & \multicolumn{2}{c}{$\log_{10} 
			(M_{\rm min}/M_{\odot})$ \tnote{c}} \\
			 & Nyx & Illustris & Nyx & Illustris \\
			 \hline
			\begin{tabular}{@{}c@{}} \cite{Font-Ribera_2013} \\  \cite{Prochaska_2013}\end{tabular} & 2.4 & 2.44 & 12.5 & 12.4 \vspace{2pt}\\
			
			\begin{tabular}{@{}c@{}c@{}} \cite{Font-Ribera_2012b} \\ \cite{Turner_2014} \\ \cite{Rubin_2015}\end{tabular} & 2.4 & 2.44 & 11.7 & 11.6 \vspace{2pt} \\
			
			\begin{tabular}{@{}c@{}} \cite{Adelberger_2003}\\ \cite{Adelberger_2005} \\ \cite{Crighton_2011}\end{tabular} & 3.0 & 3.01 & 11.5 & 11.5 \\

			 \hline
		\end{tabular}
		\begin{tablenotes}
			\item[a] Observations considered in this work.
			\item[b] Redshift of the snapshot considered to reproduce the observations in column 1.
			\item[c] Mass threshold for the halos selected to reproduce the observations in column 1.
		\end{tablenotes}
		\end{threeparttable}
	\end{center}
\end{table}

\subsection{Construction of Samples of Skewers}
\label{sec:sample}

Once we set $M_{\rm min}$, we are left with a list of halos selected from the halo catalog, corresponding to the selection criterion described in the previous subsection. We then need to extract a sample of skewers around each halo, within a certain range of the impact parameter $b$. We always adopt the same impact parameter bins as the observations that we want to reproduce, and draw $5\times10^4$ skewers for each bin. 

We randomly draw the position of the first skewer around the first halo in our list from a uniform 
distribution in $\log b$ across the impact parameter bin, and a uniform distribution in the polar angle around the halo, in the plane perpendicular to the direction of the skewer.  
We consider a regular Cartesian grid and find the nearest grid point to the position of the skewer. 
In the case of Nyx, this is simply the grid used in the simulation to describe the evolution of gas. In the case of Illustris, we treat each Voronoi cell as an SPH (Smoothed Particle Hydrodynamics) particle, and bin it into a regular grid \citep[following][]{Bird_DLA}. The size of a grid cell corresponds to the mean interparticle separation in Illustris. We verified that a finer grid would not change the conclusions of this work appreciably (see the Appendix \ref{app:cell_size} for more details).

We extract the gas density, temperature and line-of-sight velocity along the selected skewer, throughout the simulation box. To draw the second skewer of our sample, we consider the second halo in the list and repeat the aforementioned procedure. We proceed in this way until we reach the last halo of the list. Since the number of skewers that we draw is larger than the number of the selected halos\footnote{We select 3425 (1557) and 138 (72) halos in Nyx (Illustris) for the LBG/DLA and QSO hosts samples obtained as explained in \S~\ref{sec:sel_halos}, respectively.\label{foot:size_samples}}, the following skewer is again extracted around the first halo of the list. If the polar-angular distance, in the plane perpendicular to the line of sight, between this skewers and the one previously drawn around the same halo is less than a predefined threshold\footnote{The threshold is that to $2\pi/4R$, where $R$ is the ratio between the total number of skewers and the number of halos selected in the whole sample.}, the coordinates of the new skewer are re-drawn. 
This check is made any time a new skewer has to be drawn around a halo which has previously been considered to extract further skewers. With the procedure adopted, we minimize the number of skewers per halo and set a minimum transverse distance among skewers around the same halo. In this way, we avoid having two highly correlated, or even identical, skewers.

\subsection{Simulating \hi Absorption}
\label{sec:sim_absorption}

Once all density, temperature and velocity skewers are extracted, we can compute the \lya absorption spectra. To do this, we first need to determine the HI density $n_{\rm HI}$ in each pixel.

The total hydrogen number density \nh in each cell is simply obtained multiplying the gas overdensity by the cosmic fraction of hydrogen. We then compute $n_{\rm HI}$ using \cite{Rahmati_2013} fit to the relationship between hydrogen photoionization rate and \nh obtained running radiative transfer on top of EAGLE simulations. The fitting function captures the physics set by photoionization, collisional ionization, self-shielding and recombination radiation.

After computing $n_{\rm HI}$, we determine the \lya optical depth $\tau$ taking into account the thermal broadening of the absorption lines and redshift space distortions \citep[see, e.g.,][]{Lukic_2015}. The \lya flux is then simply given by $F=\exp(-\tau)$. Following the standard approach, we chose the value of the UVB such that the flux skewers are consistent with the observed mean flux of the IGM at the redshift of the observations. This is ensured by randomly drawing $10^5$ skewers from Illustris and Nyx, and tuning the UVB in both simulations such that the mean flux matches the observations by \citet{Becker_2013_mean}. The obtained values of the UVB are then used to compute the transmitted flux along the samples of skewers generated as explained in  \S~\ref{sec:sel_halos} and \S~\ref{sec:sample}. This ensures that far from the halos, at random positions in the IGM, the
mean \lya forest forest flux will be the same in both simulations, allowing for a fair comparison.

\subsection{Altering the Temperature of the CGM}
\label{sec:paint}

\begin{figure}
	\centering
	\includegraphics[width=0.5\textwidth]{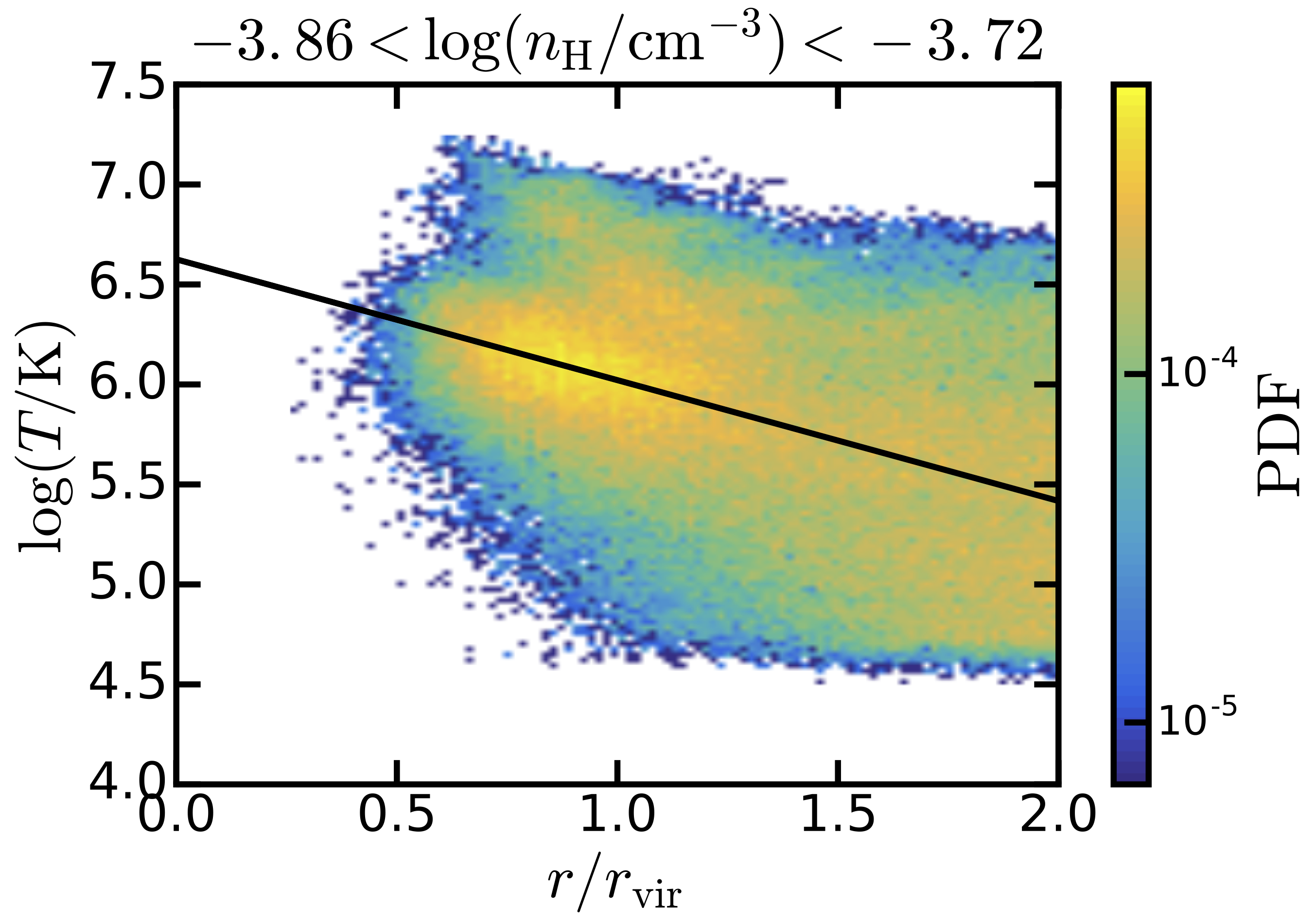}
	\caption{2D Histogram of the temperature and radial distance from the center of LBG/DLA hosting halos in Nyx, within the hydrogen density bin $(10^{-3.86},\,10^{-3.72})\, \rm cm^{-3}$. The color bar indicates the PDF and the black solid line the best fit to the temperature-radial distance relationship given by equation \eqref{eq:paint}. The fitting procedure is repeated for every density bin considered (see text for details).}\label{fig:density_bin}
\end{figure}

\begin{figure*}[]
  \centering
   \includegraphics[width=\textwidth]{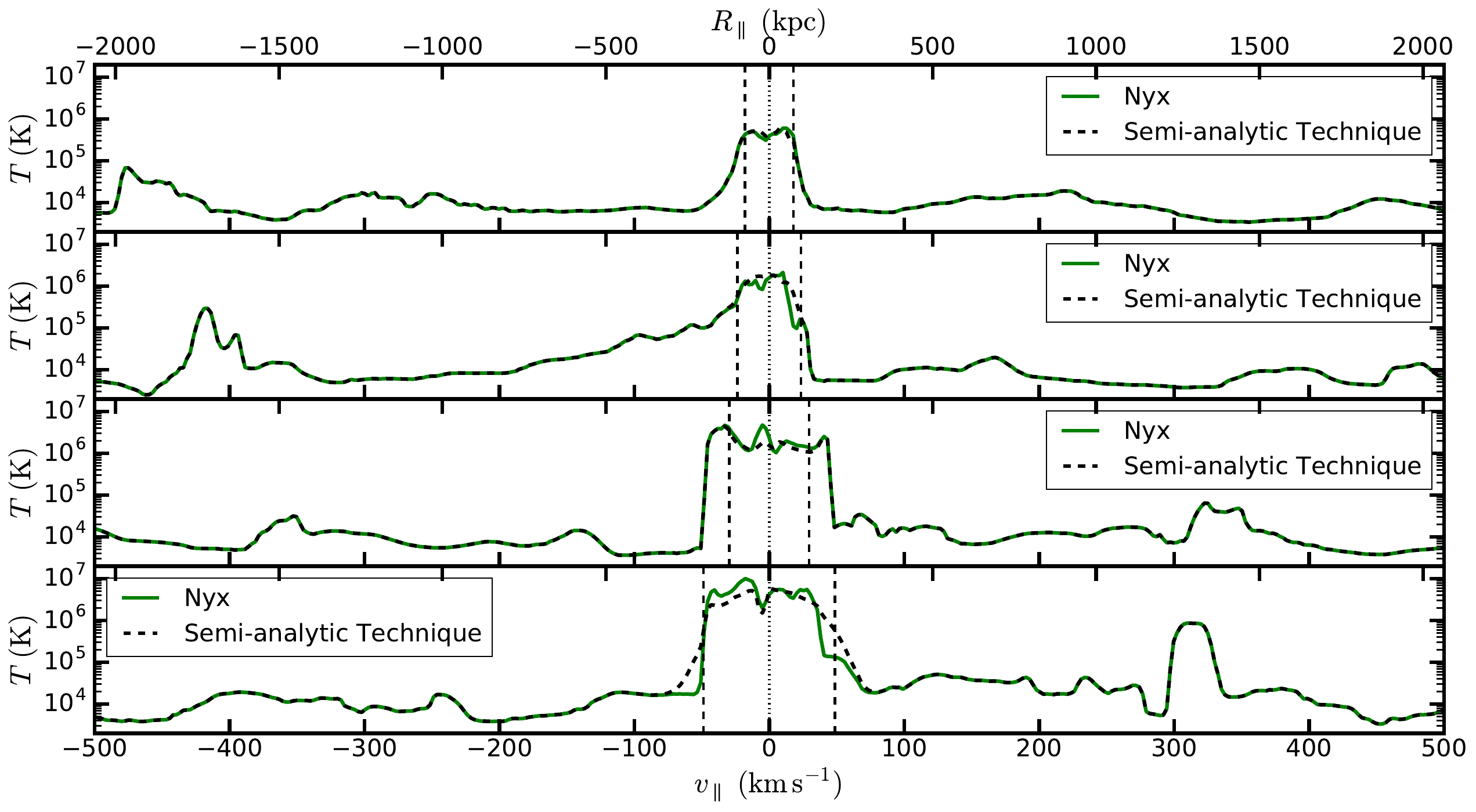}
  \caption{\textit{Top panels}: A sample of 4 temperature skewers given by Nyx (green lines) and obtained through our semi-analytic technique (black dashed lines). All skewers are taken at $50 \kpc$ impact parameter from one of the halos shown in Figure \ref{fig:slices}, ordered by increasing mass from the top to the bottom panel. All skewers are plotted such that the position of the halo lies at the center of the panel. In all panels, the vertical dashed black lines mark the boundaries of the virial radius. By construction, our semi-analytic technique matches Nyx outside the virial radius. Within the virial radius, it presents an overall good agreement with Nyx (see the main text for details). } \label{fig:paint_skewers}
\end{figure*}

\begin{figure*}[]
  \centering
   \includegraphics[width=\textwidth]{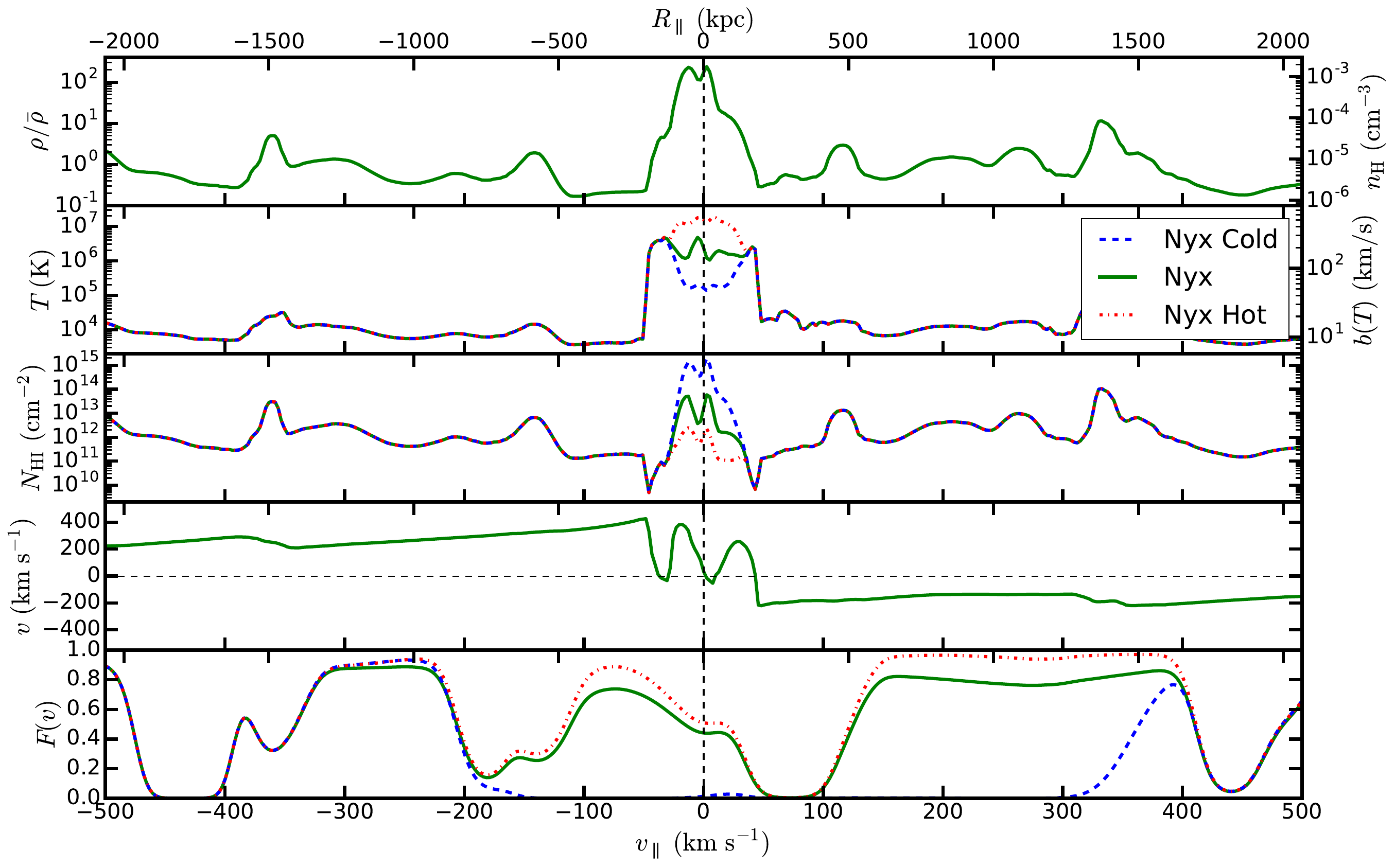}
   \includegraphics[width=\textwidth]{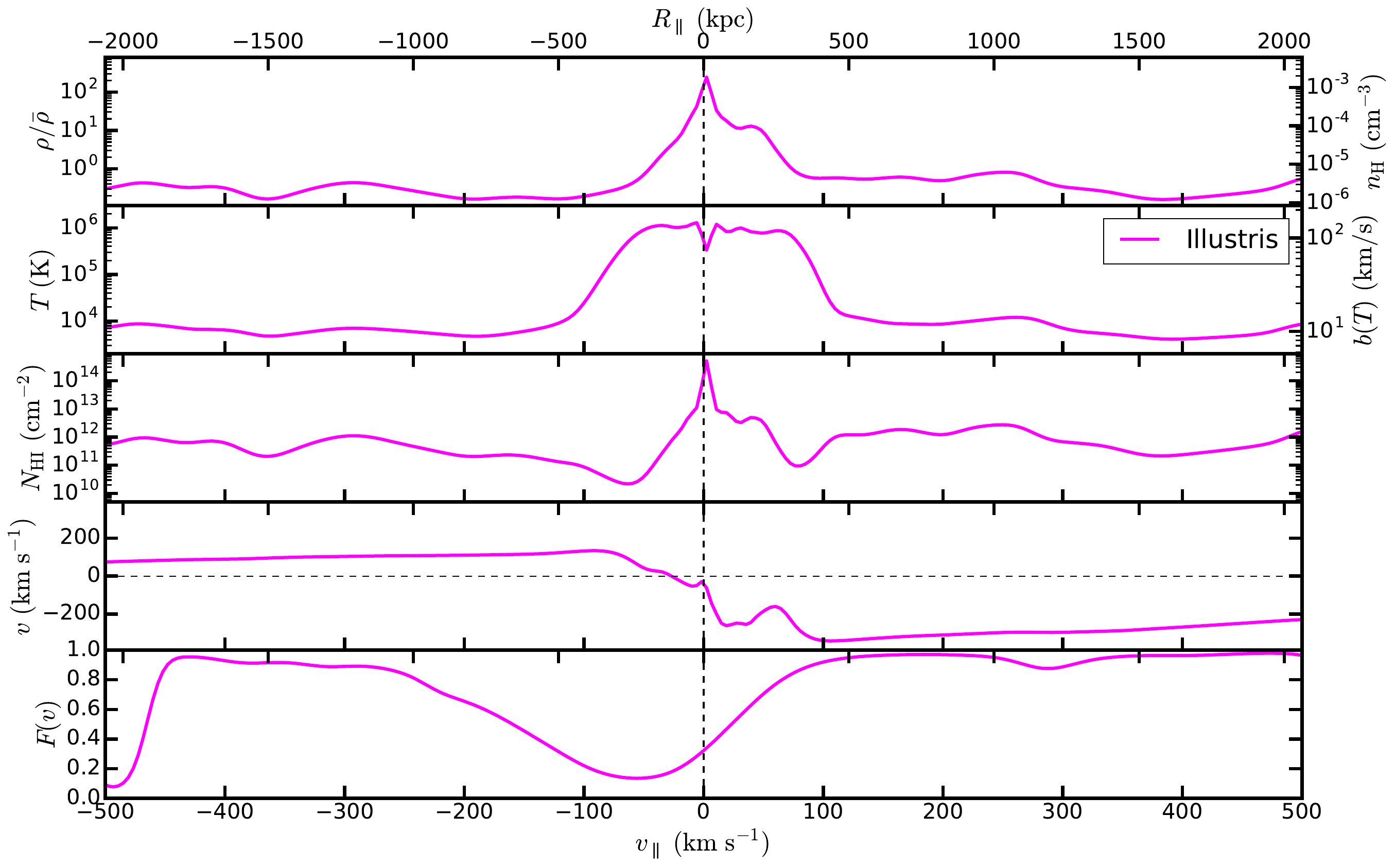}
  \caption{Different quantities along the same skewer, located at $50 \kpc$ from the Nyx and Illustris halos in the third row from the top of Figure \ref{fig:slices}. In the top set of five panels, from top to bottom, we plot the gas overdensity with the corresponding number density of hydrogen, the gas temperature with the corresponding Doppler parameter, the \hi column density, the gas velocity field and the transmitted \lya flux in a velocity window of $\pm 500 \, \rm km \, s^{-1}$ around the Nyx halo. The green solid lines refer to the pure Nyx run. The blue dashed and red dot-dashed lines refer to our Nyx Cold and Nyx Hot models altering the temperature of the CGM in Nyx, respectively (see text for details). In the bottom set of five panels, we plot exactly the same quantities as in the top set, this time for the Illustris halo.} \label{fig:skewers}
\end{figure*}

To gain insight into the connection between \hi absorption and temperature of CGM gas, we developed a semi-analytic technique to ``paint'' different temperature-density relationships in the CGM of Nyx halos. Our goal is to come up with a simple method to alter the temperature of the CGM, depending on a small number of intuitive parameters, and capable of producing physically reasonable skewers.

We begin by visually inspecting hydrogen density and temperature maps around halos. In Figure \ref{fig:slices} we show these quantities within a one-pixel-thick\footnote{Whereas for all our computations we binned Illustris on a Cartesian regular grid with a cell size of $58.5\ckpc$, for the purpose of this Figure we binned it into a grid with a cell size of $30\ckpc$. In this way, the cell size chosen for Illustris is comparable to the one of Nyx ($35.6\ckpc$).} slice centered around a sample of  4 halos in Nyx (first two columns from the left) and Illustris (last two columns from the left). Every row displays one halo from each simulation, chosen such that their masses agree within $0.2\%$. The mass of the halos increases from top to bottom. The halos in the first two rows have a mass typical of LBGs/DLAs, whereas the remaining rows contain halos with a mass characteristic of QSOs. The side of each slice shown is equal to 5 times the virial radius of the halo at its center. The virial radius $r_{\rm vir}$, marked with a black circle at the center of the slices displayed in Figure \ref{fig:slices}, was computed using the \texttt{Barak} Python package\footnote{\url{http://nhmc.github.io/Barak/index.html}} (see also, e.g., \citealt{Binney_2008}).

Let us now focus on the panels of Figure \ref{fig:slices} showing Nyx halos. We notice that, for each halo, the temperature of the CGM broadly traces the underlying hydrogen density in low-density regions, as expected given that the IGM follows a tight temperature-density relationship \citep{Hui_1997}.
On the other hand, within the halo gas cells are significantly hotter
than the temperature-density relationship would predict because of shock
heating\footnote{The only exception is the tiny central dense and overcool region. This is caused by the fact that Nyx does not convert dense gas into stars, hence gas can evolve to very high density, and for the densest cells cooling becomes more efficient.}.

Accordingly, we assume that the temperature of the CGM at a certain point in the halo can, to a first approximation, be modeled with a deterministic function $T(n_{\rm H},\,r)$ of the local hydrogen density $n_{\rm H}$ and of the distance from the center of the halo $r$ only. In general, we do not expect this function to have a trivial shape. In order to determine a physically sensible shape for $T(n_{\rm H},\,r)$, we first select a sample of halos as explained in \S~\ref{sec:sel_halos} and divide the gas cells into equally spaced logarithmic bins of $n_{\rm H}$. We consider the temperature - radial distance relationship of the gas cells within each density bin and fit it with the following function:
\begin{equation}
\label{eq:paint}
	\log_{10} ( T[{\rm K}] ) = a(n_{\rm H}) + b(n_{\rm H}) \frac{r}{r_{\rm vir}}  \, ,
\end{equation}
where $a(n_{\rm H})$ and $b(n_{\rm H})$ are the parameters of the fit, which are determined for each \nh bin. As an illustrative example, the temperature - radial distance relationship of the gas cells around LBG/DLA hosting halos ($M>10^{11.7}\, M_{\odot}$) within the density bin $(10^{-3.86},\,10^{-3.72})\, \rm cm^{-3}$ is shown in Figure \ref{fig:density_bin} as a 2D volume-weighted histogram. The color bar indicates the PDF and the black solid line is the fit given by equation \eqref{eq:paint}. 

We used equation \eqref{eq:paint} to define the aforementioned function $T(n_{\rm H},\,r)$, and used such function to re-compute the temperature of the gas within the virial radius of halos in Nyx, leaving the temperature given by Nyx outside the virial radius unmodified\footnote{The transition between the temperature given by Nyx outside the virial radius and the temperature computed through our semi-analytic technique within the virial radius is actually modulated with a smooth approximation of the step function. Given the specific shape of this function, the temperature of the skewers in the Nyx Hot and Nyx Cold models can slightly differ from the values of the pure Nyx run as far as $\sim 1.3 \, r_{\rm vir}$.\label{foot:radius}}. 

We test the validity of our approach by comparing the resulting temperature skewers with the ones given by the unmodified Nyx run. In Figure \ref{fig:paint_skewers}, we show the temperature along 4 skewers extracted from Nyx. The skewer depicted in each panel passes at an impact parameter of $50 \kpc$ from the same set of  halos shown in Figure \ref{fig:slices}, in increasing order of mass from top to bottom. The solid green lines refer to the temperature given by Nyx, while the dashed black lines indicate the temperature re-computed through our
semi-analytic 
technique.
We plot only the temperature within a velocity window of $\pm 500\; \rm km \, s^{-1}$ around the position of the halo, which is aligned at the center of every panel and marked with a vertical dotted black line. The other two vertical black dashed lines indicate the boundaries of the virial radius around the position of the halo. The discontinuities in the temperature skewers are due to the virial shocks. Our method reproduces the predictions of Nyx within a factor of $2-3$ inside the virial radius of the halos and within a factor of $\sim 5$ around the virial shocks, while Nyx is matched by construction outside the virial radius. The agreement is generally better for lower-mass halos. The overall good agreement between Nyx and our approach proves that our method is a reasonable and simple technique to semi-analytically model the temperature of the CGM. 

Having verified the robustness of our semi-analytic approach, we now use it to construct two different models for the CGM: one with an overall hotter CGM than predicted by Nyx (``Nyx Hot''), and one with a colder CGM (``Nyx Cold''). We define these models by simply adding a constant to $a(n_{\rm H})$ in Equation \eqref{eq:paint}, while keeping $b(n_{\rm H})$ unchanged. Specifically, in the Nyx Hot model, the additive constant is one, so that the temperature of the CGM within the virial radius is generally one dex higher than the predictions given by Nyx. Differences up to one dex in the temperature profiles around halos are also found in other cosmological simulations when comparing runs including feedback with feedback-free runs \citep[e.g.][]{Kollmeier_2006, Stinson_2010, Stinson_2012, Woods_2014, Nelson_2016}. Hence, although our painted Nyx Hot model applied
to the reference Nyx run is clearly not equivalent to a simulation endowed with feedback prescriptions, it does a reasonable job of mimicking the heating caused by AGN feedback.

The ``Nyx Cold'' model is defined by replacing $a(n_{\rm H})$ with $a(n_{\rm H})-1$, in analogy with the Nyx Hot model. The temperature of the CGM within one virial radius is then generally one order of magnitude lower than in Nyx. Physically, this model can be simply representative of an overall cooler CGM, but there are other possible interpretations linked to the physics which is unresolved in the simulations. In fact, clouds cooling below $10^6 \, \rm K$ tend to fragment into clumps of $\sim 20 \, \rm pc$ size, which are clearly unresolved by simulations (\citealt{McCourt_2016}; see also the Appendix \ref{sec:resolution} in this work). Evidence for $<500\,\rm pc$ clumps in the CGM of a \lya emitting galaxy at $z\approx2.5$ has been found by \cite{Crighton_2015}, and \cite{Arrigoni-Battaia_2015} observations of a $z\approx 2.3$  \lya emitting nebula imply the presence of $\sim 20 \, \rm pc$ cool and dense clumps.  
We shall discuss the connections between the predictions of \lya transmission profiles given by Nyx and Illustris and the physics of the CGM in \S~\ref{sec:discussion} (see also the Appendix \ref{app:velocity}).

\section{Results}
\label{sec:results}

\subsection{Simulated Spectra}
\label{sec:skewers}

Before computing any statistics of the \lya absorption lines, we visually inspect a sample of skewers extracted from Illustris, pure Nyx, and Nyx with modified temperature. In Figure \ref{fig:skewers}, the top set of five panels shows different physical quantities along the same skewer drawn from the Nyx run, at $50 \kpc$ from the halo shown in the third row of Figure \ref{fig:slices} ($M_{\rm halo} = 2.16 \times 10^{12}$, $r_{\rm vir} = 122 \kpc$). 
From top to bottom, we plot the gas overdensity with the corresponding number density of hydrogen, the gas temperature with the corresponding Doppler parameter, the \hi column density, the gas velocity field and the transmitted \lya flux. The solid green, dashed blue and dot-dashed red solid lines refer to the pure Nyx run, and to the Nyx Cold and Nyx Hot models, respectively. All quantities are centered at the position of the halo along the line of sight, marked with a vertical black dashed line, and plotted within a velocity window of $\pm 500 \rm km \, s^{-1}$. 

The temperature appears to trace the gas overdensity along the skewer, except for the region within $\sim 220 \kpc$ from the center of the halo. This region is delimited by a steep increase of the temperature, which is due to the virial shock. By construction, the temperature of the three models is the same far from the halo, and presents increasingly larger differences as the halo is approached. In turn, these differences impact also the \hi column density. The gas velocity field is smooth in the IGM regime. The discontinuity seen at the IGM/CGM interface is due to the virial shock. The flux is the most interesting panel, since that is the actual observable that we are interested in. It is remarkable that, even though we are changing the temperature in a small region along the line of sight, we see huge differences in the flux skewers predicted by the different models.

This happens because we observe the \lya forest in redshift space. For the Nyx Cold model dense gas is highly neutral and results in a large column density absorber over the size of the halo, which is grown into a large absorption feature (dashed blue line) by the peculiar velocities and redshift space distortions. By contrast, at the much higher temperatures found in the default Nyx ($T\sim10^6\, \rm K$) and Nyx Hot ($T\sim 10^7\, \rm K$) models, the presence of collisional ionization yields a lower $n_{\rm HI}$, thus reducing the overall absorption.

In the lower set of five panels in Figure \ref{fig:skewers} we plot the same quantities as in the top set of panels, but for a skewer drawn from the Illustris simulation, at $50 \kpc$ from the halo shown in the third row of Figure \ref{fig:slices} ($M_{\rm halo} = 2.16 \times 10^{12}$, $r_{\rm vir} = 122 \kpc$). 
Also in this case, we can distinguish between IGM and CGM regimes thanks to the virial-shock features in the temperature and gas velocity skewers, although the transition is smoother if compared with Nyx. The hot region around the halo is much more extended in the case of Illustris. Its boundaries span a length of $\sim 870 \kpc$, whereas the corresponding region for the skewer drawn form Nyx is $\sim 420 \kpc$. 
These features are not peculiar to the skewers that we showed in Figure \ref{fig:skewers} as an example. Indeed, as we shall discuss in detail in \S~\ref{sec:discussion}, the hot component of the CGM generally spreads farther from the center of halos in Illustris than in Nyx. 

\subsection{Comparison with Observations}

In this section, we compute various statistics of our sample of flux skewers, and compare them with observations of the \lya absorption in the CGM of QSOs, DLAs and LBGs. We verified that possible errors in modeling the mass and redshift distribution of foreground halos, as well as the size of the sample of observed and simulated spectra and limitations due to the resolution of the simulations would not change the main conclusions of this work. We thus focus on the presentation of our results, leaving a detailed discussion about the aforementioned systematics to the Appendices \ref{app:cell_size} and \ref{app:systematics}.

\subsubsection{Quasar Hosts}
\label{sec:qso}

\begin{figure}[]
  \centering
   \includegraphics[width=\columnwidth]{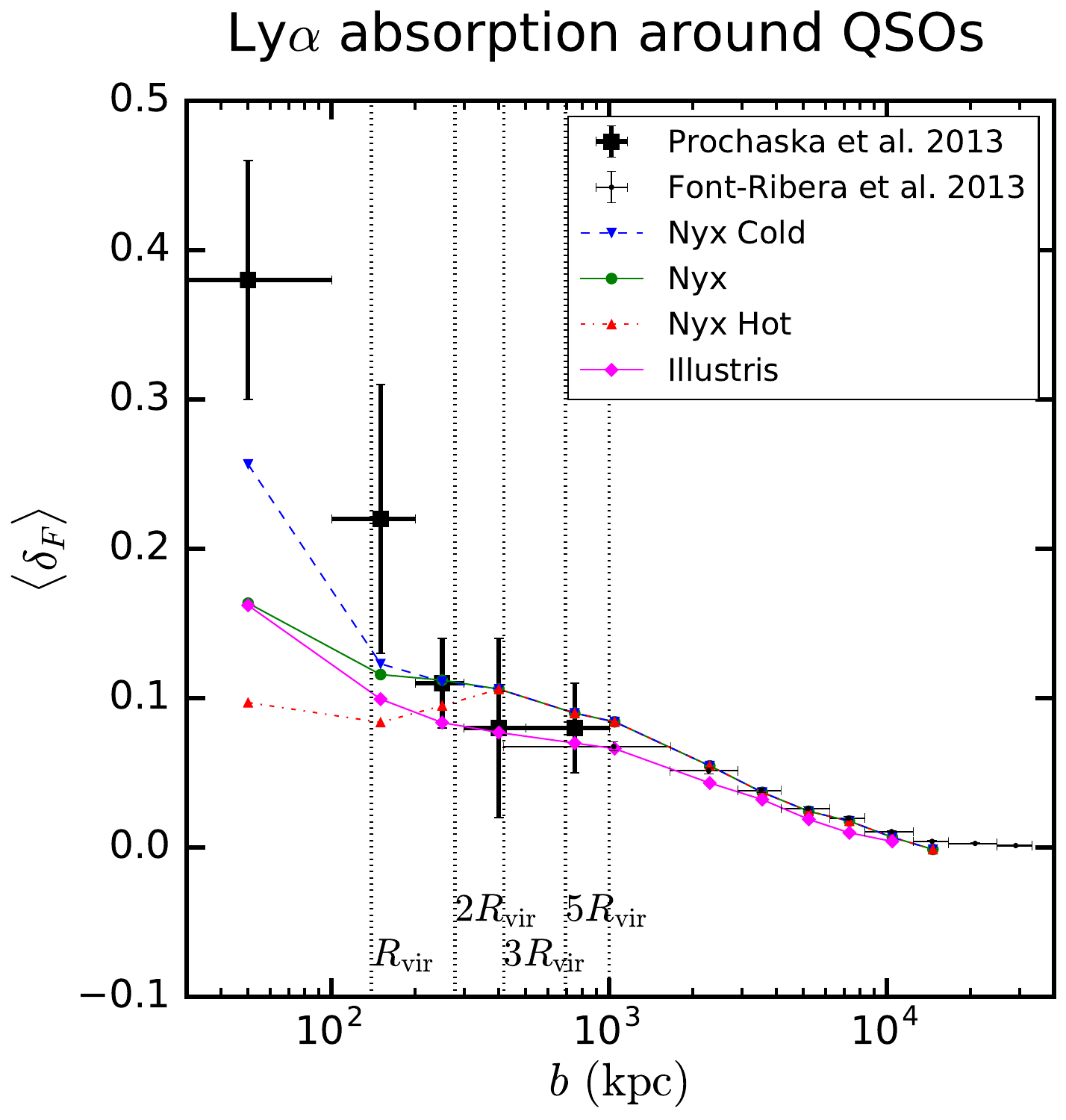}
  \caption{Mean \lya flux contrast at different impact parameter bins ($b$) from a foreground QSO (halo mass $M > 10^{12.5} \, M_{\odot}$ and $M > 10^{12.4} \, M_{\odot}$ in Nyx and Illustris, respectively), with respect to the mean flux in the IGM. The black squares are the measurements by \cite{Prochaska_2013}, while the black circles are obtained from the measurements of the QSO-\lya cross-correlation function by \cite{Font-Ribera_2013}. Vertical error bars represent the 1$\sigma$ errors of the measurements, which are smaller than the size of the marker for some of the \cite{Font-Ribera_2013} data points. The magenta diamonds and green circles, linearly interpolated with solid line with the same colors, are the results obtained with Illustris and Nyx, respectively. The red triangles connected with the dot-dashed red line and the blue reversed triangles linearly interpolated with the blue dashed line refer to the Nyx Hot and Nyx Cold models, respectively (see text for details). The vertical black dotted lines mark 1, 2, 3 and 5 times the virial radius corresponding to the minimum mass of the sample of halos considered in Nyx ($r_{\rm vir}=140\kpc$, $M_{\rm min}=10^{12.5}\, M_{\odot}$), as well as the $1 \Mpc$ boundary (see the discussion in \S~\ref{sec:T-nH_CGM}). While being consistent with the observations outside the virial radius, all simulations struggle at reproducing the data at small impact parameter, indicating a lack of \hi absorption in the CGM of quasar hosts in the simulations considered.} \label{fig:mean_flux}
\end{figure}

We want to compare the mean flux versus the impact parameter predicted by the models considered with the observations by \cite{Prochaska_2013} and \cite{Font-Ribera_2013}. The two measurements probe different ranges of impact parameter, complementing each other. 

\cite{Prochaska_2013} considered a sample of 650 projected QSO pairs in the redshift range $2< z<3$ and with a transverse separation $<1 \Mpc$. For every spectrum, they measured the \lya flux contrast, defined as
\begin{equation} \label{eq:delta_F}
	\delta_F=1-\frac{\langle F \rangle_{\Delta v} }{\bar{F}  _{\rm IGM}} \, ,
\end{equation}
where $\langle F \rangle_{\Delta v}$ is the mean flux measured within a line-of-sight velocity window of $\pm 1000 \, \rm km \, s^{-1}$ around the foreground quasar and $\bar{F} _{\rm IGM}$ is the mean flux of the IGM at the redshift of the foreground object. Thus, they divided the spectra into five bins of impact parameter and determined the mean \lya flux contrast across all spectra in each bin, $\langle \delta_F \rangle$. In this way, they obtained $\langle \delta_F \rangle$ as a function of the impact parameter. We report these measurements in Figure \ref{fig:mean_flux} as black squares. The vertical error bars are the 1$\sigma$ errors of the measurements, whereas the horizontal bars show the bin widths.

\cite{Font-Ribera_2013} considered a sample of $\sim 6 \times 10^4$ QSO spectra from BOSS. They measured the QSO-\lya cross-correlation function in different bins of transverse and line-of-sight separation. This is actually equivalent to measuring the mean \lya flux profile, as in \cite{Prochaska_2013}. Indeed, averaging their estimate of the cross-correlation over the the line-of-sight bins corresponding to a velocity window of $\pm 1000 \, \rm km \, s^{-1}$, we can infer the corresponding mean flux 
contrast 
as a function of the transverse separation between QSOs, allowing us
to directly compare their large-scale measurement with the equivalent small-scale
observations from \cite{Prochaska_2013}.
We plot the resulting $\langle \delta_F \rangle$ profile in Figure \ref{fig:mean_flux} with black circles. The vertical bars are the 1$\sigma$ errors of the measurements, whereas the horizontal bars show the bin widths. For some of the \cite{Font-Ribera_2013} data points the error bars are smaller than the size of the marker. We list \cite{Font-Ribera_2013} data in Table \ref{tab:Font-Ribera13} of the Appendix \ref{app:BOSS_conversion}, where we also provide the details of our conversion. For convenience, we also list \cite{Prochaska_2013} data points in Table \ref{tab:Prochaska13} of the Appendix \ref{app:BOSS_conversion}. 
 For the first time, we show that \cite{Font-Ribera_2013} measurements are consistent with \cite{Prochaska_2013}, extending the dynamic range probed by the \lya absorption lines from the CGM out to the IGM. As such, these measurements have the potential to jointly constrain the physics of both CGM and IGM.

We plot the results of the simulations in Figure \ref{fig:mean_flux}, with a point in the center of each impact parameter bin. To guide the eye, we linearly interpolate between the points. The magenta diamonds and the green circles, connected with solid lines of the same colors, refer to Illustris and the pure Nyx run, respectively. The red triangles connected with a dot-dashed red line and the blue reversed triangles interpolated with a dashed blue line represent the Nyx Hot and Nyx Cold models, respectively. The simulations do not extend to the outermost data points because the size of their boxes is not large enough. The vertical dotted black lines correspond to one, two, three and five times the virial radius corresponding to the minimum mass of the sample of halos considered in Nyx ($10^{12.5}\, M_{\odot}$), as well as the $1 \Mpc$ boundary (it will be useful in the discussion in \S~\ref{sec:T-nH_CGM}). 
The mean flux contrast 
predicted by the simulations in each impact parameter bin is obtained by averaging the values of $\delta_F$ obtained from $5\times 10^4$ skewers, selected as explained in \S~\ref{sec:sample}, at the median redshift of the observations $(z \approx 2.4$). We verified that the scatter in the
predictions given by samples of this size results in a relative error of
$< 3\%$. 

All models considered are broadly consistent with the observations outside the virial radius. On scales $> 2\Mpc$, data prefer the Nyx model. The good but not perfect agreement between the simulations and BOSS data should not be worrisome, since the predictions of $\langle \delta_F \rangle$ depend on the exact modeling of the foreground objects, in terms of mass cut and redshift of the halos extracted from the simulations (see the Appendices \ref{sec:halo_mass} and \ref{sec:redshift}): in fact, we verified that when these uncertainties in the modeling are taken into account the predictions of simulations are compatible with most data points, which tend to prefer Nyx over Illustris (Figure \ref{fig:largeb} in the Appendix \ref{sec:redshift}). Furthermore, making slightly different assumptions than the ones adopted in this work when computing the mean \lya flux contrast profile from \cite{Font-Ribera_2013} measurements would impact the values inferred for the $\langle \delta_F \rangle$ data points at $b> 1\Mpc$, and thus the agreement with the simulations (see the Appendix \ref{app:BOSS_conversion}).
Nevertheless, the not perfect match between the simulations and the highly precise BOSS data underlines the need for even further improvements in cosmological simulations. 

For $b\gtrsim 400 \kpc$, the Nyx Hot and Nyx Cold models give exactly the same predictions as Nyx because, by construction, the temperature-density relationship is modified only within the virial radius of the foreground object\footnote{The minimum mass of the sample is $10^{12.5}\, M_{\odot}$, corresponding to $140 \kpc$. Since the sample contains halos as massive as $10^{13.7}\, M_{\odot}$, the Nyx Hot and Nyx Cold models can differ with respect to Nyx up to  $b\sim350\kpc$. As we mentioned in Footnote \ref{foot:radius}, our semi-analytic technique can affect the temperature in the virial radius of the selected halos up to $\sim 1.3$, and this explains the differences observed in Figure \ref{fig:mean_flux} up to $b\sim 400 \kpc$.}. No Nyx-based model nor Illustris reproduce the mean flux contrast within the virial radius. This result underlines the fact that simulations considered do not produce enough \hi absorption in this range of distance. One possible reason is that the simulated gas in the CGM is too hot. Indeed, the Nyx Cold model is much closer to the data than the Nyx and Nyx Hot ones. Nevertheless, this is only one possibility to explain the discrepancy. In \S~\ref{sec:discussion} we shall discuss other possible solutions.

Nyx and Illustris give different predictions for the mean flux
contrast between $200 \kpc$ and $2 \Mpc$ from the foreground
QSO, well beyond the virial radii of the halos in the sample considered ($140 \kpc - 350 \kpc$).
This stems  from the different temperature-density relationship
of the gas in the CGM of Illustris and Nyx halos, which
we discuss in detail in \S~\ref{sec:discussion}. Whereas for $200 \kpc \lesssim b \lesssim 500
\kpc$ the error bars in \cite{Prochaska_2013} measurements are
too large to rule out either simulations, the data point at $1 \Mpc$ validates Illustris, while being discrepant with Nyx. 
This result shows that, with the current exquisitely precise \lya absorption BOSS data, it is already possible to tightly constrain physical models implemented in simulations (e.g. feedback) at relatively large separations from foreground QSOs.
One possible explanation for the discrepancy is that the impact of radiation from nearby AGN on self-shielding and ionization (transverse proximity effect; \citealt{Dobrzycki_1991, Adelberger_2004}) is neglected in the simulation, whereas this effect is included in Illustris. This radiation increases the transmitted flux (i.e. decreases $\langle \delta_F \rangle$), therefore Nyx overpredicts the \lya flux contrast.

We stress that at large impact parameters the observations tend to an asymptotic trend, i.e. to the mean \lya flux in the IGM. The simulations considered reproduce this trend accordingly, underscoring the success of the $\Lambda$CDM model in describing the IGM. The physics underlying the \lya forest is simple and well-understood, so one should take these large-scale measurements as a starting point to extend the investigation of \lya absorption towards smaller impact parameters.

\subsubsection{Damped \lya Absorbers}
\label{sec:dla}

\begin{figure}[]
  \centering
   \includegraphics[width=\columnwidth]{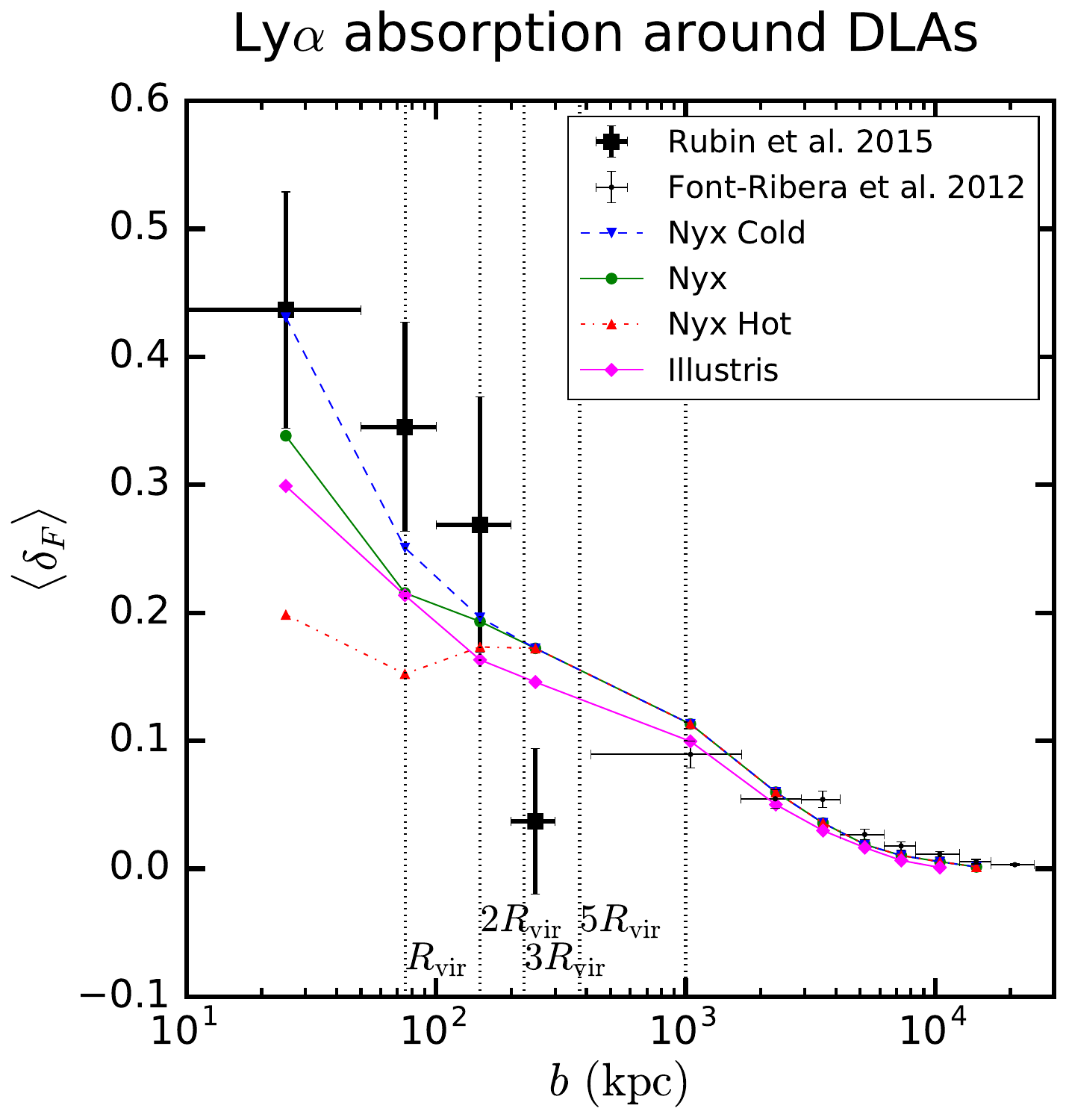}
  \caption{Mean \lya flux contrast at different impact parameter bins ($b$) around DLAs (halo mass $M > 10^{11.7} \, M_{\odot}$ and $M > 10^{11.6} \, M_{\odot}$ in Nyx and Illustris, respectively). The black squares are the measurements by \cite{Rubin_2015}, while the black circels are obtained from the measurements of the DLA-\lya correlation function by \cite{Font-Ribera_2013}. Vertical error bars represent the 1$\sigma$ errors of the measurements, which are smaller than the size of the marker for some of the \cite{Font-Ribera_2012b} data points. The magenta diamonds and green circles, linearly interpolated with solid line with the same colors, are the results obtained with Illustris and Nyx, respectively. The red triangles connected with the dot-dashed red line and the blue reversed triangles linearly interpolated with the blue dashed line refer to the Nyx Hot and Nyx Cold, respectively (see text for details). The vertical black dotted lines mark 1, 2, 3 and 5 times the virial radius corresponding to the minimum mass of the sample of halos considered in Nyx ($r_{\rm vir}=75\kpc$, $M_{\rm min}=10^{11.7}\, M_{\odot}$), as well as the $1 \Mpc$ boundary (see the discussion in \S~\ref{sec:T-nH_CGM}). Except for the Nyx Hot model, all simulations are broadly consistent with the data. There is tension with the data in the outermost bin of \cite{Rubin_2015}.} \label{fig:mean_flux_DLA}
\end{figure}

We now compare the observations of \lya absorption around DLAs by \cite{Rubin_2015} and \cite{Font-Ribera_2012b} with the predictions given by the simulations considered in this work.

\cite{Rubin_2015} considered a sample of 40 DLAs in the redshift range $1.6<z<3.6$, intervening along the line of sight of a background QSO, and passing at different impact parameters from another 
background QSO. They stacked the absorption spectra in four impact parameter bins, and measured the equivalent width of the \lya absorption feature within a velocity window of $\pm 500 \, \rm km \, s^{-1}$ around the DLA. We converted the measured equivalent width in each bin into the corresponding mean flux contrast. The results are plotted in Figure \ref{fig:mean_flux_DLA} as black squares. The vertical bars indicate the 1$\sigma$ errors of the measurements, while the horizontal bars are the bin widths. We report the inferred mean flux contrast in Table \ref{tab:Rubin15} of Appendix \ref{app:Rubin}, where we also explain the details of our conversion of the \cite{Rubin_2015} measurement.

\cite{Font-Ribera_2012b} considered a sample of $\sim 5 \times 10^4$ QSO spectra from the 9th Data Release of BOSS, and a subsample of $\sim 10^4$ DLAs from the catalogue by \cite{Noterdaeme_2012}. They measured the cross-correlation of \lya forest absorption and DLAs, in different bins of transverse and line-of-sight separation. As we did for the QSO-\lya cross-correlation in \S~\ref{sec:qso}, we convert the measurements by \cite{Font-Ribera_2012b} into a mean flux contrast profile (see Appendix \ref{app:BOSS_conversion} for details). The results, shown as black circles in  Figure \ref{fig:mean_flux_DLA} and reported in Table \ref{tab:Font-Ribera12} of Appendix \ref{app:BOSS_conversion}, extend the observations by \cite{Rubin_2015} to $16.8 \Mpc$. The vertical error bars represent the 1$\sigma$ errors of the measurements, whereas the horizontal bars the widths of the impact parameter bins. The BOSS measurements are much more precise than \cite{Rubin_2015} observations because of the much larger DLA and QSO samples. There seems to be a statistical fluctuation in the \cite{Rubin_2015} data point at $\sim350 \kpc$, which appears to be inconsistent with the otherwise smooth trend that would be inferred connecting all other data points in Figure \ref{fig:mean_flux_DLA}.

We overplot the predictions given by our simulations with the same marker styles, line styles and colors as in Figure \ref{fig:mean_flux}. Following the same approach as in \S~\ref{sec:qso}, the mean flux contrast 
predicted by the simulations in each impact parameter bin is obtained by averaging the values of $\delta_F$ obtained from $5\times 10^4$ skewers, selected as explained in \S~\ref{sec:sample}, at the median redshift of the observations $(z \approx 2.4$).
We verified that the scatter in the
predictions given by samples of this size results in a relative error of
$< 1.5\%$. 
Nyx, Nyx Hot, and Nyx Cold give the same predictions for $b\gtrsim 300 \kpc$, by construction\footnote{The minimum mass of the sample is $10^{11.7}\, M_{\odot}$, corresponding to $75 \kpc$. Since the sample contains halos as massive as $10^{13.7}\, M_{\odot}$, the Nyx Hot and Nyx Cold can actually differ with respect to Nyx up to  $b\sim 350\kpc$.}. The Nyx Cold model matches the observations in the innermost bin of transverse distance, where Nyx is almost consistent with the $1\sigma $ error bar of the measurements. Instead, Nyx Hot and Illustris underpredict the value of $\langle \delta_F \rangle$ within the same bin. 
Apart from the already discussed data point at $\sim350 \kpc$, there appears to be tension with the 
\cite{Font-Ribera_2012b} measurement at $\sim 1 \Mpc$. 
Illustris is consistent
within $\sim1.5\sigma$, while Nyx within $\sim2.5\sigma$. Similar to what obtained for the $\langle \delta_F \rangle$ profile around QSOs, the slight discrepancies between simulations and \cite{Font-Ribera_2013} data at $b\gtrsim 5 \Mpc$ could be accounted for by the exact definition of the sample of halos extracted from the simulations and the assumptions behind the inference of the $\langle \delta_F \rangle$ profile from BOSS data (see Appendices \ref{sec:halo_mass}, \ref{sec:redshift} and \ref{app:BOSS_conversion}). 

Also our choice of placing DLAs at the center of their host halos could impact the agreement or disagreement between observations and simulations. If DLAs were given an offset from the center, as it would be the case for proto-galactic clumps, that would presumably decrease the absorption on small scales, and possibly slightly increase it on intermediate and large scales. As such, the discrepancies of Nyx and Illustris with \cite{Rubin_2015} within the virial radius may be underestimated, underscoring one more time the challenges for models of galactic physics posed by small-scale measurements.

We notice that the largest differences between the predictions of Nyx and Illustris arise in the range $100 \kpc \lesssim b \lesssim 1 \Mpc$. Improving the precision of the measurements of the $\delta_F$ in this range with future observations would allow setting meaningful constraints on the physics of the gas in the CGM and at the CGM/IGM interface, as well as on feedback prescriptions implemented in simulations. In this case, a more refined criterion for assigning DLAs to their host halos may be necessary.

\subsubsection{Mean \lya Transmission Profile around LBGs}
\label{sec:lbg}

\begin{figure*}[]
\begin{center}
	{\Large \lya absorption around LBGs}
\end{center}
  \centering
   \includegraphics[width=0.49\textwidth]{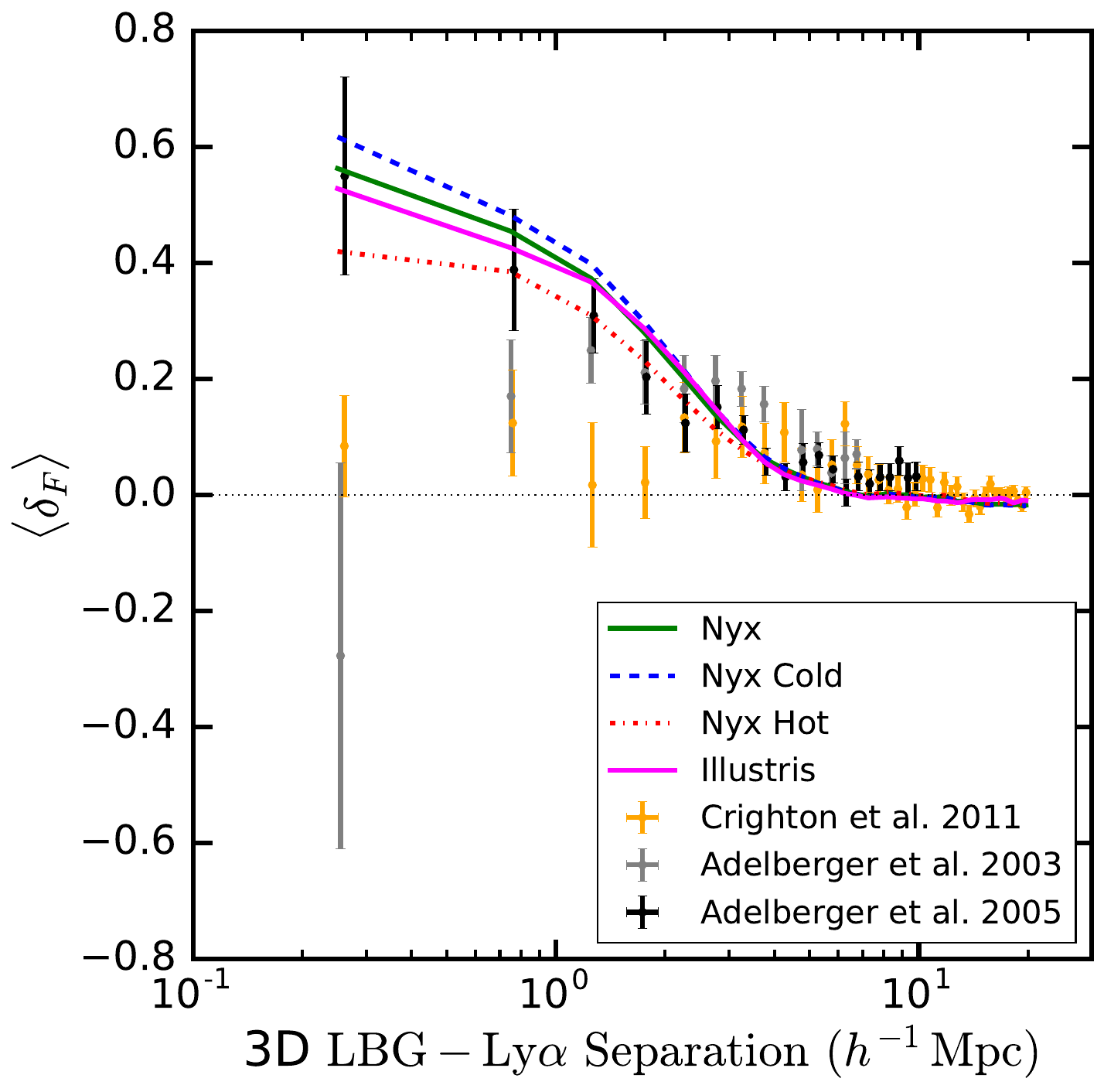}
   \includegraphics[width=0.49\textwidth]{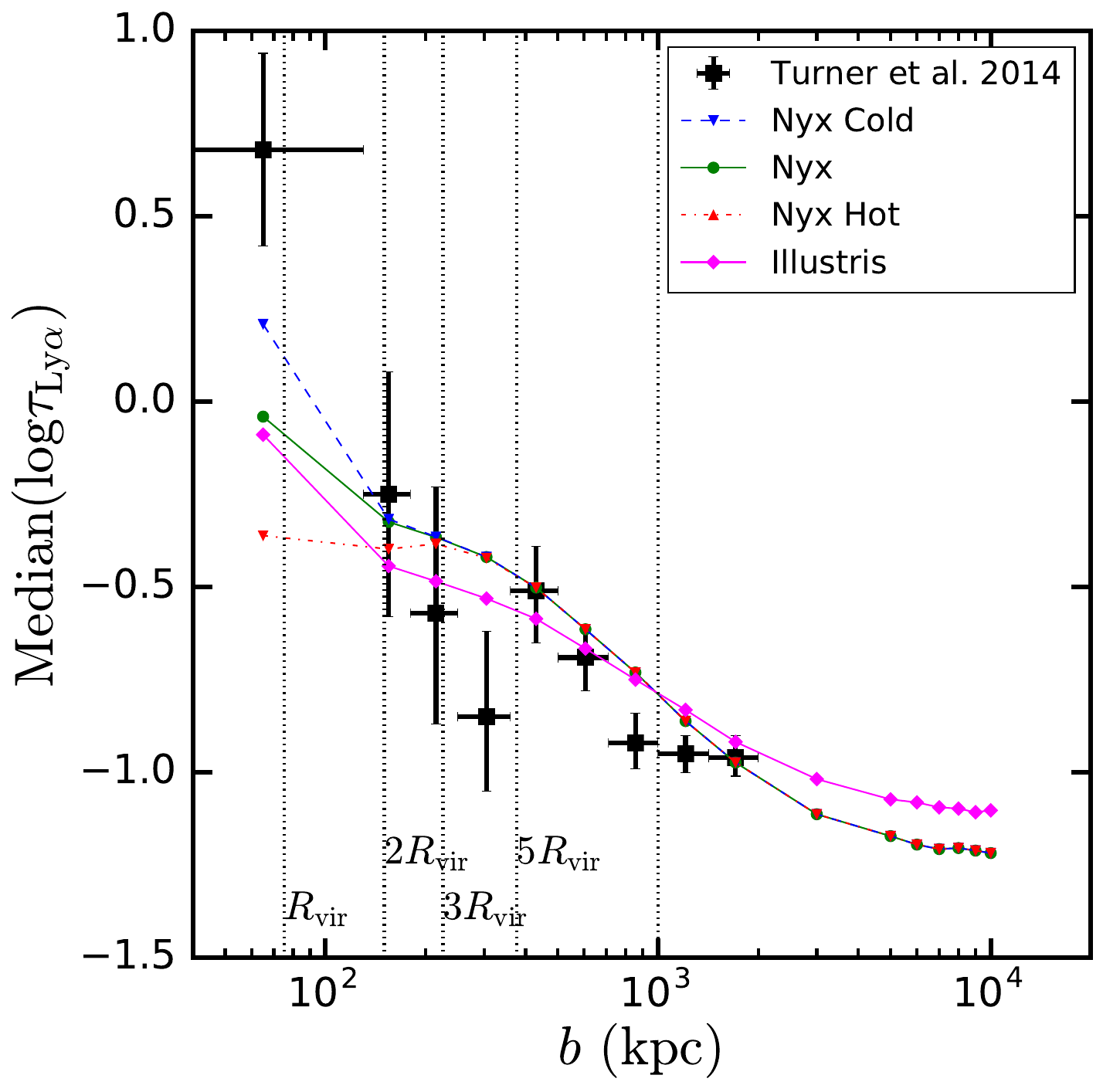}
  \caption{\textit{Left panel}: Mean \lya flux contrast at different 3D separations from the foreground LBG (halo mass $M > 10^{11.7} \, M_{\odot}$ and $M > 10^{11.6} \, M_{\odot}$ in Nyx and Illustris, respectively). The orange squares, grey and black points are the measurements by \cite{Crighton_2011}, \cite{Adelberger_2003} and \cite{Adelberger_2005}, respectively. The solid magenta, solid green, dot-dashed red and the dashed blue lines Illustris, Nyx, Nyx Hot and Nyx Cold models, respectively (see text for details). 
  \textit{Right panel}: Median \lya optical depth of HI in the CGM of foreground LBGs (halo mass $M > 10^{11.7} \, M_{\odot}$ and $M > 10^{11.6} \, M_{\odot}$ in Nyx and Illustris, respectively). The black squares are the measurements by \cite{Turner_2014}. The magenta diamonds and green circles, linearly interpolated with solid lines of the same colors, are the results obtained with Illustris and Nyx, respectively. The red triangles connected with the dot-dashed red line and the blue reversed triangles linearly interpolated with the blue dashed line refer to the Nyx Hot and Nyx Cold models, respectively (see text for details). 
The vertical black dotted lines mark 1, 2, 3 and 5 times the virial radius corresponding to the minimum mass of the sample of halos considered in Nyx ($10^{11.7}\, M_{\odot}$), as well as the $1 \Mpc$ boundary (see the discussion in \S~\ref{sec:T-nH_CGM}).
} \label{fig:tau_median}
\end{figure*}

We consider the measurements of the HI \lya transmission profile in the CGM of LBGs by \citealt{Adelberger_2003} ($z\sim3$), \citealt{Adelberger_2005} (mean redshift $z\approx2.5$) and \citealt{Crighton_2011} ($z\sim3$). Since these measurements are at different redshifts, we renormalize the transmission profiles so that they asymptote to the mean flux of the IGM ($\bar{F}_{\rm IGM}=0.76$, following \citealt{Crighton_2011}).

We convert the transmissivities into $\langle \delta_F \rangle$ profiles, and report them in the left panel of Figure \ref{fig:tau_median}. Grey circles, black circles, and orange squares refer to the measurements by \cite{Adelberger_2003}, \cite{Adelberger_2005} and \cite{Crighton_2011}, respectively. Unlike Figures \ref{fig:mean_flux} and \ref{fig:mean_flux_DLA}, the $x$-axis represents the 3D distance from the LBG, and not the impact parameter. In the observations, the 3D distance between each LBG and \lya absorption feature is determined from their measured angular separation, and their comoving distances from the observer. The latter are inferred from the measurement of the redshifts of the LBG and the absorption line, under the assumption of a pure Hubble flow. Therefore, the 3D distance estimated in the measurements differs from the real distance, due to velocity flows and redshift space distortions.

To reproduce the observations, we considered the halos with mass $M>10^{11.5} \, M_{\odot}$ in the Nyx (Illustris) snapshot at $z=3$ ($z=3.01$). The mass threshold has been determined as explained in \S~\ref{sec:sel_halos}. Around the selected halos, we considered the same impact parameter bins adopted to reproduce the measurements by \cite{Font-Ribera_2012b} and \cite{Rubin_2015}. We drew a sample of $5\times 10^4$ skewers in each impact parameter bin as explained in \S~\ref{sec:sample}, and re-normalized the mean flux of each sample to $\bar{F}_{\rm IGM} =0.76$ (following \citealt{Crighton_2011}). We binned the pixels in all skewers according to the their radial distance from the foreground object, using the same binning adopted in the observations that we want to reproduce. We then computed the mean flux contrast in each radial bin.

We plot the predictions given by the simulations and models considered
in this work in the left panel of Figure \ref{fig:tau_median}. The
results of Illustris, Nyx, Nyx Cold and Nyx Hot are plotted with the
solid magenta, solid green, dashed blue and dot-dashed red lines,
respectively. We verified that the scatter in $\langle \delta_F
\rangle$ arising from the choice of different samples of skewers is
negligible. The Nyx Cold and Nyx Hot models give different predictions
than Nyx at distances $\lesssim 4 \, h^{-1} \Mpc$. This may seem
somewhat puzzling, since in the Nyx-painted models the temperature of
the CGM is altered only within the virial radius. However, we reiterate 
that the statistic considered is different from the one discussed in \S~\ref{sec:qso} and
\S~\ref{sec:dla}. Even if in the Nyx Cold and Nyx Hot models we change the temperature
of the default run within the virial radius, peculiar velocities and virial motions in the halo of order $\rm km \, s^{-1}$ 
extend the influence of the galaxies environment out to $\rm Mpc$ scales (see upper axis of Figure
\ref{fig:skewers}).

All models are consistent with all observations at separations $\gtrsim 5 h^{-1}  \cMpc$, except for the tension with \cite{Adelberger_2005} at $\gtrsim 7 h^{-1} \cMpc$. Between $3 h^{-1}  \cMpc$ and $5 h^{-1}  \cMpc$ the observations by \cite{Adelberger_2003} are harder to reproduce, while below $2 h^{-1} \cMpc$, all models struggle
to reproduce the \cite{Crighton_2011} measurements. The innermost bin of the observations by \cite{Adelberger_2005} is consistent with all simulations, while the data between $700 h^{-1}  \ckpc$ and $2 h^{-1}  \cMpc$ are best reproduced by the Hot Nyx model.

Overall, the results in the left panel of Figure \ref{fig:tau_median} may seem somewhat in contradiction with the findings discussed in \S~\ref{sec:qso} and \S~\ref{sec:dla}, which were generally favoring a cooler CGM. However, it is hard to compare the radial profile of the \lya transmission with the mean flux contrast as a function of the transverse separation. Indeed, at a fixed 3D distance $R$, one probes the \lya absorption of HI at all impact parameters $b<R$. Because gas at small impact parameters can impact the absorption at larger radii due to Hubble flow velocities, the radial profile of $\langle \delta_F \rangle$ does not separate the physical effects occurring transverse and parallel to the line of sight. 
Moreover, a well-posed comparison among the data in Figures \ref{fig:mean_flux}, \ref{fig:mean_flux_DLA} and \ref{fig:tau_median} is not really possible unless we know the distribution of impact parameters in the observations. 

A further reason why it is hard to interpret the results shown in this section is that the measurements considered are not everywhere consistent with one another within the error bars, and it is not obvious to understand which one is the most reliable.  As pointed out by \cite{Crighton_2011}, their measurements are more precise at large separations, while the data by \cite{Adelberger_2003, Adelberger_2005} should me more reliable at small distance from the LBG. The high transmission in the innermost bin in \cite{Adelberger_2003} was interpreted as a bubble of ionized gas around the foreground LBG, but that result was retracted by \cite{Adelberger_2005}. Furthermore, \cite{Crighton_2011} claimed that the error on their measurement in the innermost bin may be underestimated. In conclusion, given these limitations it seems that the $\langle \delta_F \rangle$ profile as a function of the 3D separation from the LBG is not the optimal statistic to use in order to constrain simulations.

\subsubsection{Median \lya Optical Depth around LBGs}

Building on previous work by \cite{Rakic_2012}, \cite{Turner_2014} considered a sample of 854 foreground LBGs at redshift $z\approx2.4$, and studied the \lya and metal absorption in their CGM exploiting spectra of background QSOs. They determined the median \lya pixel optical depth within a velocity window of $\pm170 \, \rm km \, s^{-1}$ around the LBG, as a function of the impact parameter. We report their measurements as black squares in the right panel of Figures \ref{fig:tau_median}. The vertical bars indicate the 1$\sigma$ errors in the measurements, whereas the horizontal bars the bin width.

We overplot the results of the models considered in this work at the same redshift of the observations, with the same colors, markers and line styles as in the left panel of Figure \ref{fig:tau_median}. Nyx and Illustris are generally consistent with the observations. Both simulations underpredict the optical depth in the innermost bin. The point at $\sim300 \kpc$ falls $1.3\sigma$ and $2\sigma$ below Illustris and Nyx models, respectively, and there is some tension between simulations and measurements in the range $(800,\,1000)\kpc$. 
Once again, the observations within the virial radius favor a cooler CGM, even though our Cold Nyx model still underpredicts the data. 
The Nyx Hot and Nyx Cold models give the same predictions as Nyx at impact parameters $b\gtrsim 300 \kpc$, by construction.

At impact parameters larger than $2 \Mpc$, the simulations tend to an asymptotic value, which is the median optical depth of the IGM. At first glance, it might seem puzzling that these asymptotic values are different, given that we have tuned the UVB to give the same mean flux for Illustris
and NyX (see Figures \ref{fig:mean_flux} and \ref{fig:mean_flux_DLA}, and the discussion in \S~\ref{sec:sim_absorption}). 
However,  the flux PDFs (in the ambient IGM) for Illustris and Nyx are sufficiently
different that, although they give the same mean flux, the median optical
depths are in fact slightly different.

\section{Discussion}
\label{sec:discussion}

In the previous section, we highlighted how the profile of the mean \lya flux contrast can constrain simulations over a large dynamic range in impact
parameter ( $20 \kpc \lesssim b\lesssim  20 \Mpc$) spanning the CGM, the CGM-IGM
interface, and the IGM.
We also pointed out that both Nyx and Illustris underpredict the \lya absorption within the virial radius of QSOs and LBGs. 
To gain physical insight into the reasons behind this, we painted different temperature-density relationships on top of Nyx, hence obtaining significantly different
predictions of the mean flux contrast within the virial radius. Clearly, the temperature-density relationship in the CGM has a strong impact on the resulting \lya absorption, and is worth discussing further.

On the one hand, at higher temperatures ($T > 10^5\, \rm K$) collisional ionization 
becomes very important, decreasing neutral fraction and the corresponding \lya 
absorption. On the other hand, the density of gas in the CGM can overcome the nominal 
self-shielding threshold ($\sim 6.0 \times 10^{-3} \, \rm cm^{-3}$; \citealt{Rahmati_2013}), 
leading to increased neutral fraction and higher Ly-a absorption. 
Since we consider photoionization, collisional ionization and self-shielding in our modeling of 
the \lya absorption, investigating the temperature-density relationship of the gas within and outside 
the virial radius will provide us with good insight in the physics captured by the simulations. 

\subsection{Temperature-Density Relationship in the IGM}
\label{sec:T-nH_IGM}

 \begin{figure*}[]
  \centering
   \includegraphics[width=\textwidth]{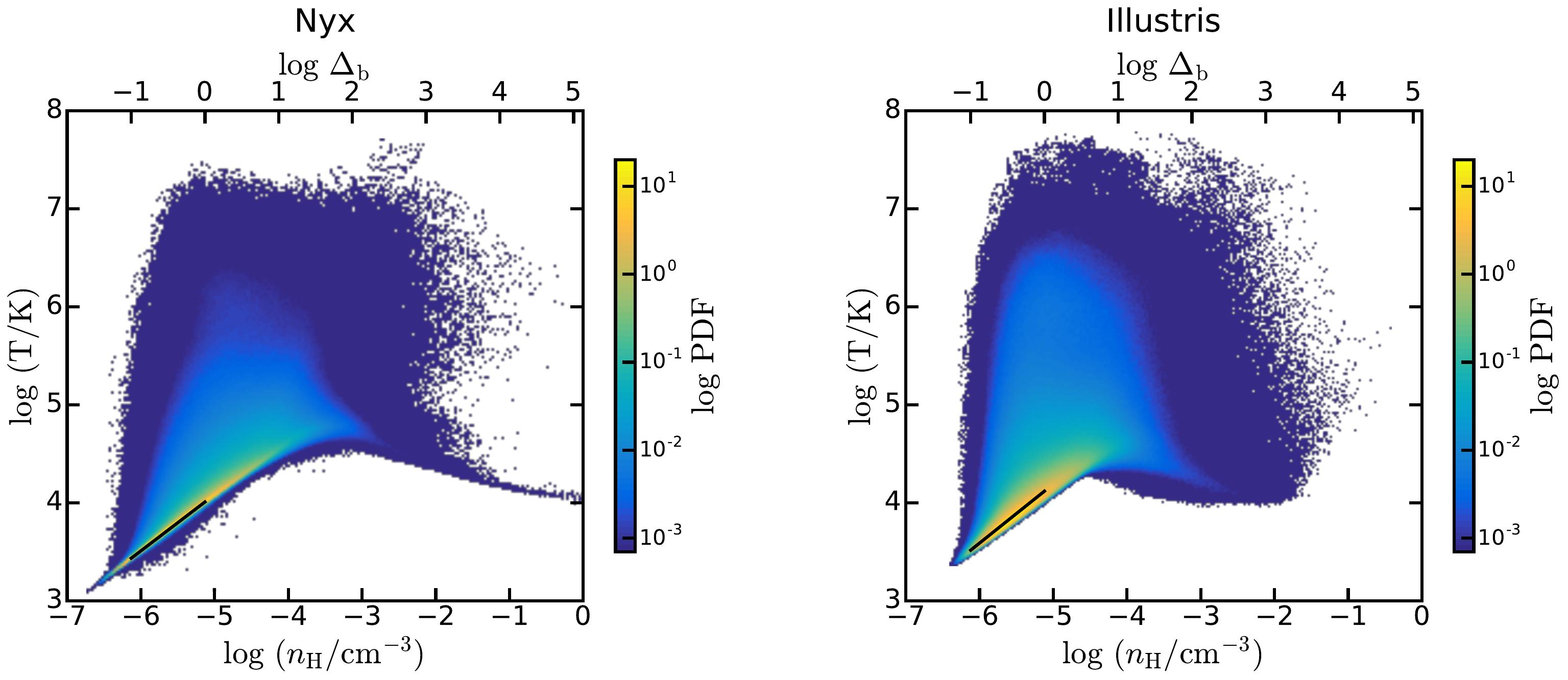}
  \caption{2D histogram of hydrogen temperature and density for a sample of $10^5$ skewers located at random positions in the Nyx (left panel) and Illustris (right panel) runs at $z=2.4$ and $z=2.44$, respectively. The color bar shows the logarithm of the resulting PDF. The $x$-axis reports the gas overdensity (top) and the corresponding neutral hydrogen density (bottom). The black line is the best-fit power-law $T=T_0 \Delta_{\rm b} ^{\gamma-1}$ across the region of the plot in the IGM regime ($-1<\log_{10} \Delta_{\rm b} <0$). The temperature-density relationships of the IGM in the two simulations are consistent with each other, meaning that they give a similar description of the IGM.} \label{fig:T-nH_IGM}
\end{figure*}

We want to check whether Nyx and Illustris give a consistent description of the temperature-density relationship of the IGM.
 We start by
extracting $10^5$ skewers in each simulation, at random positions
and
parallel to one sides of the box. We then construct the volume-weighted
2D-histogram of density and temperature of hydrogen. In the left and
right panels of Figure \ref{fig:T-nH_IGM} we plot the results obtained
for Nyx and Illustris, respectively. Although the global shape of the
temperature-density relationship is similar in both simulations,
Illustris produces a larger amount of hot ($10^{5.5} \, {\rm K} < T <
10^7 \, {\rm K}$), rarefied ($10^{-6}\,{\rm cm^{-3}} < n_{\rm H} <
10^{-5} \, {\rm cm^{-3}}$) gas. 

We then want to find the best-fit power law describing the temperature-density relationship \citep{Hui_1997}. We divide the gas cells into two bins centered at the gas overdensities $\Delta_{\rm b\,1}=10^{-1}$ and $\Delta_{\rm b\,2}=1$, with a bin width of $5\%$ around the central value. We then compute the median temperature $T^{\rm med}_1$ and $T^{\rm med}_2$ of the gas cells in the two overdensity bins centered in $\Delta_{\rm b\, 1}$ and $\Delta_{\rm b\,2}$, respectively. We determine the power law $T=T_0 \Delta_{\rm b} ^{\gamma-1}$ passing through the points $(\Delta_{\rm b\,1},\,T^{\rm med}_1)$ and $(\Delta_{\rm b\,2},\,T^{\rm med}_2)$. 

We obtain $(T_0,\,\gamma)=(10^{4.01} \K,\,1.57)$ and
$(T_0,\,\gamma)=(10^{4.12} \K,\,1.60)$ for Nyx and Illustris,
respectively. 
These subtle differences in the temperature-density relationship arise from minor
differences in the UVB model adopted and the respective reionization histories
of the simulations \citep{Onorbe_2017}.  Although Illustris and Nyx do not adopt the same model for the UVB, the values of the photoionization rate after matching the mean flux in the IGM are quite similar. At $z=2.4$, it is $1.2 \times 10^{-12}\,\rm s^{-1}$ and $1.1 \times 10^{-12}\,\rm s^{-1}$ for Nyx and Illustris, respectively.

The fact that the temperature-density relationships are very similar in Nyx and Illustris means that the temperature-density structure of the IGM is well matched between the two simulations. However the different predictions of the mean \lya flux for impact parameters $\lesssim 5\Mpc$ suggest that the temperature and density structure
of the CGM and CGM-IGM interface are in fact different between the two simulations.

\subsection{Radial Temperature and Density Profiles}
\label{sec:profiles}

\begin{figure*}[]
\centering
   \includegraphics[height=5.3cm]{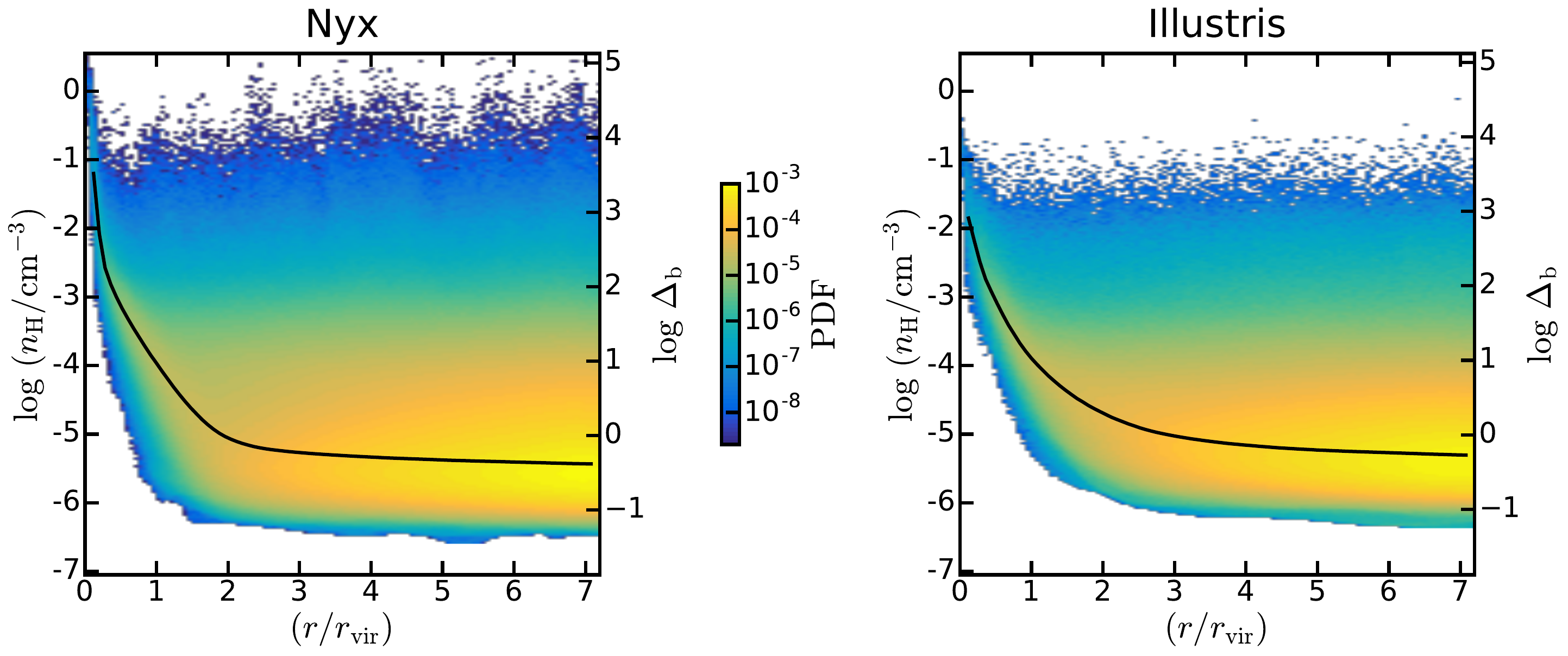}
   \hfill
   \includegraphics[height=5.3cm]{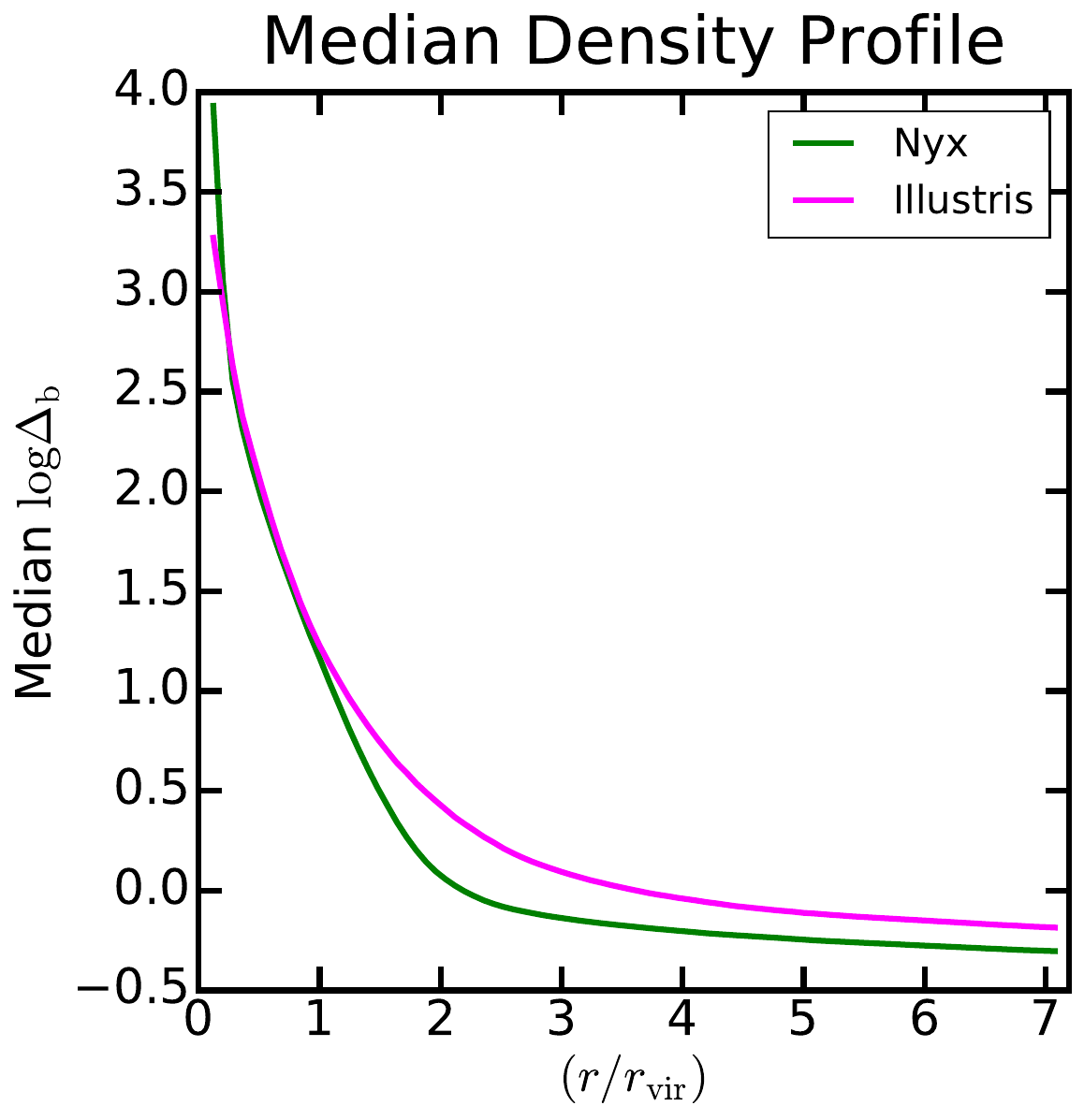}\\
  \centering
   \includegraphics[height=5.3cm]{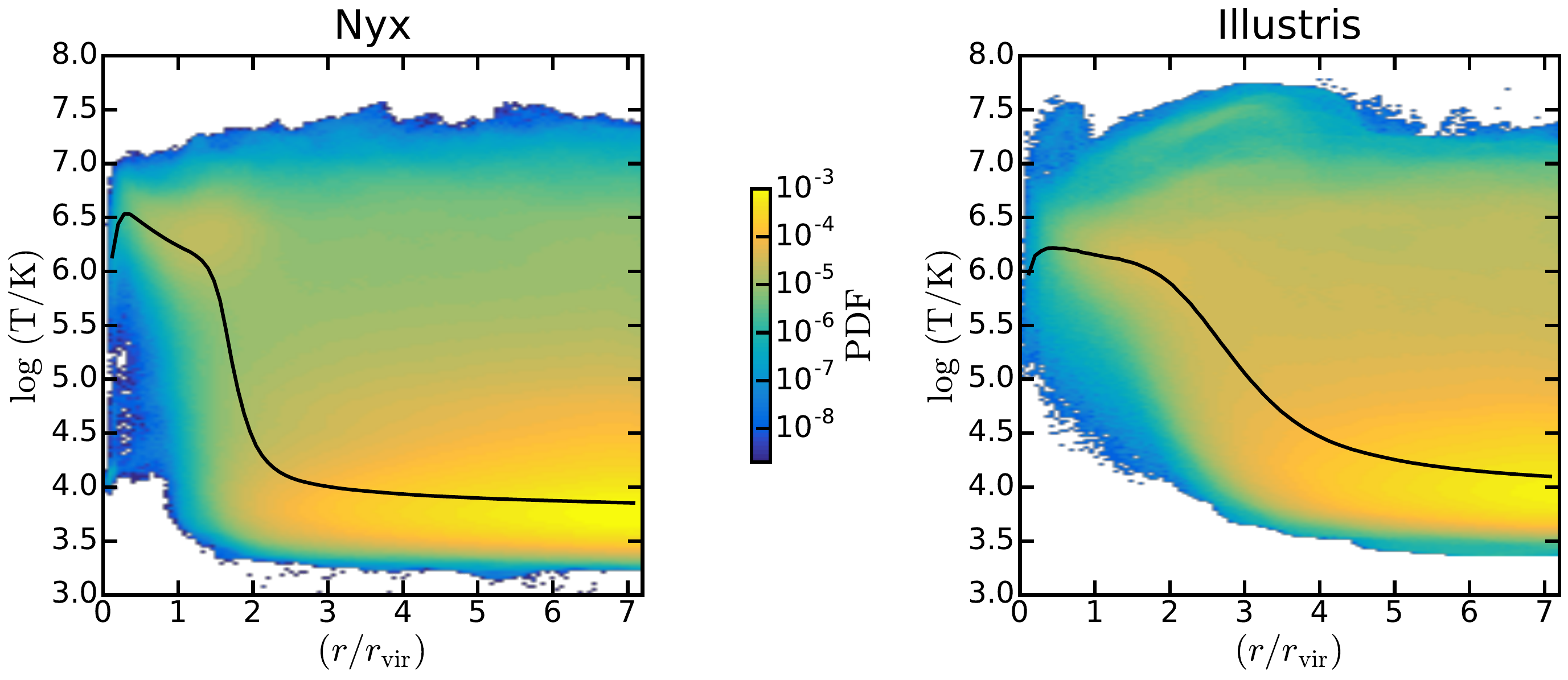}
   \hfill
   \includegraphics[height=5.3cm]{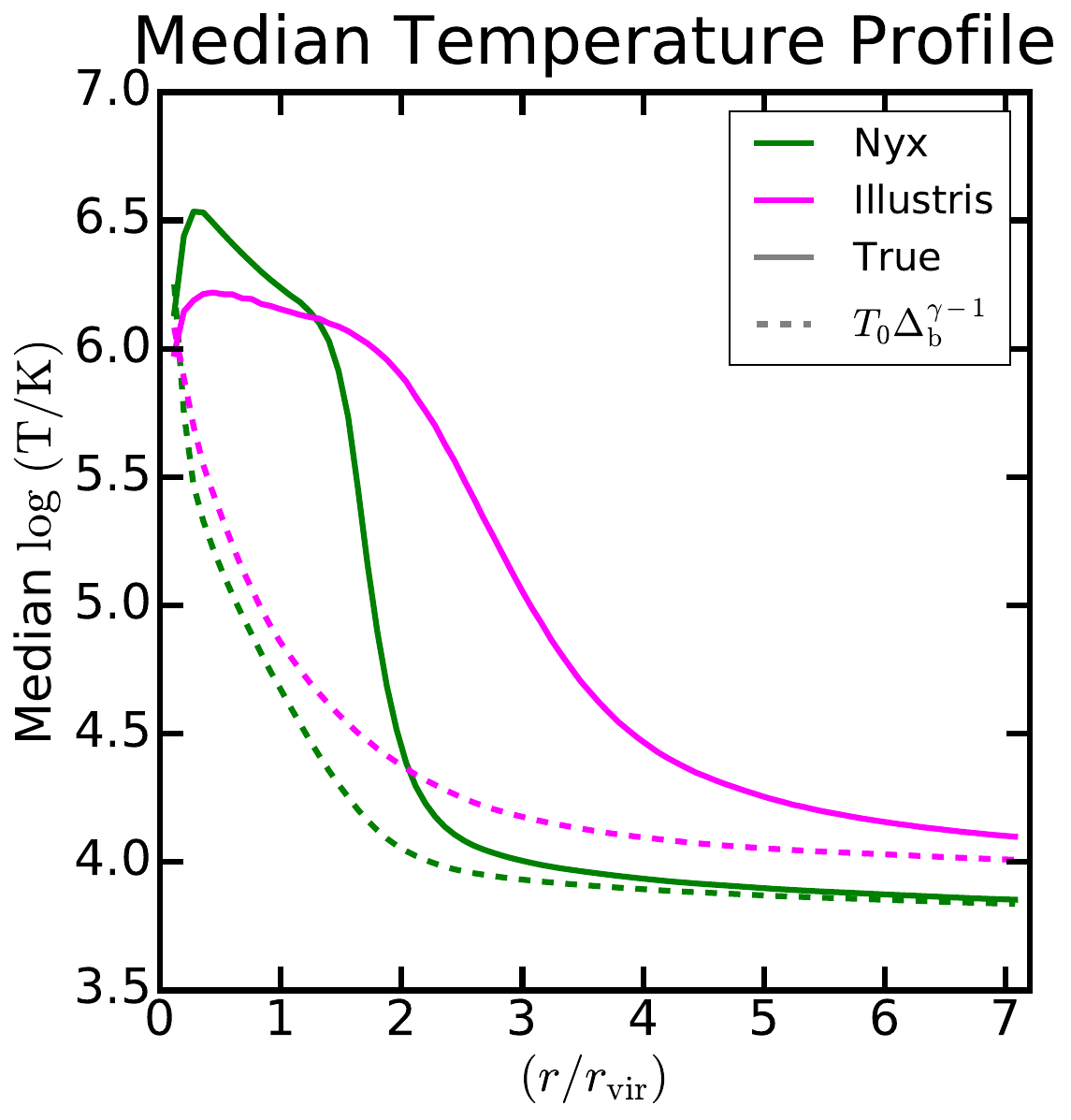}

  \caption{\textit{Top panels}: Radial hydrogen density profile obtained stackin 100 QSO-hosting halos in Nyx ($M>10^{12.5}\,M_{\odot}$, left panel) and all 72 Illustris QSO-hosting halos ($M>10^{12.4}\,M_{\odot}$, middle panel). The solid black line is the median hydrogen density profile, and the color bar gives the PDF in every radial bin within which the median is computed (see text for details). The right $y$-axis of the left and middle panels show the logarithm of the baryon overdensity (see text for details). The median baryon overdensity profiles are plotted together in the right panel. The green and magenta solid lines refer to Nyx and Illustris, respectively. Within the virial radius, Illustris is less dense than Nyx, but it is denser at larger distances. \textit{Bottom panels}: Same as in the top panels, but for the temperature profile. In the right panel, the solid magenta and green lines show the median temperature in Nyx and Illustris, while the corresponding dashed lines the predictions of the temperature-density relationship of the IGM given by the two simulations. Illustris forms halos that are cooler than Nyx within the virial radius, but hotter than Nyx at larger distances.} \label{fig:profiles_qso}

\centering
   \includegraphics[height=5.3cm]{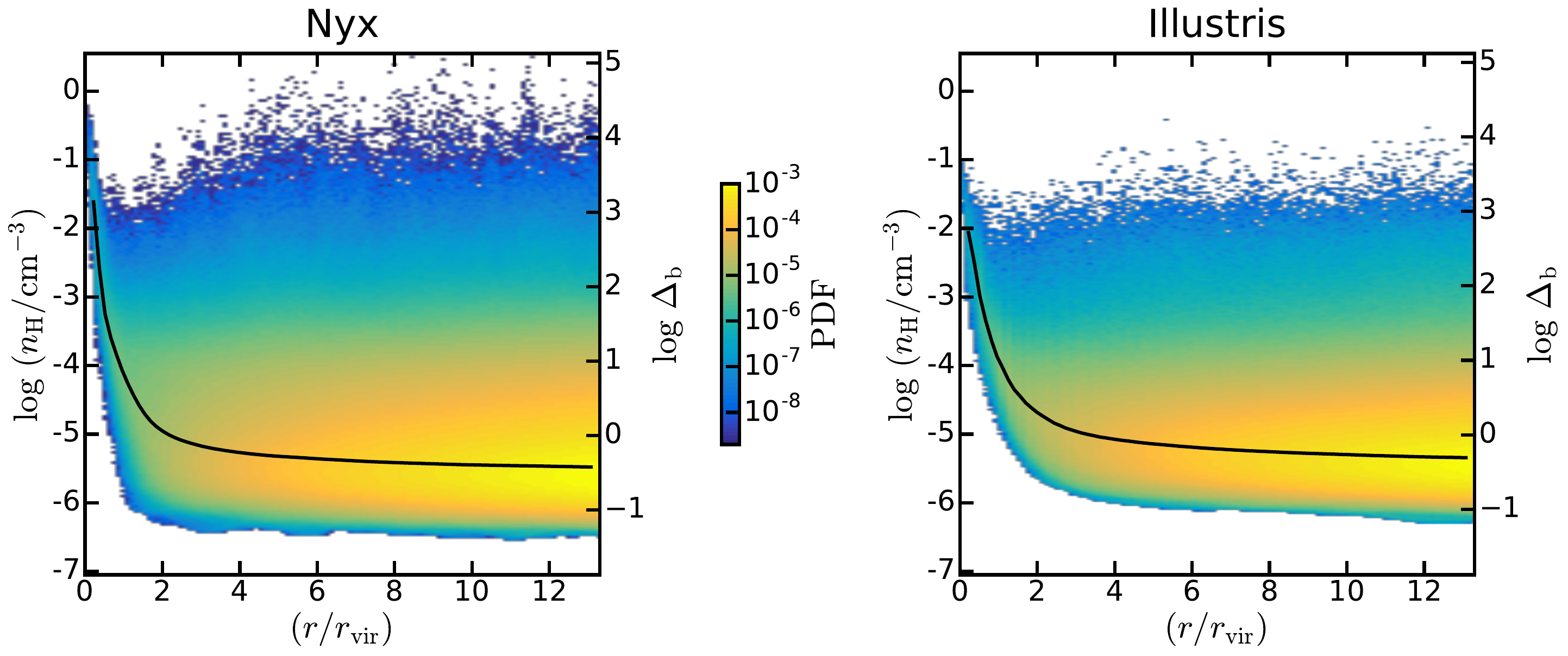}
   \hfill
   \includegraphics[height=5.3cm]{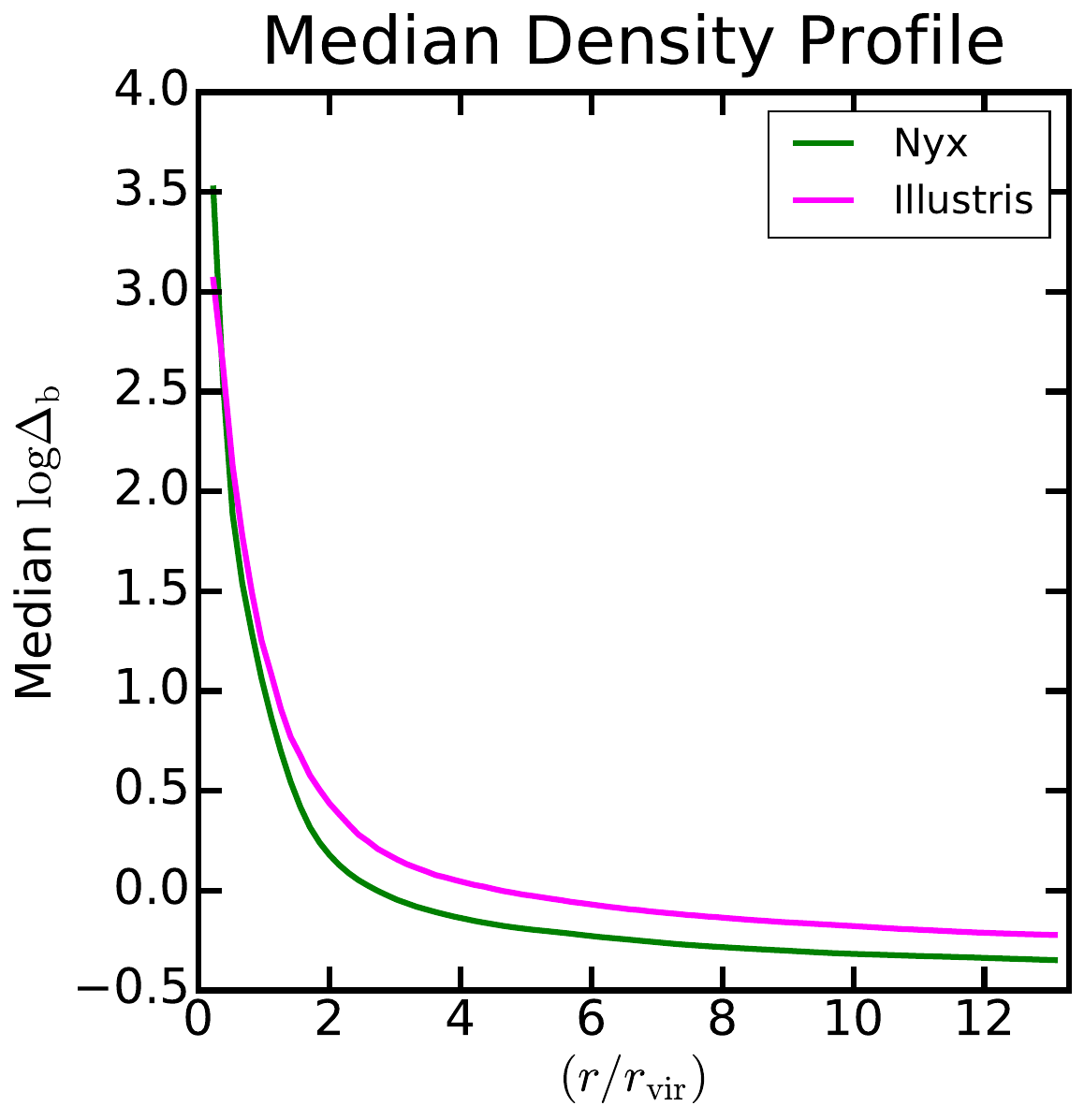}\\
  \centering
   \includegraphics[height=5.3cm]{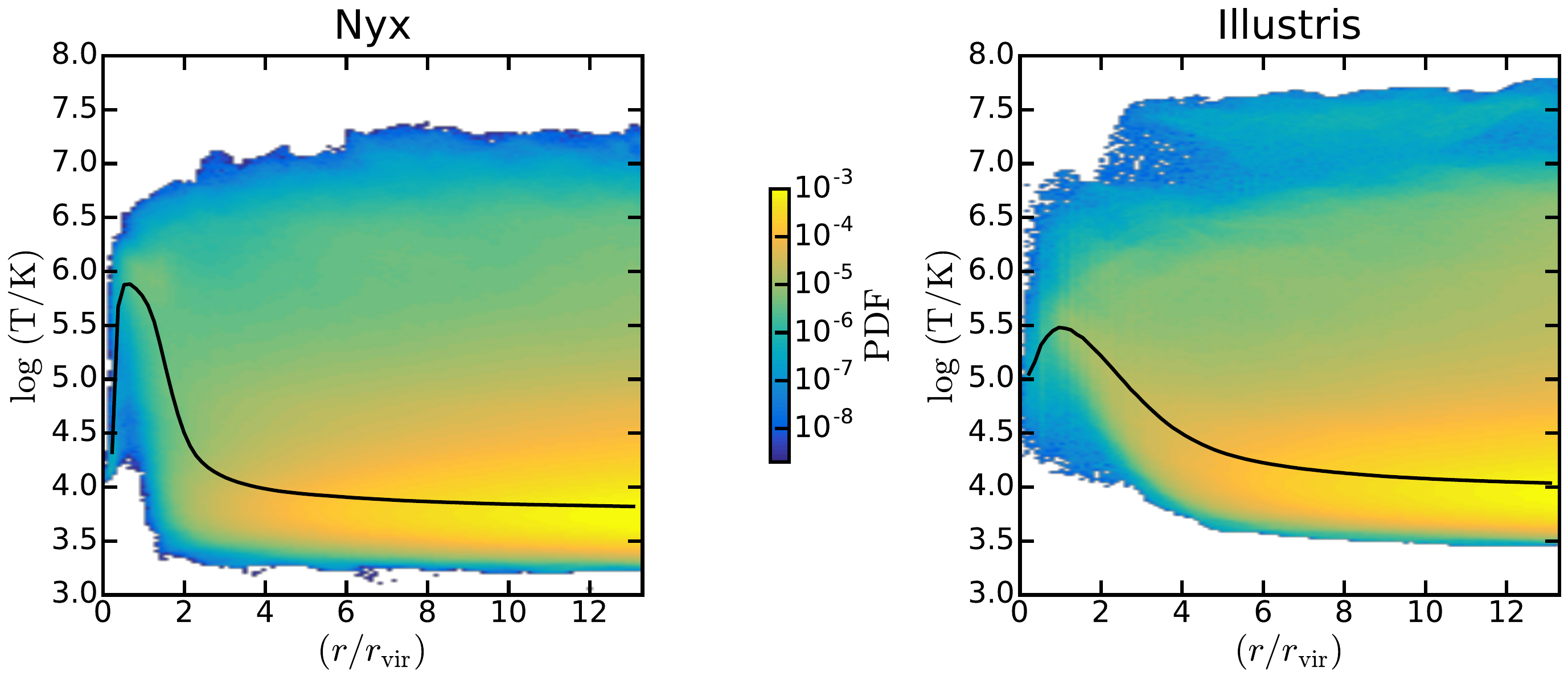}
   \hfill
   \includegraphics[height=5.3cm]{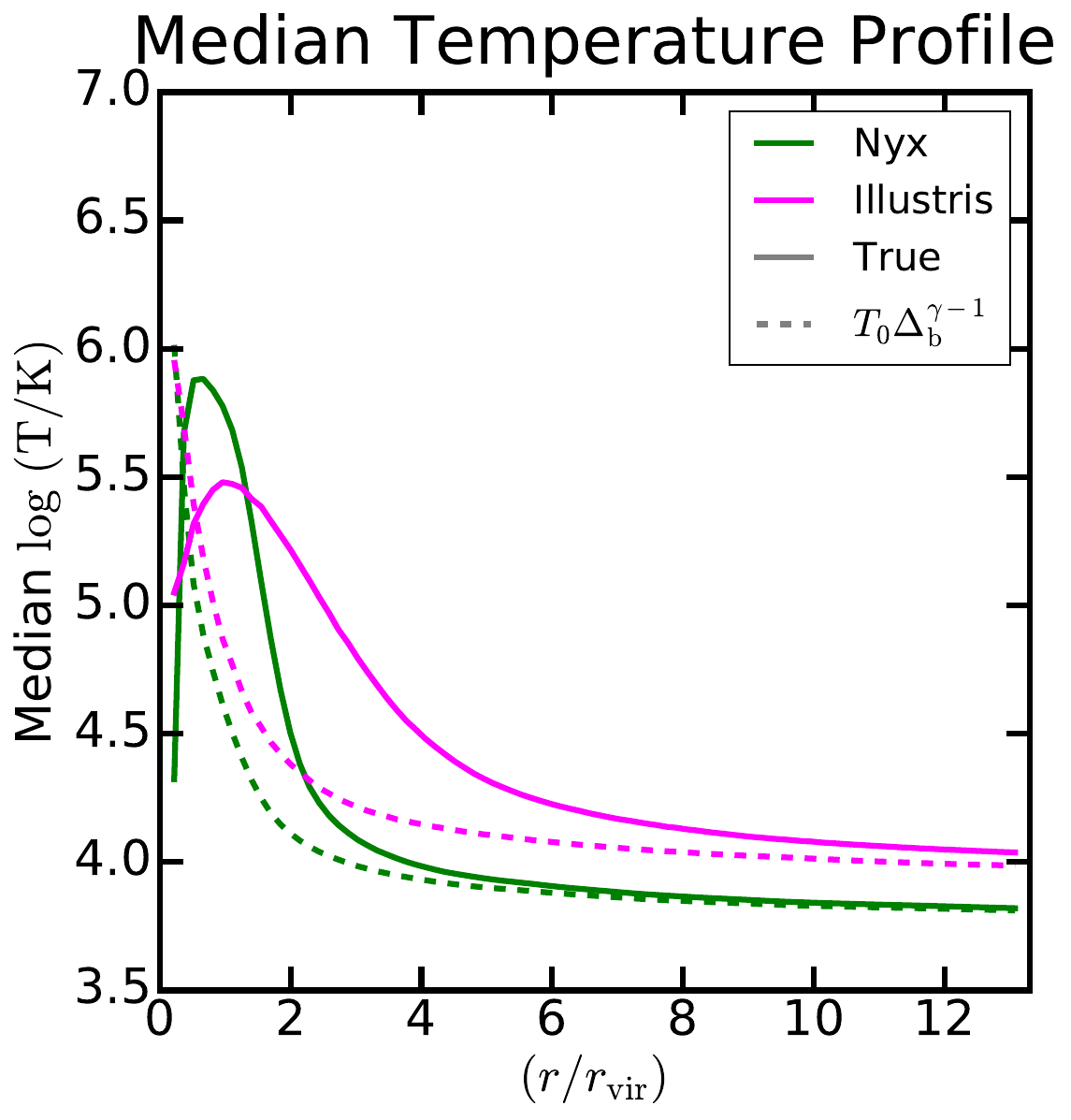}
  \caption{Same as Figure \ref{fig:profiles_qso}, but for 100 LBG/DLA-hosting halos randomly chosen from Nyx ($M>10^{11.7}\,M_{\odot}$) and Illustris ($M>10^{11.6}\,M_{\odot}$). Also in this case, Illustris forms halos that are cooler and less dense than in Nyx within the virial radius, but hotter and denser than in Nyx at larger distances.} \label{fig:profiles_lbg}
\end{figure*}

To understand the root of these differences we start by visually comparing
density and temperature slices around halos from both simulations (Figure \ref{fig:slices}). 
Massive halos drawn from Illustris are surrounded by a bubble of hot gas extending well 
beyond the virial radius, whereas the size of the shock-heated gas is generally smaller in Nyx halos.

To quantify the extent of the hot CGM in Nyx and Illustris, and the correlation
between the gas temperature with the underlying hydrogen density distribution, we 
investigated radial profiles in the two simulations. We construct the hydrogen
density profiles by choosing 100 from the sample of QSO hosts drawn from Nyx and all 72 halos from the corresponding Illustris sample, 
which were used to reproduce the observations discussed in \S~\ref{sec:qso}. We then
stack all gas cells within $1\Mpc$ from the centers of the halos, and
compute the median temperature within 90 equally extended bins of
radial distance, normalized to the virial radius of each halo. The
resulting median hydrogen density profiles for Nyx and Illustris are shown
with the black solid line in the top-left and top-middle panels of
Figure \ref{fig:profiles_qso}, respectively. In both panels, the color
bar indicates the PDF of \nh within each radial bin, while the right vertical axis displays the logarithm of the baryon overdensity, i.e. $\log \Delta_{\rm b}=\log(\rho_{\rm b} / \bar{\rho}_{\rm b})$, where $\rho_{\rm b}$ is the baryon density and $\bar{\rho}_{\rm b}$ its mean value in the whole box. 

The median
baryon overdensity profiles given by Nyx and Illustris are plotted together
in the upper-right panel of Figure \ref{fig:profiles_qso} (green and magenta solid lines, respectively).
Within the virial radius, Nyx and Illustris exhibit similar
overdensity profiles, except for the peak within $\sim0.2\,r_{\rm
  vir}$, which is $\sim 0.7\,\rm dex$ larger in Nyx than in
Illustris. This is caused by the absence of star formation in Nyx,
that allows gas density to increase without being converted into
stars.  Outside the virial radius, Illustris is systematically denser
than Nyx; the difference between the simulations reaches its maximum ($0.4\, \rm dex$) at $2r_{\rm
  vir}$, and reduces to $0.15 \, \rm dex$ at $7r_{\rm vir}$.
  
The median baryon overdensity far from halos does not necessarily have to be the same in the two simulations. Unlike the mean baryon overdensity, which should converge to $\Delta_b=1$ at large distances from the halos, the exact value of the median depends on the PDF of the baryon overdensity, which is different in Nyx and Illustris. Nevertheless, one expects that both in Nyx and Illustris the gas density will resemble the typical values of \nh in the IGM ($n_{\rm H} \sim 10^{-5}\, \rm cm^{-3}$), as it is the case for $r>3 \, r_{\rm vir}$.

In the lower panels of Figure \ref{fig:profiles_qso}, we plot the 
temperature profiles around QSO hosts in Illustris and Nyx, using the same
structure and color coding adopted in the upper panels. The median temperature can be up to
$0.3 \, \rm dex$ higher in Nyx (solid green line) than in Illustris
(solid magenta line) within the virial radius, but in the outer region
Illustris generates systematically hotter gas. The difference with
respect to Nyx reaches $1.5 \, \rm dex$ in the range $(2r_{\rm
    vir},\,3r_{\rm vir})$, and is still as large as $0.3\,\rm dex$ out
to $7\,r_{\rm vir}$. In the lower-right panel
 we also plot the temperature
that follows applying the IGM temperature-density
relationship given by Nyx and Illustris to the underlying median baryon overdensity
(dashed green and magenta lines, respectively). The median temperature profile given by
Nyx matches the values implied by the IGM temperature-density relationship for $r>3r_{\rm vir}$; the deviation
from this regime around $\sim 2\,r_{\rm vir}$ is due to virial shocks.
On the contrary, Illustris produces much hotter gas out to $3-4\,r_{\rm vir}$ and does not
asymptote to the temperature-density relationship even at $7\, r_{\rm vir}$. We argue that this behavior
can be explained by the extra physics implemented in Illustris: the strong radio-mode AGN feedback prescription
\citep{Sijacki_2007, Vogelsberger_2013} acts as an extra source of heating well outside the
virial radius. 

Our interpretation is in line with the findings of previous works \citep{Genel_2014, Haider_2016}, evidencing that Illustris underestimates the observed baryon mass fraction within $M\gtrsim 10^{13} \, M_{\odot}$ galaxy clusters at $z\sim0$ \citep{Giodini_2009, McGaugh_2010, Gonzalez_2013, Sanderson_2013}. According to the authors of such studies, this would be caused by the depletion of gas well outside the virial radius, induced by the violent radio-mode AGN feedback prescription. Albeit at higher redshifts, the temperature slices (Figure \ref{fig:slices}) and profiles (Figure \ref{fig:profiles_qso}) around $M \geq 10^{13} \, M_{\odot}$ halos obtained in this work corroborate this scenario. Since the overzealous AGN feedback model of Illustris has been tempered in the more recent IllustrisTNG simulations \citep[se, e.g.,][]{Pillepich_2017}, it will be interesting to compute the mean \lya flux profiles around halos from this new simulation.

We also point out that the temperature profiles in Nyx and Illustris
differ from each other to a much larger extent than the corresponding
baryon overdensity profiles. This justifies our approximation of
modifying only the temperature of the CGM, and not its density,
to mimic the effects of feedback within the semi-analytic technique
explained in \S~\ref{sec:paint}.

We repeat the same analysis shown in Figure \ref{fig:profiles_qso} also for the LBG/DLA-hosting halos extracted from Nyx and Illustris, this time randomly drawing 100 halos from both samples \footnote{Unlike the QSO-host sample, the LBG/DLA-hosts sample of Illustris contains more than 100 halos (see also Footnote \ref{foot:size_samples}).}. We show the results with analogous plots in Figure \ref{fig:profiles_lbg}, which has the same structure and color coding as Figure \ref{fig:profiles_qso}. The behavior of the two simulations is qualitatively the same as in the case of the QSO-hosting halos. Both Nyx and Illustris exhibit colder and less dense radial profiles with respect to what we found for the QSOs, but the relative differences in temperature and hydrogen density at the same virial radii are very similar. Similarly, the deviation from the predictions of the temperature-density relationship of the IGM occurs at comparable radial distances. Therefore, we can once again argue that the differences between the temperature in the two simulations out to $\sim 3\, r_{\rm vir}$ is mainly driven by feedback mechanisms. In particular, we think that also in this case the main driver of the observed differences is the radio-mode AGN feedback prescription of Illustris, consistently with the findings of previous works \citep{Genel_2014, Haider_2016}. Nevertheless, the most reliable way to disentangle the effect of quasar-mode and radio-mode AGN feedback would be repeating our analysis on simulations implementing only one of the two processes. Furthermore, it would also be relevant to assess the roles of stellar and AGN feedback mechanisms separately, by running a suite of hydrodynamic simulations. We plan to perform such a study in the near future.

We now discuss the mean \lya flux contrast profiles shown in \S~\ref{sec:qso} and \S~\ref{sec:dla} in the context of these different temperature profiles. 
Within the virial radius, Nyx is hotter and denser than Illustris. A higher temperature would yield a lower $\langle \delta_F\rangle$ (i.e. less \lya absorption) whereas a higher density would tend to increase  $\langle \delta_F\rangle$ (more
\lya absorption).  The fact that the two simulations give similar predictions for $\langle \delta_F \rangle$ implies that these two effects roughly cancel out. Outside the virial radius, Illustris yields higher temperatures, thus collisional ionization increases the ionization rate of hydrogen. Hence, Nyx always predicts a larger $\langle \delta_F\rangle$ (more \lya absorption)
than Illustris in this regime.

From the foregoing discussion, we can conclude that the main driver of the differences between the $\langle \delta_F \rangle$ predicted by Nyx and Illustris is the temperature of the gas. In general, we can say that the mean \lya flux contrast as a function of the impact parameter represents an excellent probe of the physics of the CGM and of the CGM-IGM interface, being closely related to the density and temperature profiles of the foreground halos. As such, the $\langle \delta_F \rangle$ profile can be used as a further test for feedback implementations in simulations. 

\subsection{The Halocentric Temperature-Density Relationship}
\label{sec:T-nH_CGM}

\begin{figure*}
\begin{adjustbox}{addcode={\begin{minipage}{\textheight}}{\caption{%
      Temperature - density relationship of hydrogen at different radial bins from a foreground QSO. The volume-weighted 2D histograms are plotted upon stacking 100 halos randomly chosen among the QSO-hosting Nyx halos ($> 10^{12.5}\,M_{\odot}$, top panels) and all 72 Illustris QSO-hosting halos ($> 10^{12.4}\,M_{\odot}$, bottom panels), in different bins of distance form the center of the halos. From left to right, the histograms refer to the radial bins $(0,\, r_{\rm vir})$, $(r_{\rm vir},\, 2r_{\rm vir})$, $(2r_{\rm vir},\, 3r_{\rm vir})$, $(3r_{\rm vir},\, 5r_{\rm vir})$, and $(5r_{\rm vir},\, 1 \Mpc)$. As a reference, the temperature-density relationship in the IGM (already shown in Figure \ref{fig:T-nH_IGM}) is plotted in the rightmost panel for both simulations. The two simulations yield similar temperature-density diagrams within the virial radius, while they strongly differ in the other bins. Whereas Nyx converges to the temperature-density relationship of the IGM already in the bin $( r_{\rm vir},\,2r_{\rm vir})$, Illustris begins producing the IGM feature for $r>3\,r_{\rm vir}$.
      }\label{fig:T-nH_CGM}\end{minipage}},rotate=90,left}
     \hspace{0.04\textheight}
     \includegraphics[width=0.92\textheight]{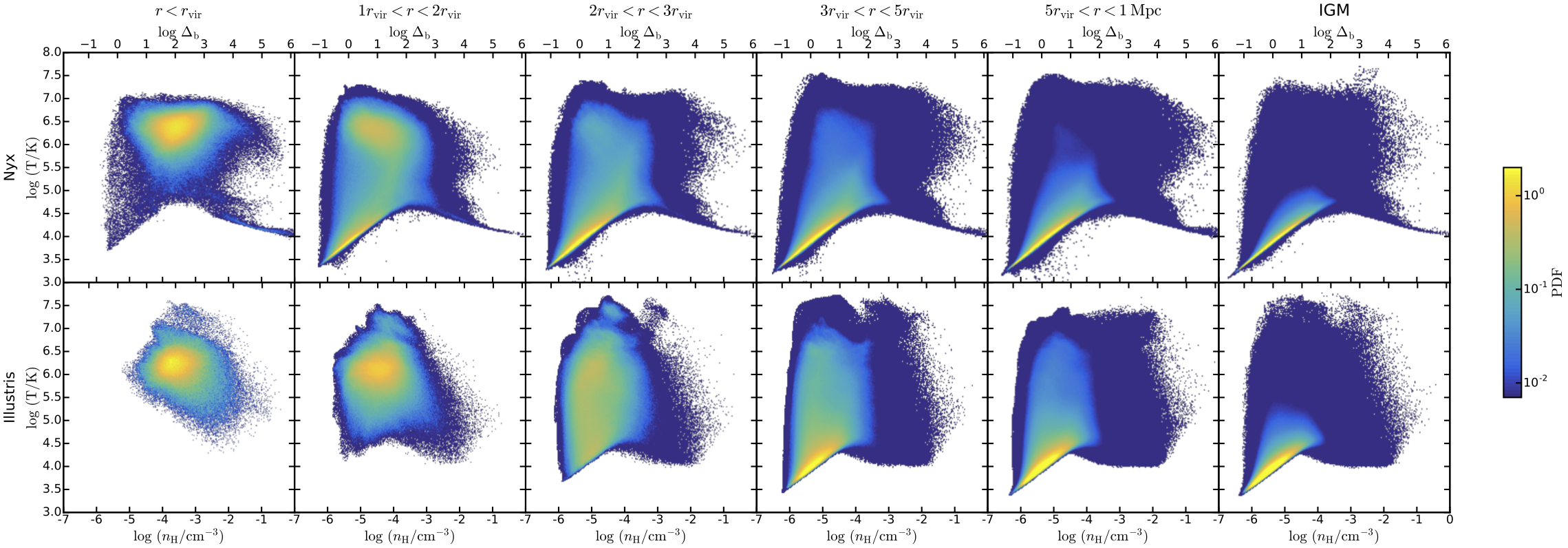}%
\end{adjustbox}

\vspace{-\textheight}
\vspace{-4mm}
  \begin{adjustbox}{addcode={\begin{minipage}{\textheight}}{\caption{%
      Same as in Figure \ref{fig:T-nH_CGM}, but for 100 LBG/DLA-hosting halos randomly chosen from Nyx ($> 10^{11.7}\,M_{\odot}$) and Illustris  ($> 10^{11.6}\,M_{\odot}$). Whereas Nyx converges to the temperature-density relationship of the IGM already in the bin $( r_{\rm vir},\,2r_{\rm vir})$, Illustris begins producing the IGM feature for $r>3\,r_{\rm vir}$.
      }\label{fig:T-nH_CGM_LBG}\end{minipage}},rotate=90,right}
      \hspace{0.04\textheight}
     \includegraphics[width=0.92\textheight]{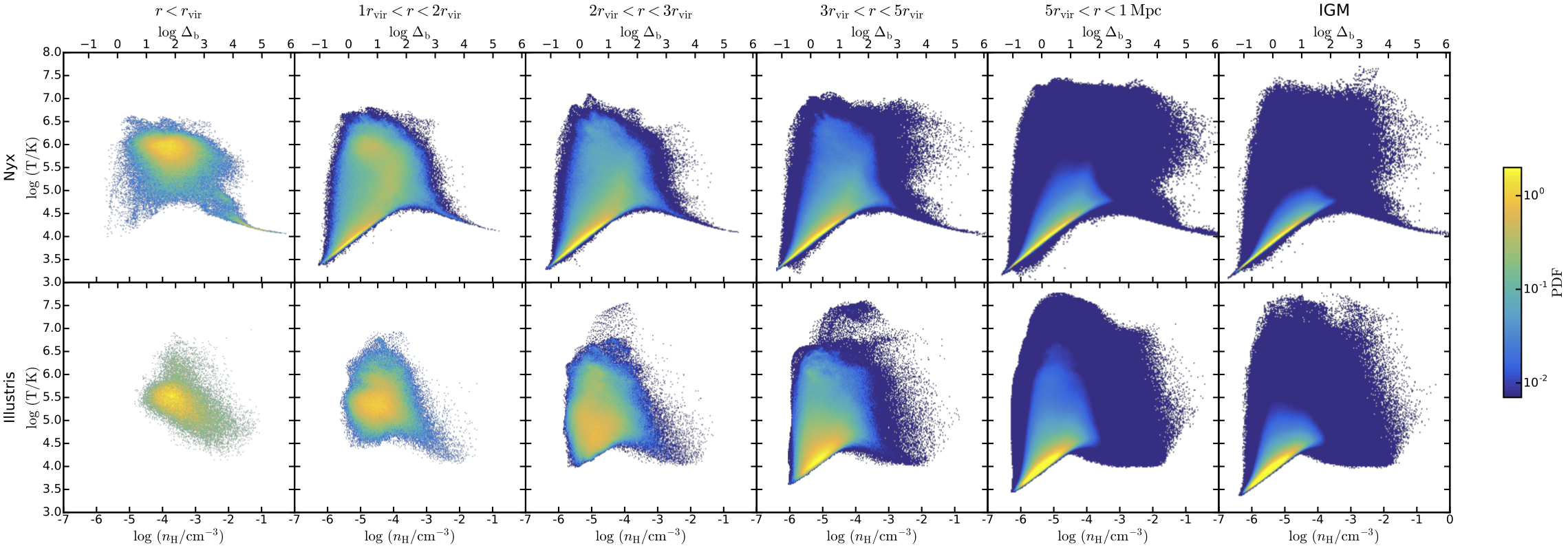}%
\end{adjustbox}
\end{figure*}

To interpret the differences in the temperature and density profiles of halos in Nyx and Illustris, shown in \S~\ref{sec:profiles}, in terms of the physics implemented in the two simulations, we repeatedly used arguments based on the temperature-density relationship of the IGM. In this section, we want to complete our discussion by investigating such relationship within different spherical shells around the center of the halos selected from the two simulations. Specifically, we consider the intervals $(0,\,  r_{\rm vir})$, $(r_{\rm vir},\, 2 r_{\rm vir})$, $(2r_{\rm vir},\, 3r_{\rm vir})$, $(3r_{\rm vir},\, 5r_{\rm vir})$, and $(5r_{\rm vir},\, 1 \Mpc)$, corresponding to the regions delimited with the vertical dotted black lines in Figures \ref{fig:mean_flux}, \ref{fig:mean_flux_DLA} and in the right panel of Figure \ref{fig:tau_median}. We shall then compare the temperature-density relationships in such radial bins with the \lya flux contrast predicted in the same intervals.

The temperature-density relationship of the hydrogen within one virial
radius from the center of the QSO-hosting halos obtained in Nyx and Illustris ($> 10^{12.5}\, M_{\odot}$ and $> 10^{12.4}\, M_{\odot}$, respectively) can
be seen in the top-leftmost and bottom-leftmost panels of Figure
\ref{fig:T-nH_CGM}, respectively. We plot the volume-weighted 2D
histograms resulting from stacking a subsample of halos randomly chosen from the QSO-hosting halos drawn from Nyx and Illustris. We took 100 halos from Nyx and all 72 halos from Illustris. 
For both simulations, the temperature-density
relationship is profoundly different from the one of the IGM (Figure
\ref{fig:T-nH_IGM}), as expected.
Shock heating and/or feedback, coupled with the impact of collisional
ionization, and self-shielding, change the structure of the
temperature-density distribution around galactic halos.
Both Nyx and Illustris exhibit a cloud of high-temperature gas
spanning the density range $(10^{-5},\,10^{-1})\,\rm cm^{-3}$ in the
$T-n_{\rm H}$ diagram, in excess of the densities typical of the IGM
($n_{\rm H} \lesssim 10^{-5}\, \rm cm^{-3}$). Nyx exhibits
a high-density ($>10^{-3}\,\rm cm^{-3}$),
low-temperature ($<10^{-4.5}$) locus,
arising from the aforementioned overcooling due
to the absence of star formation. 
Such a feature is of course absent
in Illustris, which does include star formation. In addition, a minimal fraction of the gas within the virial radius in Nyx follows a low-density ($<10^{-4}\,\rm cm^{-3}$), low-temperature ($<10^{-4.5}$) line, which cannot be found in Illustris. This gas has not been shock heated, thus it lies along the temperature-density power law of the IGM. Nevertheless, the temperature-density relationship
within the virial radius is qualitatively similar in the two
simulations. Likewise, they predict a similar $\langle \delta_F \rangle$ in the same
region (see Figure \ref{fig:mean_flux}).

In the second panels from the left in Figure \ref{fig:T-nH_CGM}, we show the temperature-density relationship in the interval $(r_{\rm vir},\, 2 r_{\rm vir})$. The results from Nyx and Illustris are reported in top and bottom panels, respectively. There is a huge difference between the two simulations. Many Nyx pixels fall on the
power law temperature-density relation of the IGM
(see the left panel of Figure \ref{fig:T-nH_IGM}), but there is still a considerable fraction of gas cells in a hot phase ($n_{\rm H}\sim 10^{-4.5} \, \rm cm^{-3}$, $T\sim 10^{6.3} \, \rm K$). Illustris does not exhibit any power-law
feature; instead its temperature-density relationship is still dominated by a cloud of hot gas ($n_{\rm H}\sim 10^{-4.5} \, \rm cm^{-3}$, $T\sim 10^{6.3} \, \rm K$) analogous to that seen within one virial radius. As previously argued, the radio-mode AGN feedback prescription in Illustris is probably responsible for heating the gas as far as two virial radii, erasing the IGM power-law feature in the temperature-density relationship in the range $(r_{\rm vir},\, 2 r_{\rm vir})$. We notice that, in the same range, the predictions of $\langle \delta_F \rangle$ given by Nyx and Illustris differ by $\sim 30\%$ (see Figure \ref{fig:mean_flux}). Illustris predicts less absorption, due to the larger amount of hot gas.

We plot the temperature-density relationship in the range $(2\,r_{\rm vir},\,3 \, r_{\rm vir})$ in the third panels from the left in Figure \ref{fig:T-nH_CGM}. Nyx (top panel) shows the typical power-law feature of the IGM. Such feature begins to appear also in lllustris (bottom panel), but the majority of the pixels still lie in a
hot phase ($n_{\rm H}\sim 10^{-5} \, \rm cm^{-3}$, $T\sim 10^6\, \rm K$).  This indicates that the AGN feedback prescription in Illustris dominates the temperature-density relationship even in the range $(2\,r_{\rm vir},\,3 \, r_{\rm vir})$. The large amount of gas leads to predicting a lower absorption compared with Nyx (see Figure \ref{fig:mean_flux}).

The fourth and fifth panels from the left in Figure \ref{fig:T-nH_CGM} show the temperature-density relationship in the radial bins $(3\,r_{\rm vir},\,5 \, r_{\rm vir})$ and $(5\,r_{\rm vir},\,1 \Mpc)$, respectively. In the former, Illustris still exhibits a larger amount of hotter gas than Nyx. In the latter, the diagrams of the two simulations look similar. As a reference, in the rightmost panels of Figure  \ref{fig:T-nH_CGM} we plot the temperature-density relationship of the IGM, already shown in Figure \ref{fig:T-nH_IGM}. While Nyx starts qualitatively resembling the shape of the IGM temperature-density relationship for $r> r_{\rm vir}$, Illustris does so for $r>3 r_{\rm vir}$. This is consistent with our estimate of the distance from the center of the halos up to which the AGN feedback prescription in Illustris seems to dominate the thermal state of the CGM (see \S~\ref{sec:profiles}).

We repeat the same analysis discussed above for the halos hosting LBGs and DLAs in both simulations. The corresponding temperature-density relationships are reported in Figure \ref{fig:T-nH_CGM_LBG}. The panels report the volume-weighted 2D histograms resulting from stacking 100 halos, randomly drawn from the LGB-hosting ($>10^{11.7}\,M_{\odot}$ and $>10^{11.6}\,M_{\odot}$ for Nyx and Illustris, respectively) used to reproduce the observations of \lya absorption around DLAs and LBGs.
Qualitatively, the diagrams present the same differences observed for the QSO-hosting halos (Figure \ref{fig:T-nH_CGM}). Likewise, the predictions of $\langle \delta_F \rangle$ in the innermost bins around DLAs differ by $\lesssim 13\%$ (see Figure \ref{fig:mean_flux_DLA}). The difference increases up to $\sim30\%$ in the interval $(2\,r_{\rm vir},\,3 \, r_{\rm vir})$, and decreases again at farther distances. 

Clearly, the different predictions of the $\langle \delta_F \rangle$ given by the two simulations are manifest as differences between the
corresponding temperature-density relationships. Therefore, also the temperature-density diagrams within different radial shells, presented in the current section, are a potentially excellent method to visualize the impact of feedback implementations on the physics of the CGM and CGM-IGM interface.

As a caveat, we point out that the \lya optical depth does not depend only on the temperature-density relationship of the gas, but also also on its peculiar velocity. In principle, Nyx and Illustris may exhibit different gas velocity fields, and this may also have an impact on the \lya absorption profile. We verified that the radial velocity-density relationship is very similar in the two simulations, confirming that the differences in the $\langle \delta_F \rangle$ profiles are driven by the temperature-density relationship (see the Appendix \ref{app:velocity} for details).

Furthermore, we verified that uncertainties on the predictions of $\langle \delta_F \rangle$ that could derive from possible systematic uncertainties 
in our analysis would not change the main conclusions of this work (see the Appendix \ref{app:systematics} for a detailed discussion). 

\section{Comparison with Previous Work}
\label{sec:previous_work}

There is a large body of work considering the covering factor of optically thick absorbers, absorption profiles of metal lines, or the \lya column density distribution. Nevertheless, as in this study we primarily focus on the mean \lya flux contrast and optical depth, we restrict our discussion to works that considered a similar statistic.
 
\cite{Meiksin_2015} reproduced observations of \lya transmission profiles around QSOs and LBGs to the predictions of hydrodynamic cosmological simulations. We compare our findings with a later similar work, based on more recent state-of-the-art simulations \citep{Meiksin_2017}. \cite{Meiksin_2017} compared the \lya absorption profile around QSOs measured by \cite{Prochaska_2013} with two runs of the Sherwood suite of hydrodynamic simulations. Galactic winds were implemented in only one of the two runs. The predictions of $\langle \delta_F \rangle$ around QSOs given by the runs with and without feedback differ up to $\sim 700 \kpc$. In our work, we find discrepancies between Nyx and Illustris out to larger impact parameters, i.e. $\sim 2 \Mpc$. 

The larger radius to which we see differences between the simulations considered with respect to \cite{Meiksin_2017} can be due to the strong AGN feedback prescription in Illustris up to $\sim 3\, r_{\rm vir}$ (as discussed in \S~\ref{sec:discussion}). On top of that, we use a different criterion to select the QSO-hosting halos in the simulations.
We select halos above a certain mass threshold ($10^{12.5} \, M_{\odot}$ for Nyx and $10^{12.4} \, M_{\odot}$ for Illustris)  because, from their observations, \cite{Prochaska_2013} could not set an upper limit to the mass of the halos hosting the QSOs of their sample, but only a lower limit. Consequently, whereas most of the halos have a mass around $\sim 10^{12.5} \, M_{\odot}$, our sample includes also halos as massive as $10^{13.7}\, M_{\odot}$. Instead, \cite{Meiksin_2017} consider halos with mass between $10^{12.2}\, M_{\odot}$ and $10^{12.8}\, M_{\odot}$. The median virial radius of \cite{Meiksin_2017} sample of halos is then smaller than ours, therefore it is reasonable that their signature of their feedback prescription extends out to smaller impact parameters.

The simulations considered by \cite{Meiksin_2017} reproduce \cite{Prochaska_2013} observations outside the virial radius, but underpredict the observations within the virial radius. This is analogous to our findings. In addition, we compared Nyx and Illustris out to $\sim 20 \Mpc$ with BOSS data and verified that the simulations asymptote to \cite{Font-Ribera_2013} measurements. We also examined the \lya transmission profile around DLAs, considering \cite{Rubin_2015} and \cite{Font-Ribera_2012b} observations.

\cite{Turner_2017} compared the observations of the median \lya optical depth due to \hi around LBGs by \cite{Turner_2014} with the predictions of a run of the EAGLE suite of hydrodynamic simulations. The run considered included both stochastic thermal stellar feedback and AGN feedback. To reproduce the observations, they considered samples of halos in different mass bins. The best match with the data occurs for the bin $(10^{11.5},\, 10^{12.0})\,M_{\odot}$, which is consistent with the halo mass threshold for Nyx and Illustris derived in this work ($10^{11.7}\,M_{\odot}$ and $10^{11.6}\,M_{\odot}$, respectively). Whereas the simulation yields good agreement with the observations, there is some tension with the data points at a transverse distance of $\sim 70 \kpc$, $\sim 300 \kpc$ and $\sim 800 \kpc$ from the foreground LBG. These data points are problematic to reproduce also for Nyx and Illustris (see Figure \ref{fig:tau_median}), so our findings are consistent with \cite{Turner_2017}. However, we notice that the median logarithm of the \lya optical depth at $\sim 70 \kpc$ obtained in \cite{Turner_2017} simulations
underpredicts the measurements by $\sim0.3$, whereas we find that the discrepancy with Nyx and Illustris is $\sim0.7$. Analysing the temperature and density profiles, as well as the halocentric teperature-density relationship on the EAGLE simulation, similarly to what we did for Nyx and Illustris in this work, may shed light on the reasons for the better match with the data.

\section{Conclusions and Perspectives}
\label{sec:conclusions}

We compared state-of-the-art hydrodynamic cosmological simulations with observations of \lya absorption around QSOs, DLAs, and LBGs, with the
goal of characterizing the physical state of their surrounding CGM and IGM
and constraining feedback models implemented in galaxy formation simulations.
Specifically, we focused on observations of the average \lya transmission
profiles around galactic halos, and showed that existing measurements
from BOSS \citep{Font-Ribera_2012b, Font-Ribera_2013} and QSO pairs \citep{Prochaska_2013, Rubin_2015} can be combined to tightly constrain this quantity over a large dynamic range in impact parameter ($20 \kpc \lesssim b \lesssim 20\Mpc$). We compared observations to the publicly available Illustris simulation, based on the \texttt{Arepo} code, and a hydrodynamic cosmological
simulation run with
the Nyx code. The former models star formation, stellar/AGN feedback, and metals, while the latter contains none of these prescriptions.
Below we list the primary conclusions of this work: 

\begin{enumerate}

\item At large separations ($b > 2 \Mpc$) BOSS data are well reproduced by both Illustris and
  Nyx, meaning that they both asymptote to a correct description of the ambient IGM
far from halos. This is a result of the success of the $\Lambda$CDM model in
describing the Ly$\alpha$ forest.

\item At intermediate separations ($r_{\rm vir} < b < 2\Mpc$)
  Illustris and Nyx give different predictions for the mean \lya flux
  contrast around QSOs and DLAs (Figures \ref{fig:mean_flux} and
  \ref{fig:mean_flux_DLA}), which results from underlying differences
  between the two simulations in the temperature-density relationships
  of the gas in the CGM and CGM-IGM interface. We showed that
  Illustris exhibits large bubbles of hot gas extending to $\sim 3- 4
  \, r_{\rm vir}$ that are absent in Nyx, which we attribute to the
  star-formation and radio-mode AGN feedback prescriptions (Figures
  \ref{fig:profiles_qso}-\ref{fig:T-nH_CGM_LBG}) in Illustris. 
  
\item At small separations ($b<r_{\rm vir}$) Illustris and Nyx
  underpredict the \lya absorption around QSOs, DLAs and LBGs (Figures
  \ref{fig:mean_flux}, \ref{fig:mean_flux_DLA} and \ref{fig:tau_median}), reflecting the
  challenges of accurately modeling the physics of galaxy
  formation. Using a novel semi-analytic technique to alter the
  temperature within the CGM, we showed that the discrepancy with the data could be mitigated
  if simulations produced
  a colder CGM.

  \item We find broad agreement between the simulations and observations of the median \lya optical depth around LBGs \citep{Turner_2014}, except for the underprediction of \lya absorption within the virial radius. Also in this case, the discrepancy could be alleviated by invoking a cooler CGM (right panel of Figure \ref{fig:tau_median}).
	
 \end{enumerate}
 
We verified that the impact of several possible sources of systematic errors in our analysis (redshift distribution and sample size of the simulated and observed spectra, uncertainties in the mass of the halos selected in the simulations, resolution of the simulations considered, and our approach to construct mock spectra from results of the \textsc{Arepo} moving-mesh code) does not change our conclusions (see Appendices \ref{app:cell_size} and \ref{app:systematics}).
 
We reiterate that although both simulations predict significantly different \lya
transmission profiles between the virial radius and $b\sim 500 \kpc$, the measurements
are still too noisy to discriminate between different models. Thus, increasing the precision of the measurements in this crucial range with future observations would enable 
much stronger constraints on the physical state of gas in the CGM and the CGM-IGM interface,
and as a result on the feedback prescriptions of different simulations.
A promising approach for obtaining much higher precision at these small impact parameters
is  \lya forest tomography \citep{Tomography_2,Tomography_1,Tomography_3,Tomography_4}, whereby a dense ensemble
of background galaxies are used as background sources instead of quasars. Higher-precision
measurements will soon be provided by the ongoing COSMOS Lyman Alpha Mapping And Tomography Observations (CLAMATO) survey \citep[][Lee et al. in prep.]{Clamato}. 

The exquisite high precision of the large-scale BOSS measurements of the \lya
transmission around QSOs and DLAs motivates further numerical investigation,
since their agreement with the state-of-the-art simulations considered here
is good but not yet perfect.
Besides comparing different feedback prescriptions to the observations of $\langle \delta_F \rangle$, these BOSS measurements may be able to discriminate between small variations of $\langle \delta_F \rangle$ given by different cosmological models, or even from different DM models \citep{Fuzzy, Warm}, primordial magnetic fields \citep{B-Field}, and any other physics that
is expected to have an impact on large-scale structure. 

Although this paper focused on the \lya
transmission profiles as a function of impact parameter, 
it would also be fruitful to  extend our study by comparing the predictions of \lya
transmission as a function of both transverse and line-of-sight
(i.e. redshift space) separation from foreground halos with BOSS measurements, similar to the
comparison of the EAGLE simulations with the \cite{Turner_2014} data, performed by
\cite{Turner_2017}.

Finally, this study focused solely on the mean (or median) \lya transmission around
galactic halos, but there is clearly much more information available from the full \lya
flux distribution and its correlations. Other statistics quantifying \lya absorption, such as
the PDF
\citep[see also][]{Kollmeier_2003}, the $N_{\rm HI}$ CDDF \citep{Prochaska_2011, Stinson_2012, Tumlinson_2013, Gutcke_2017}, or the line-of-sight power spectrum at different impact parameters from halos, can
provide additional constraints on gas in the CGM and CGM-IGM interface. However, our work
demonstrates that before investigating these other statistics, it is important to ensure that
simulations can reproduce the mean \lya flux profile over the full range of scales probed
around galaxies.

\acknowledgments 

We are thankful to Volker Springel for sharing his code to bin SPH outputs into Cartesian grids and to Andreu Font-Ribera for sharing the values of the data points and error bars published in \cite{Font-Ribera_2012b} and \cite{Font-Ribera_2013}. We thank Andrea Macci\`o, Annalisa Pillepich, Lars Hernquist and the members of the ENIGMA group at the
Max Planck Institute for Astronomy (MPIA) for helpful comments and discussions. We are grateful to the anonymous referee, whose comments improved the quality of this scientific article. DS thanks Emanuele Paolo Farina, Hector Hiss and Michael Walther for useful comments on a draft of this manuscript. 
This research used resources of the National Energy Research Scientific Computing Center (NERSC), which is supported by the Office of Science of the U.S. Department of Energy under Contract no. DE-AC02-05CH11231.  ZL acknowledges support from the Scientific Discovery through Advanced Computing (SciDAC) program funded by U.S. Department of Energy Office of Advanced Scientific Computing Research and the Office of High Energy Physics.
This work made extensive use of the NASA Astrophysics Data System and of the astro-ph preprint archive at arXiv.org.

\appendix

\section{Calibration of Halo Masses}
\label{app:halo_masses}

\begin{figure*}[]
  \centering
   \includegraphics[width=0.45\textwidth]{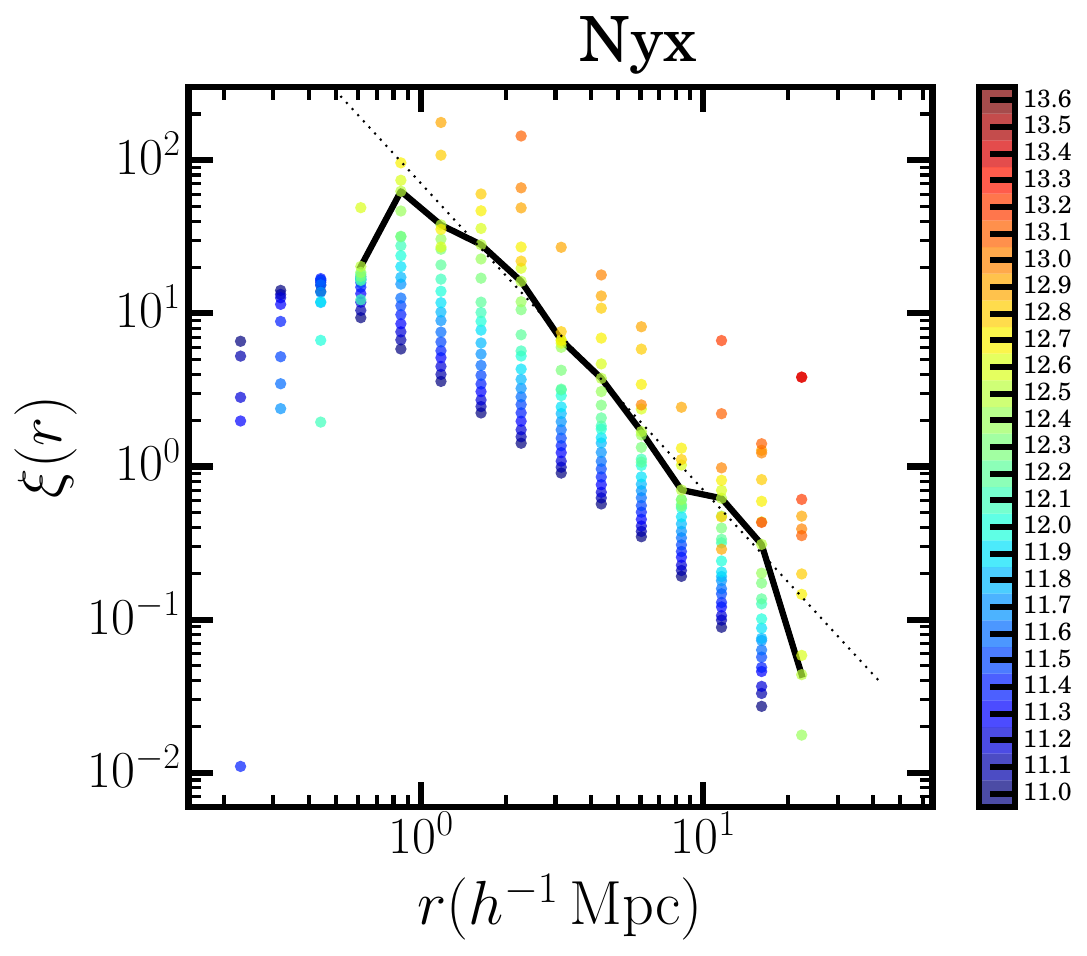}\hfill
   \includegraphics[width=0.45\textwidth]{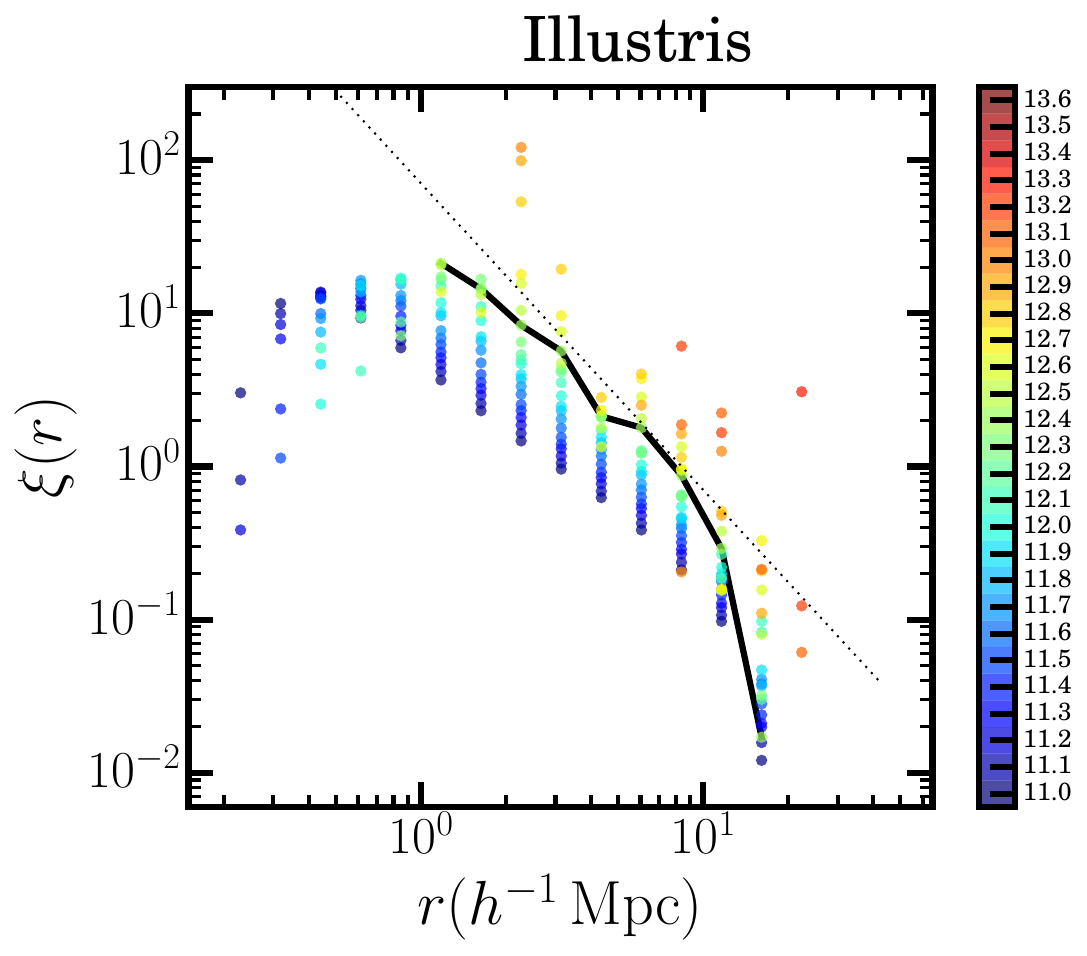}
    \includegraphics[width=0.45\textwidth]{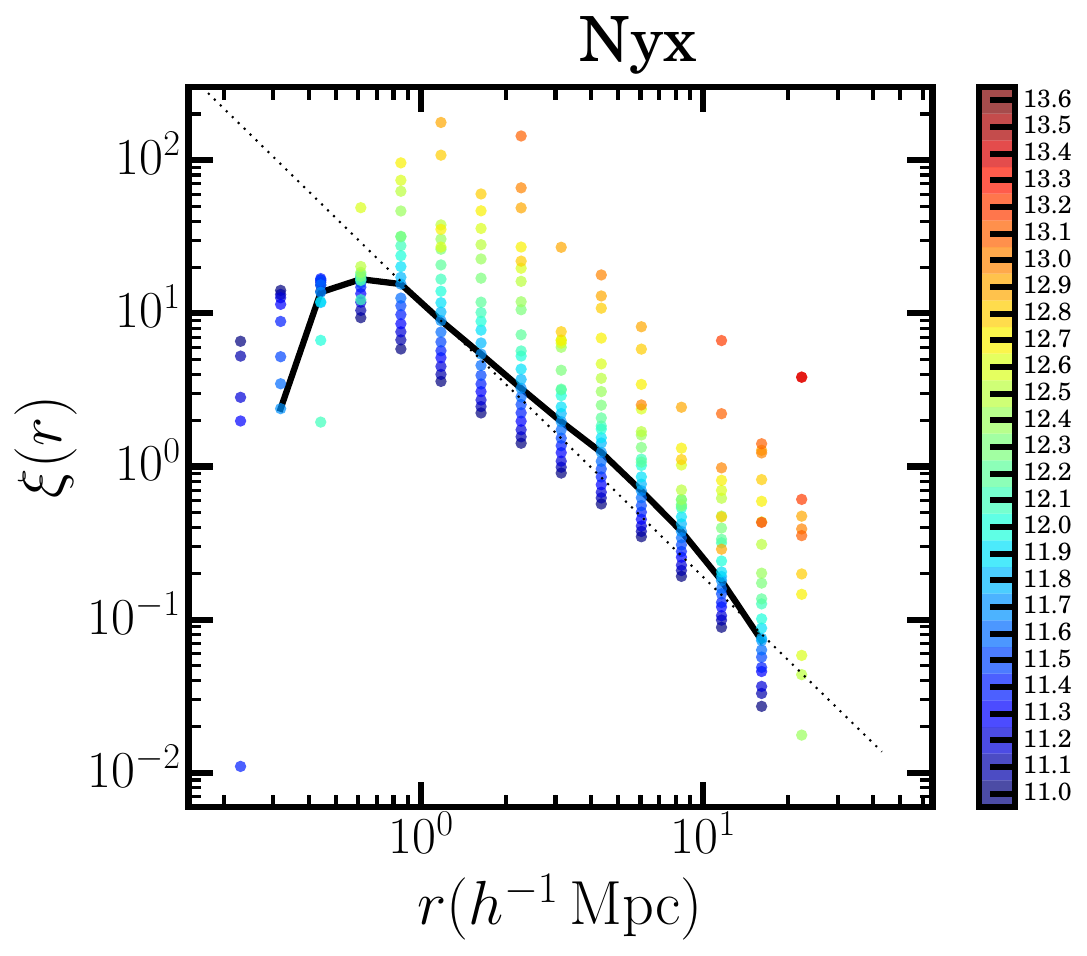}\hfill
   \includegraphics[width=0.45\textwidth]{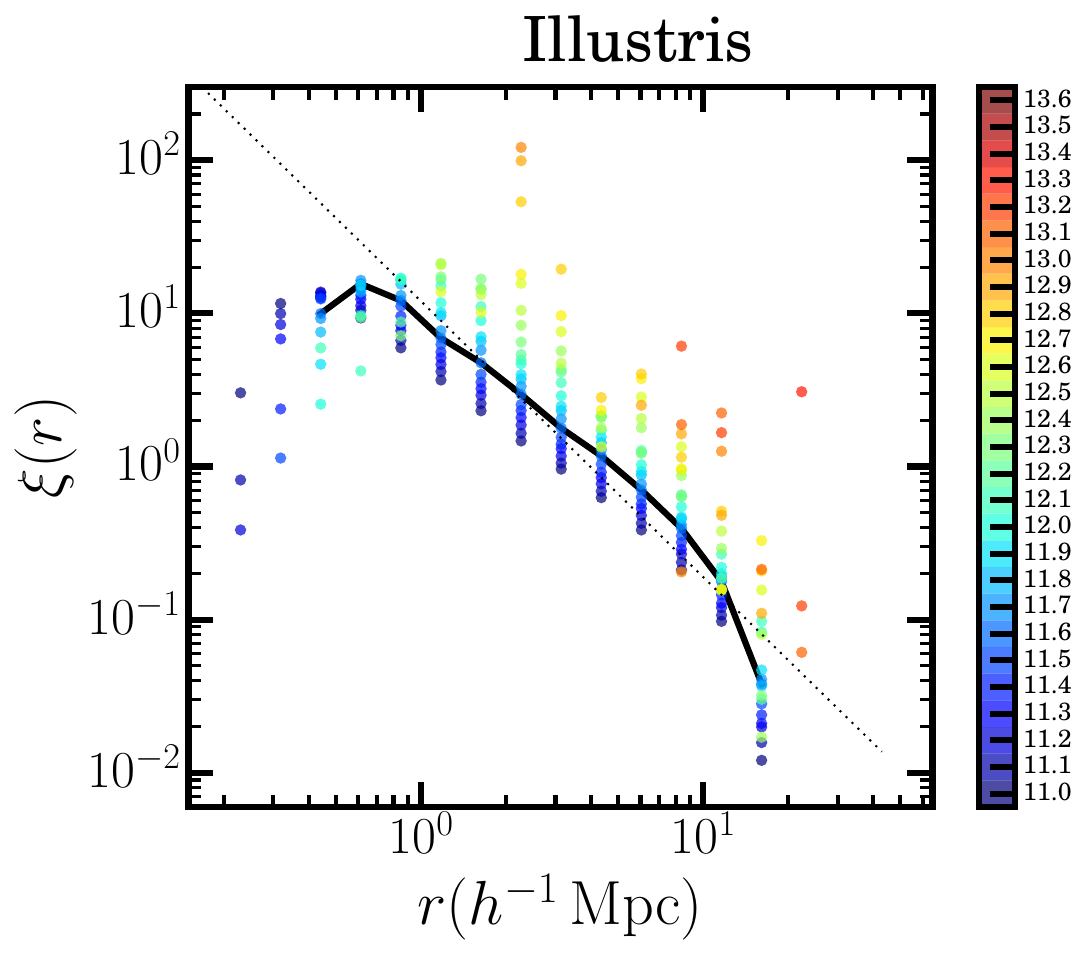}
       \includegraphics[width=0.45\textwidth]{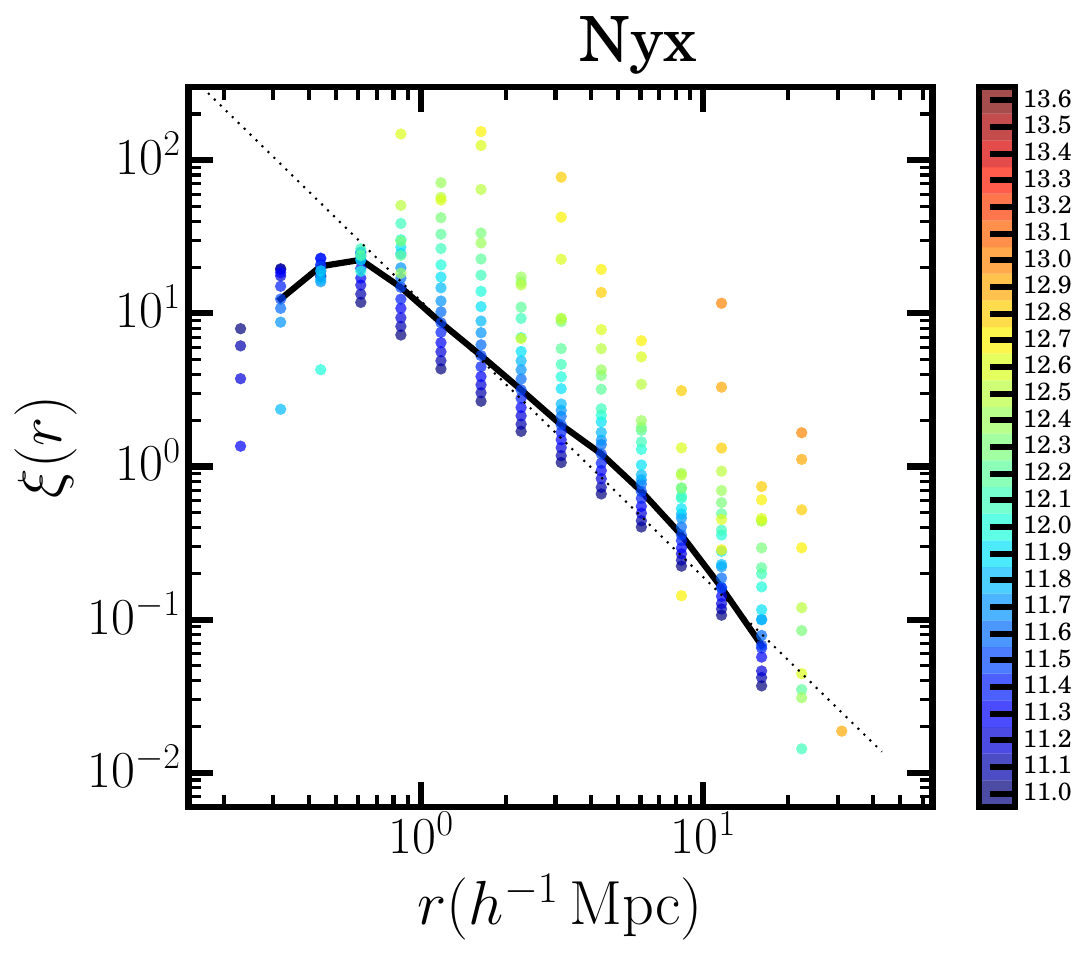}\hfill
   \includegraphics[width=0.45\textwidth]{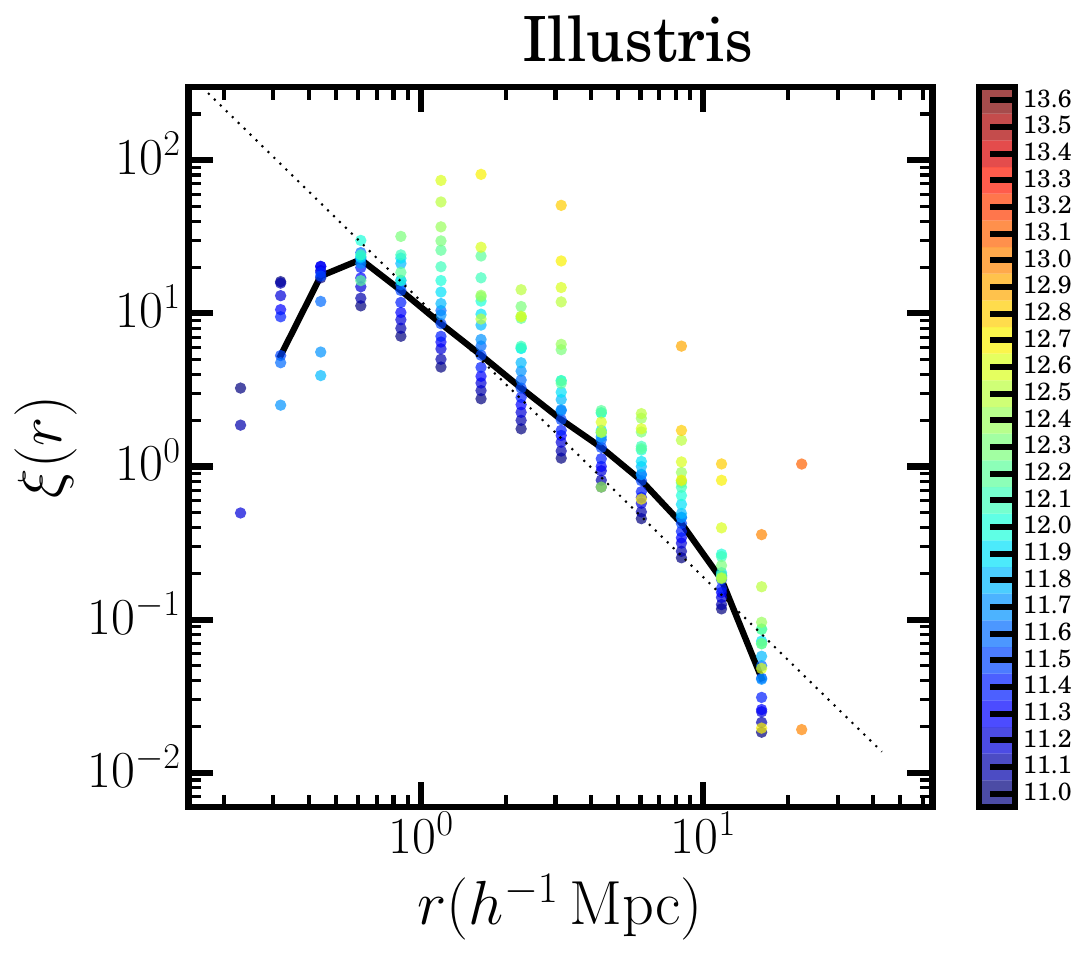}
  \caption{\textit{Top panels:} Correlation function of halos in Nyx (left panel) and Illustris (right panel), at redshift $z=2.4$ and $z=2.44$, respectively. Dots of the same color represent the correlation function of halos with mass larger than the one corresponding to that color. The black dotted line is the analytic fit to the correlation function of quasars measured by \cite{White_2012}. The black solid line is the correlation function of halos above the mass threshold that best fits those observations (see text for details). \textit{Middle panels:} As top panels, except that the black dotted line is the analytic fit to the correlation function of galaxies measured by \cite{Bielby_2011}, and the black solid line is the correlation function of halos above the mass threshold that best fits these observations (see text for details).  \textit{Bottom panels:} As middle panels, except that the correlation function of halos refers to the Nyx and Illustris snapshots at redshift $z=3$ and $z=3.01$, respectively (see text for details).} \label{fig:corr_func}
\end{figure*}

In this work we are interested in studying the physics of the CGM around QSOs, LBGs and DLAs, which reside in halos with different mass. Hence, we model these objects by selecting samples of halos above a certain mass threshold $M_{\rm min}$ from Nyx and Illustris, which depends on the nature of the foreground object considered. In principle, the halo masses in the two simulations may not be calibrated in the same way, so we cannot assume the same mass threshold for both simulations. 

When creating mock 
samples of foreground QSOs, we determine $M_{\rm min}$ for Illustris and Nyx such that the resulting sample of halos reproduces the 3D two-point correlation function of quasar-hosting halos in the redshift quitrange $2.2<z<2.8$, measured by \cite{White_2012}. 

For each simulation, we select all halos such that $M_{\rm halo}>M_{\rm th}$, where $M_{\rm halo}$ is the halo mass as reported in the halo catalog of that simulation and $M_{\rm th}$ is a threshold, which is fixed to an arbitrary value in the first place. We then compute the 3D two-point correlation function of the halos into 20 equally extended logarithmic bins in the range of distance $(0.1,\,50)\, h^{-1} \cMpc$. We compare the correlation function with the measurements by \cite{White_2012} and calculate the $\chi ^2$ within the range of distance $(4,\,13)\, h^{-1} \cMpc$. We repeat the steps just described increasing $M_{\rm th}$ by $0.1\, \rm dex$, until there is no halo with mass larger than $M_{\rm th}$. At this point, we are left with a family of correlation functions depending on $M_{\rm th}$, and look for the value of the $M_{\rm th}$ minimizing the $\chi ^2$. This value is the minimum mass $M_{\rm min}$ that we need to set when we select halos from that simulation, such that their 3D correlation function will be as close as possible to the one observed by \cite{White_2012}. In other words, we assume a step function halo occupation distribution, i.e. the number of QSOs per halo is zero below $M_{\rm min}$ and unity above such threshold.

The top panels of Figure \ref{fig:corr_func} illustrate the procedure for the calibration of $M_{\min}$ for quasar hosts. In the top-left panel, the correlation function given by all Nyx halos with $M>M_{\rm th}$ from the snapshot at redshift $z=2.4$ is represented by circles, color-coded according to the corresponding value of $M_{\rm th}$. The black dotted line is the analytic fit to the measurements by \cite{White_2012}, and the black solid line the correlation function computed from the Nyx halos, corresponding to the value of $M_{\rm th}$ minimizing the $\chi ^2$. The top-right panel shows exactly the same quantities for the snapshot at $z=2.44$ of the Illustris simulation. We obtained $M_{\rm min}=10^{12.5} \, M_{\odot}$ for Nyx and $M_{\rm min}=10^{12.4} \, M_{\odot}$ for Illustris. The samples of halos selected according to these thresholds have been utilized to reproduce the measurements by \cite{Font-Ribera_2013} and \cite{Prochaska_2013}.

To simulate a sample of foreground LBGs, we apply the same procedure described for the QSOs. Instead of using observations of QSO clustering, we determine the $M_{\rm min}$ that gives the best match with the observations of the 3D two-point correlation function of LBGs in the redshift range $2<z<4$, by \cite{Bielby_2011}. The value of $M_{\rm min}$ is obtained minimizing the $\chi^2$ in the distance range $(1,\,10)\,h^{-1}\cMpc$.

The middle panels of Figure \ref{fig:corr_func} are analogous to the top panels, except that the black dotted lines now represent the analytic fit to the measurements by \cite{Bielby_2011}. We obtained $M_{\rm min}=10^{11.6} \, M_{\odot}$ for the $z=2.44$ Illustris snapshot and $M_{\rm min}=10^{11.7} \, M_{\odot}$ for the $z=2.4$ Nyx snapshot. These values of $M_{\rm min}$ will be used to select the halos for the comparison of the simulations with the observations by \cite{Turner_2014}, who considered a sample of LBGs with $z=2.4$ as median redshift. 

We shall also compare the simulations with the data by \cite{Adelberger_2003}, \cite{Adelberger_2005} and \cite{Crighton_2011}, who measured the radial mean flux profile around LBGs at redshifts $z\approx3$, $z\approx 2.5$ and $z\approx 3$, respectively. Despite not being centered at the same redshift, the mean flux profiles measured by such observations are all normalized to $z=3$. Therefore, to compare them with the simulations, we consider the Nyx snapshot at $z=3$ and the Illustris snapshot $z=3.01$. We determined the value of $M_{\rm min}$ for such snapshots, obtaining $10^{11.5}\, M_{\odot}$ in both cases. The corresponding best fits to the correlation function measured by \cite{Bielby_2011} are reported in the bottom panels of Figure \ref{fig:corr_func}. 

We notice that the values of $M_{\rm min}$ inferred for our simulated LBG samples are consistent with the typical mass of LBG-hosting halos, $\sim10^{12}\, M_{\odot}$, deduced by various authors \citep{Adelberger_2005a, Conroy_2008, Trainor_2012, Rakic_2013, Turner_2014} 
for the KBSS survey. 
As discussed in \S~\ref{sec:sample}, we used the same halos selected for the foreground LBGs, and assumed that the DLAs lie at the center of the halos. The values of $M_{\rm min}$ obtained for the LBGs are also of the same order of the characteristic mass of DLAs estimated by \citealt{Font-Ribera_2012b} ($10^{12} \, M_{\odot}$) from BOSS quasar spectra.

In \S~\ref{sec:sel_halos} we explained that the Nyx halo finding algorithm is equivalent to a FOF algorithm with a linking length of 0.168 times the mean interparticle separation, whereas the Illustris halo finder adopts a linking length of 0.2 times the interparticle separation. Since we calibrate the mass of both Nyx and Illustris halos with the same observations, the differences between the halo finding algorithms is not really an issue for the analysis presented in this paper. Nevertheless, we verified that, tuning the Nyx halo finding algorithm to produce results compatible with a FOF algorithm with the same linking length as the one used in Illustris, the values of $M_{\rm min}$ obtained with Nyx would differ by 0.1 dex from the ones adopted in the analysis of this work. As we further demonstrate in the Appendix \ref{sec:halo_mass}, this difference would not change the main conclusions of this work.

\section{Generating Mock Spectra from a Moving-Mesh Code}
\label{app:cell_size}

We simulate the absorption spectra extracting skewers on a regular grid. Since in Nyx a gas element is a cell of a Cartesian grid, the cell size of the skewers is simply given by the cell size of the simulation. Instead, Illustris treats gas on a moving mesh, constructed with a Voronoi tessellation. When we draw skewers form Illustris, we bin the gas cells into a regular grid. For this purpose, we treated each gas element as an SPH particle \citep[following][]{Bird_DLA}. The smoothing length of the SPH kernel for a certain gas cell is chosen to be the maximum radius of all Delaunay tetrahedra with that cell at a vertex \citep[see][for more details]{Springel_2010}. We construct in this way the gas density, temperature and 3D velocity fields.

We need to choose the cell size of our grid so that we can have reliable predictions of the mean \lya absorption around galaxies, which is the main goal of the paper. If the grid is too coarse, we may not be able to resolve the small-scale density and temperature fluctuations of the CGM. For example, if the cell size is as big as the typical virial radius, the CGM would be represented as a uniform gas cell with the average temperature and density of the CGM. 

We choose a cell size equal to the mean separation of the gas cells in Illustris ($58.5 \ckpc$), corresponding to a $1820^3$ grid. We verified that, with a $3550^3$ grid, corresponding to a cell size of $30 \ckpc$, the predicted mean \lya flux contrast within the virial radius of QSOs and DLAs (see \S~\ref{sec:results}) increases by only 1.5\%. Outside the virial radius, the difference with respect to the predictions obtained on a $1820^3$ grid is even smaller. Since using the finest grid does not change the main results of our work, we show all predictions given by Illustris utilizing the $1820^3$ grid. Regarding Nyx, the run that we are using has already the finest currently available grid.

\section{Assessment of Systematics in the Analysis}
\label{app:systematics}

In this appendix, we show that possible systematic errors in our analysis do not impact the main conclusions of our work. In \S~\ref{sec:halo_mass} and \S~\ref{sec:redshift}, we quantify the uncertainty on the predictions of $\langle \delta_F \rangle$ deriving from possible errors in the calibration of the halo masses in the two simulations and in the redshift of the foreground objects, respectively. In \S~\ref{sec:sample_size} we discuss the effect of the sample size of quasar-galaxy pairs on the estimate of $\langle \delta_F \rangle$. Finally, in \S~\ref{sec:resolution} we study the convergence of the Illustris results. 

\subsection{Halo Mass} 
\label{sec:halo_mass}

\begin{figure*}
\centering
\includegraphics[width=0.49\textwidth]{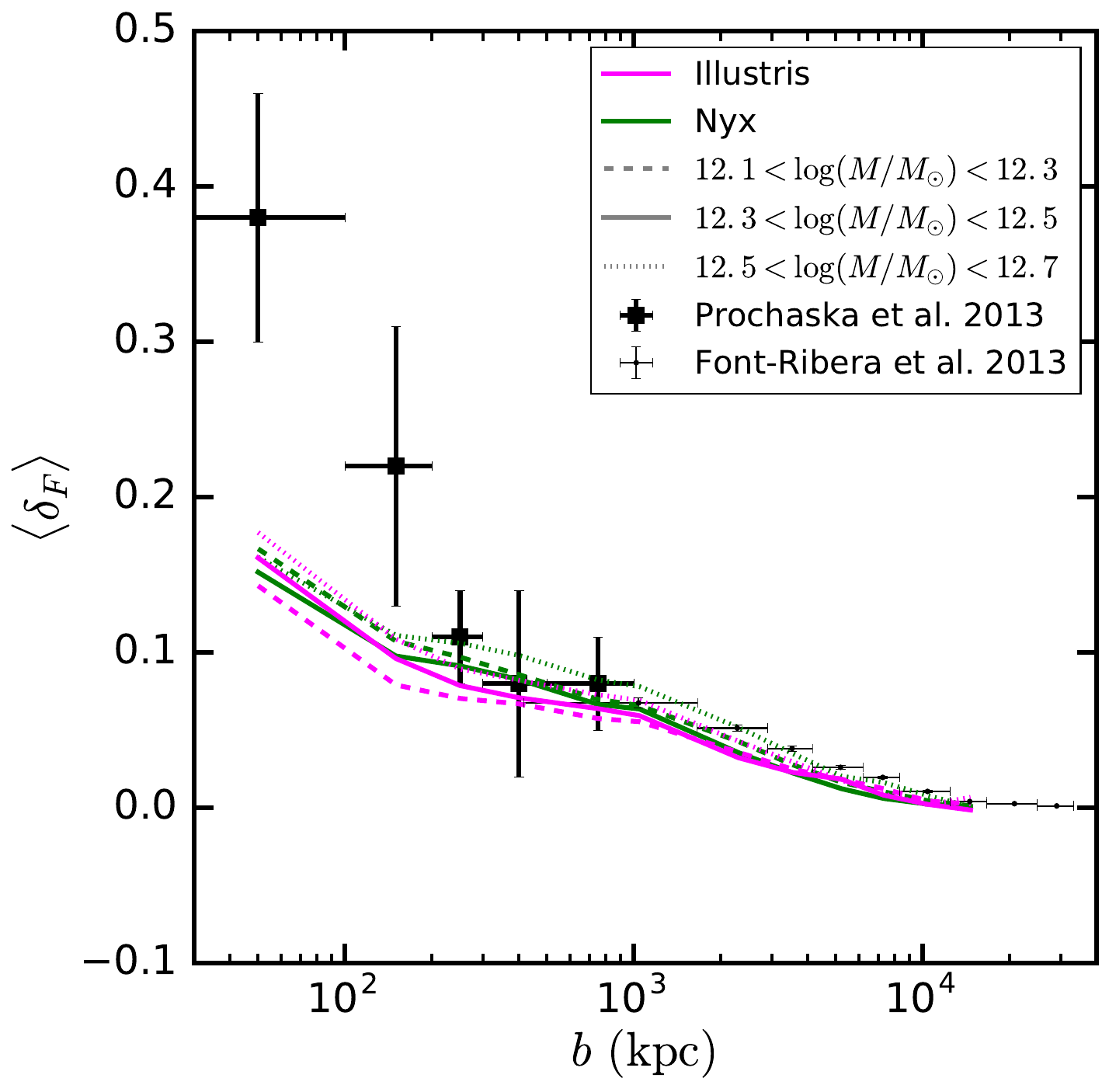}
\includegraphics[width=0.49\textwidth]{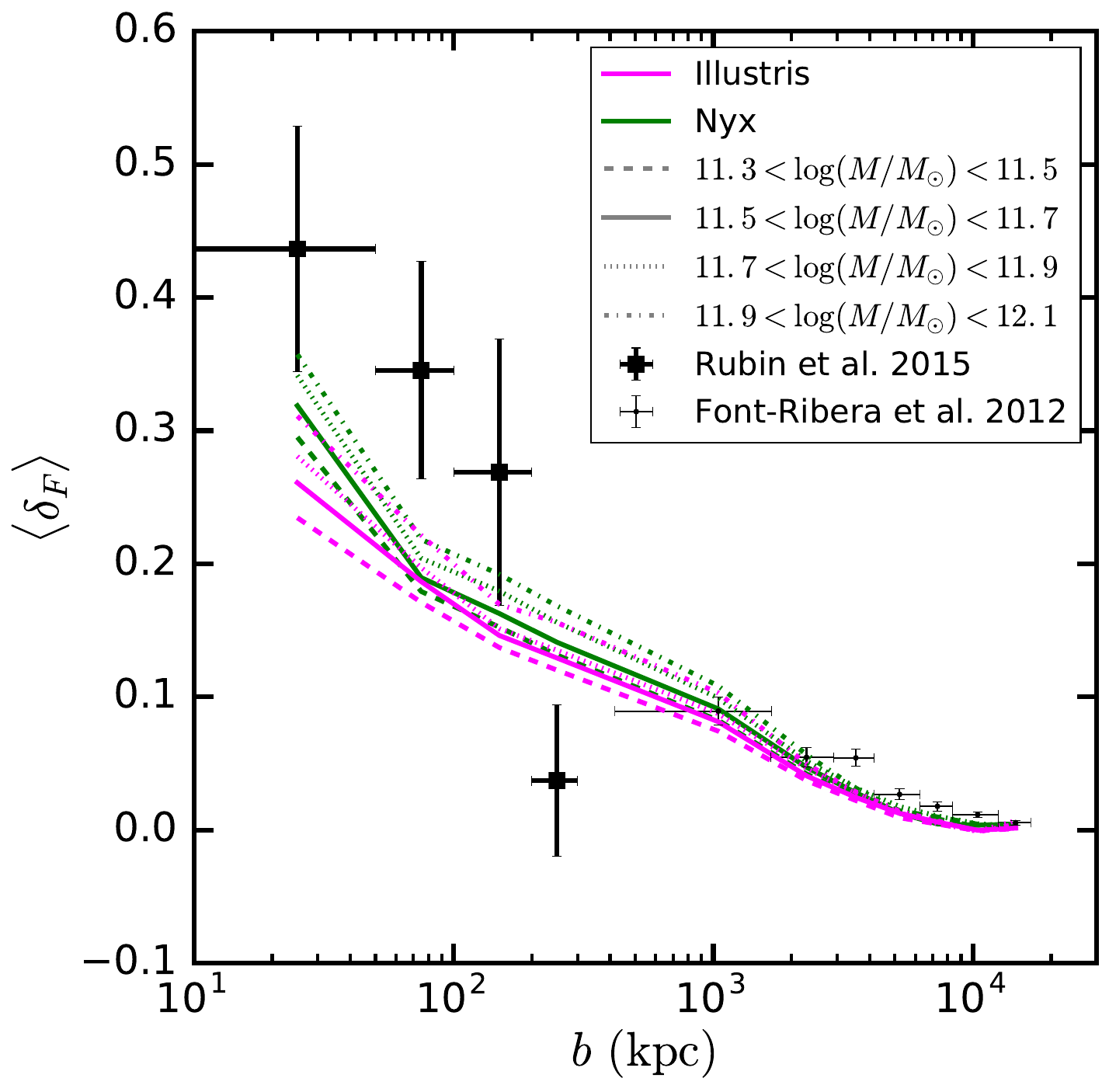}
\caption{Mean \lya flux contrast around QSOs (left panel) and DLAs (right panel) predicted by Nyx (green lines) and Illustris (magenta lines) for different halo mass bins of 0.2 dex width (indicated with different line styles). In the right panel, black squares and black circles indicate the measurements by \cite{Prochaska_2013} and \cite{Font-Ribera_2013}, respectively. In the left panel, the same symbols refer to the observations by \cite{Rubin_2015} and \cite{Font-Ribera_2012b}, respectively.}\label{fig:mass}
\end{figure*}

We want to investigate how the mass of a halo impacts the \lya absorption in the CGM. More massive halos reside in denser regions, which would yield more absorption. On the other hand, more massive halos contain hotter gas, and a higher temperature causes less absorption. It is not obvious which effect should prevail.

We reproduce once again the measurements of $\langle \delta_F \rangle$ around QSOs by \cite{Prochaska_2013} and \cite{Font-Ribera_2013}, selecting a sample of halos within a different mass range. Instead of setting a minimum halo mass as explained in \S~\ref{sec:sel_halos}, we consider different mass bins with an extension of 0.2 dex, centered in $10^{12.2}\, M_{\odot}$, $10^{12.4}\, M_{\odot}$, and $10^{12.6}\, M_{\odot}$. Likewise, to mock the observations of \lya absorption around DLAs \citep{Font-Ribera_2012b, Rubin_2015}, we define different mass bins centered in $10^{11.4}\, M_{\odot}$, $10^{11.6}\, M_{\odot}$, $10^{11.8}\, M_{\odot}$, and $10^{12}\, M_{\odot}$, all with a width of 0.2 dex. 

We plot the predictions of the $\langle \delta_F\rangle$ profile around QSOs and DLAs given by the simulations in the left and right panels of Figure \ref{fig:mass}, respectively. Green and magenta lines refer to Nyx and Illustris, respectively, while different line styles distinguish the aforementioned mass bins. The data are plotted with the same symbols and color coding as in Figures \ref{fig:mean_flux} and \ref{fig:mean_flux_DLA}.

Both for the QSO and DLA measurements, we find out that $\langle \delta_F \rangle$ slightly increases at larger halo masses. 
This trend is in agreement with the results obtained by \cite{Meiksin_2017} with the Sherwood suite of hydrodynamic simulations. At impact parameters $\lesssim 1 \Mpc$, the values of $\langle \delta_F \rangle$ predicted by Nyx or Illustris in two adjacent mass bins differ by $\lesssim 0.03$. At larger impact parameters, the effect of the halo mass is less pronounced.
The differences in the predictions of Nyx and Illustris for different halo masses at small impact parameters cannot explain the discrepancies between simulations and data within the virial radius of QSOs (see Figure \ref{fig:mean_flux}). Therefore, a possible systematic error of 0.2 dex in our procedure to calibrate the halo masses (see \S~\ref{sec:sel_halos}) would not affect the main conclusions of this work. We also point out that the good, but not perfect agreement between the simulations and BOSS data at large scales (see Figures \ref{fig:mean_flux} and \ref{fig:mean_flux_DLA}) should not be too worrisome, as small systematic errors in the determination of halo masses can partially account for the discrepancies with the data (Figure \ref{fig:mass}). On the other hand, the high precision of BOSS data should motivate further improvements of cosmological simulations.

\cite{Font-Ribera_2012b} estimated the typical mass of DLAs to be $10^{12} \, M_{\odot}$ fitting the DLA-\lya cross-correlation measurements with a model based on linear theory. This result was somewhat controversial, as DLA-hosting halos were thought to be less massive. For the first time, we compared \cite{Font-Ribera_2012b} measurements with a fully non-linear model, using cosmological hydrodynamic simulations. Whereas the two innermost BOSS data points favor smaller masses, it is hard to draw robust conclusions at larger separations, because the differences between the predictions of simulations within different mass bins are not pronounced enough.

We remind the reader that, for all foreground objects, we assumed the HOD to be a step function, which might be a somewhat questionable choice. However, recent observations of QSO clustering show that the parameter dominating the HOD is the mean mass of the sample \citep{Rodriguez-Torres_2017}, which in our context is effectively set by the mass threshold $M_{\rm min}$. Thus, given the overall mild dependence of the $\langle \delta_F \rangle$ profile on $M_{\rm min}$ evidenced by Figure \ref{fig:mass}, we expect even smaller variations upon implementing a more refined HOD model. We also point out that a mass-independent HOD in the halo mass and redshift range considered in this work is consistent with current uncertainties on the parameters of HOD models \citep{Rodriguez-Torres_2017}. Hence, a more detailed investigation of the impact of the exact shape of the HOD on the mean \lya flux profile around QSOs is beyond the scope of this work.

\subsection{Redshift of Foreground Objects}
\label{sec:redshift}

\begin{figure*}
\centering
\includegraphics[width=0.49\textwidth]{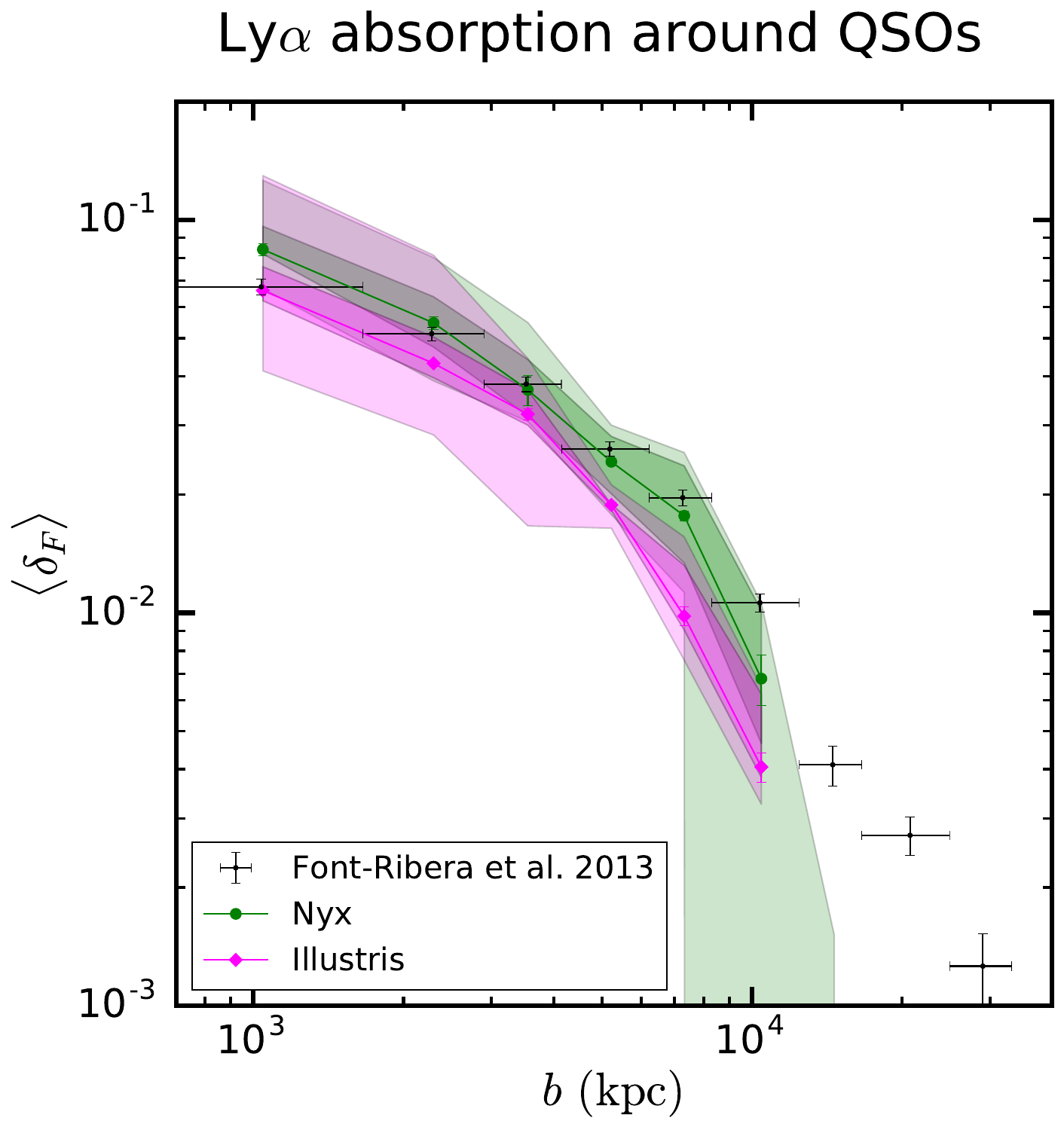}
\includegraphics[width=0.49\textwidth]{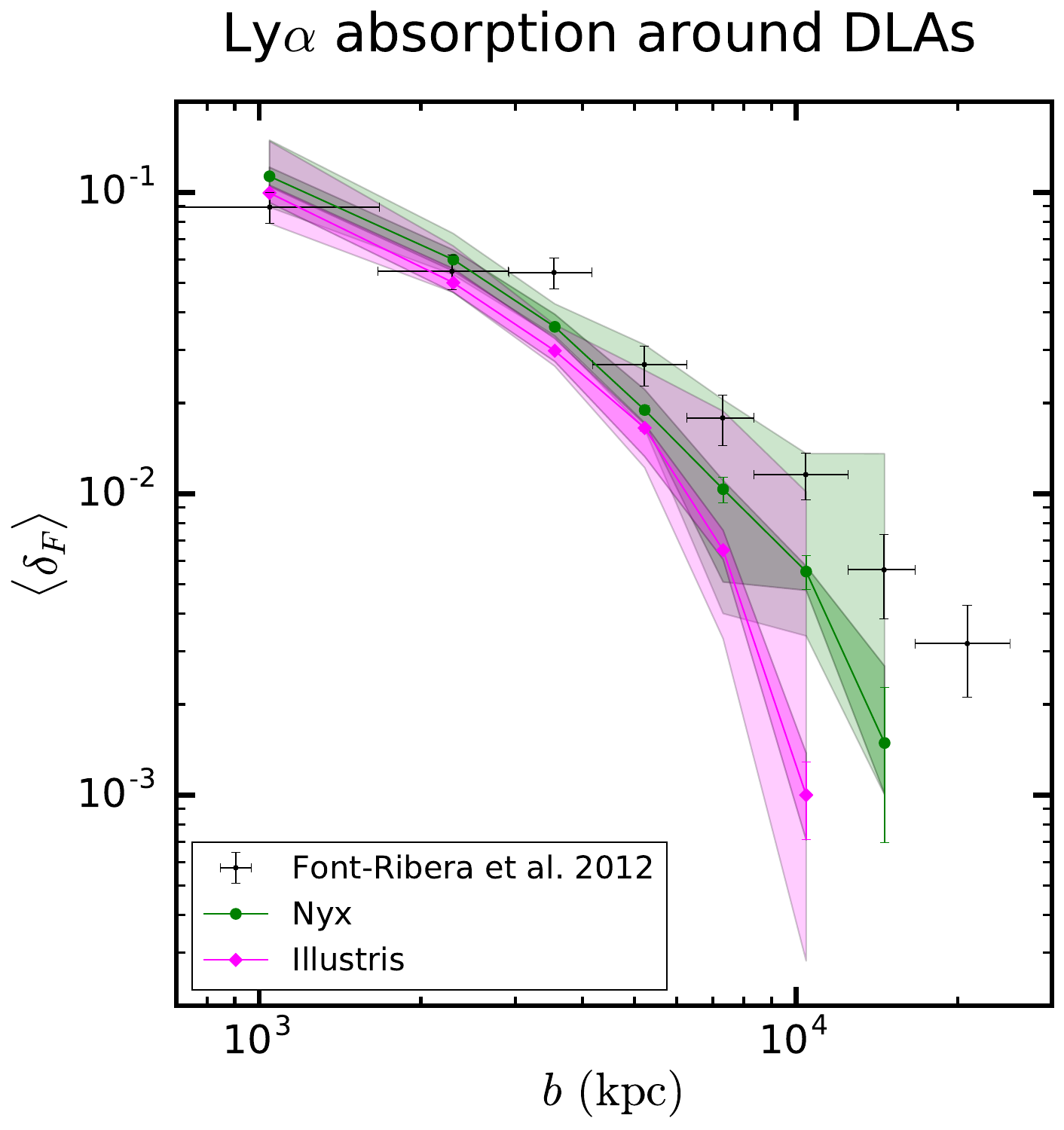}
\caption{Mean \lya flux contrast around QSOs (left panel) and DLAs (right panel) predicted by Nyx (green circles connected with green solid line) and Illustris (magenta circles connected with magenta solid line) for transverse distances $b> 1 \Mpc$. The green and magenta dark shaded areas show the spread in the prediction of $\langle \delta_F \rangle$ due to the uncertainty in the mass cut (see \S~\ref{app:halo_masses}) for Nyx and Illustris, respectively. Likewise, the light shaded areas indicate the spread in the predicted $\langle \delta_F \rangle$ because of the approximations in the modeling of the redshift distribution of the foreground objects (see \S~\ref{sec:redshift}). The vertical green and magenta error bars, where appreciable, show the 1$\sigma$ error in the predicted $\langle \delta_F \rangle$  given by the sample size of the simulated spectra. In the left and right panel, the black circles indicate the measurements by \cite{Font-Ribera_2013} and \cite{Font-Ribera_2012b}, respectively.}\label{fig:largeb}
\end{figure*}

All observations mentioned in this work have been reproduced with Nyx and Illustris taking all foreground objects at the median redshift of the corresponding data sets, effectively neglecting their spread in redshift.  Although the mean flux of the IGM evolves across the redshift range, this should not represent a big issue, since the quantity provided by the observations is not the mean flux profile, but the $\langle \delta_F \rangle$ profile. Despite the definition of $\langle \delta_F \rangle$ normalizes out the mean flux of the IGM, there may still be some residual redshift-dependence in the $\langle \delta_F \rangle$ profile predicted by the simulations, which needs to be evaluated. 
For this purpose, we reproduced \cite{Prochaska_2013} measurements taking all foreground QSOs at redshifts $z=2$ and $z=3$, which bracket the redshift range of the observations. As expected, at $z=3$ we have more absorption, because the neutral fraction of hydrogen is higher at earlier redshifts. 
Nevertheless, in the innermost bin, $\langle \delta_F \rangle$ increases by $\sim0.1$ for Nyx and $\sim 0.05$ for Illustris, which is still not enough to reproduce \cite{Prochaska_2013} data. 
Therefore, even if we unrealistically modeled all foreground QSOs at the upper bound of the redshift range of the observations, we could not explain the underprediction of \lya absorption by the simulations. We ran an analogous test for the measurements by \cite{Rubin_2015}, and our conclusions did not change with respect to what stated in \S~\ref{sec:lbg}. 

On the scales probed by the BOSS measurements, $\langle \delta_F \rangle$ varies by $\lesssim 0.03$
 when snapshots at $z=2$ and $z=3$ instead of $z\approx 2.4$ are considered. These differences are thus smaller than the ones observed at $b<1 \Mpc$, but large enough to account for most of the slight discrepancies between the simulations and BOSS data.
This is shown in Figure \ref{fig:largeb}, where we plot the $\langle \delta_F \rangle$ profile around QSOs and DLAs, for transverse distances $b>1 \Mpc$. The black data points represent BOSS measurements, and the green and magenta markers connected with solid lines the predictions of Nyx and Illustris, respectively. The light shaded green and magenta regions show the uncertainty on the predictions of Nyx and Illustris, respectively, due to the spread in redshift of the foreground objects. The dark shaded regions around each simulated mean \lya flux profile delimits the maximum change to be expected in the mean \lya flux profile upon a different choice of the halo mass of 0.2 dex, as explained in \S~\ref{sec:halo_mass}. The error bars on the profiles given by the simulations indicate the statistical $1\sigma$-error due to the sample size of the simulated spectra.

Although the mean \lya flux profiles given by both simulations tend to generally underpredict the observations, they result compatible with them once all sources of uncertainty intrinsic in the modeling are considered. As far as the absorption profiles around DLAs are concerned, the only data point posing mild tension with the theoretical predictions is the one at $\sim 3.5 \Mpc$. This data point seems to be slightly off the trend suggested by the other measurements, thus it is not clear whether it exposes limitations of the simulations and our modeling. Regarding the observations of the $\langle \delta_F \rangle$ around QSOs, all data points except the one at $\sim 14 \Mpc$ are compatible with the predictions of Nyx. On the contrary, Illustris seems to underestimate the observations at $b> 5 \Mpc$, and appears to provide systematically lower values of $\langle \delta_F \rangle$ with respect to Nyx. Also in this case, we are reluctant to interpret this mild discrepancy with the observations as a failure of Illustris. In fact, the conversion of \cite{Font-Ribera_2013} measurements into a $\langle \delta_F \rangle$ incorporates the assumption of equally weighting the \lya flux measured in each pixel of the data used to estimate of the DLA-\lya and QSO-\lya cross-correlation  (see the Appendix \ref{app:BOSS_conversion}). However justified, this assumption might exacerbate the aforementioned discrepancies. Furthermore, in assessing the uncertainties underlying the theoretical modeling, we focused on what we view as the main potential sources of error.

Our general conclusion is that the high precision of BOSS observations grant tight constraints on galaxy formation, and potentially on the cosmological models underlying simulations. In the latter case, a more careful modeling of the redshift distribution and, in second instance, of the mass cut of foreground objects, is necessary, and we wish  to do it in future work. On the contrary, the redshift distribution of foreground LBGs in \cite{Adelberger_2003}, \cite{Adelberger_2005}, \cite{Crighton_2011} and \cite{Turner_2014} is so narrow with respect to the snapshots available for the simulations, that the test discussed in this section becomes superfluous.

There is another effect connected to the redshift of the foreground objects. The velocity windows considered to reproduce the observations are centered around the systemic velocity of the foreground galaxies or quasars. Following \cite{Meiksin_2017}, we modeled the typical observational errors in the redshifts of the foreground halos by adding a Gaussian-distributed random component to their velocities, with a variance of $130 \, \rm km \, s^{-1}$ and $520 \, \rm km \, s^{-1}$ for LBG/DLA and quasar hosts, respectively. We found that introducing such scatter has a marginal ($<1.2\%$) 
effect on the prediction of the \lya absorption profiles. This is not surprising, because the data are already averaged over a large velocity window along the line of sight, exactly for the purpose of dealing with the errors on the redshifts of the foreground objects. 

\subsection{Sample Size of Observed Spectra}
\label{sec:sample_size}

\begin{figure*}
	\centering
		\includegraphics[width=0.49\textwidth]{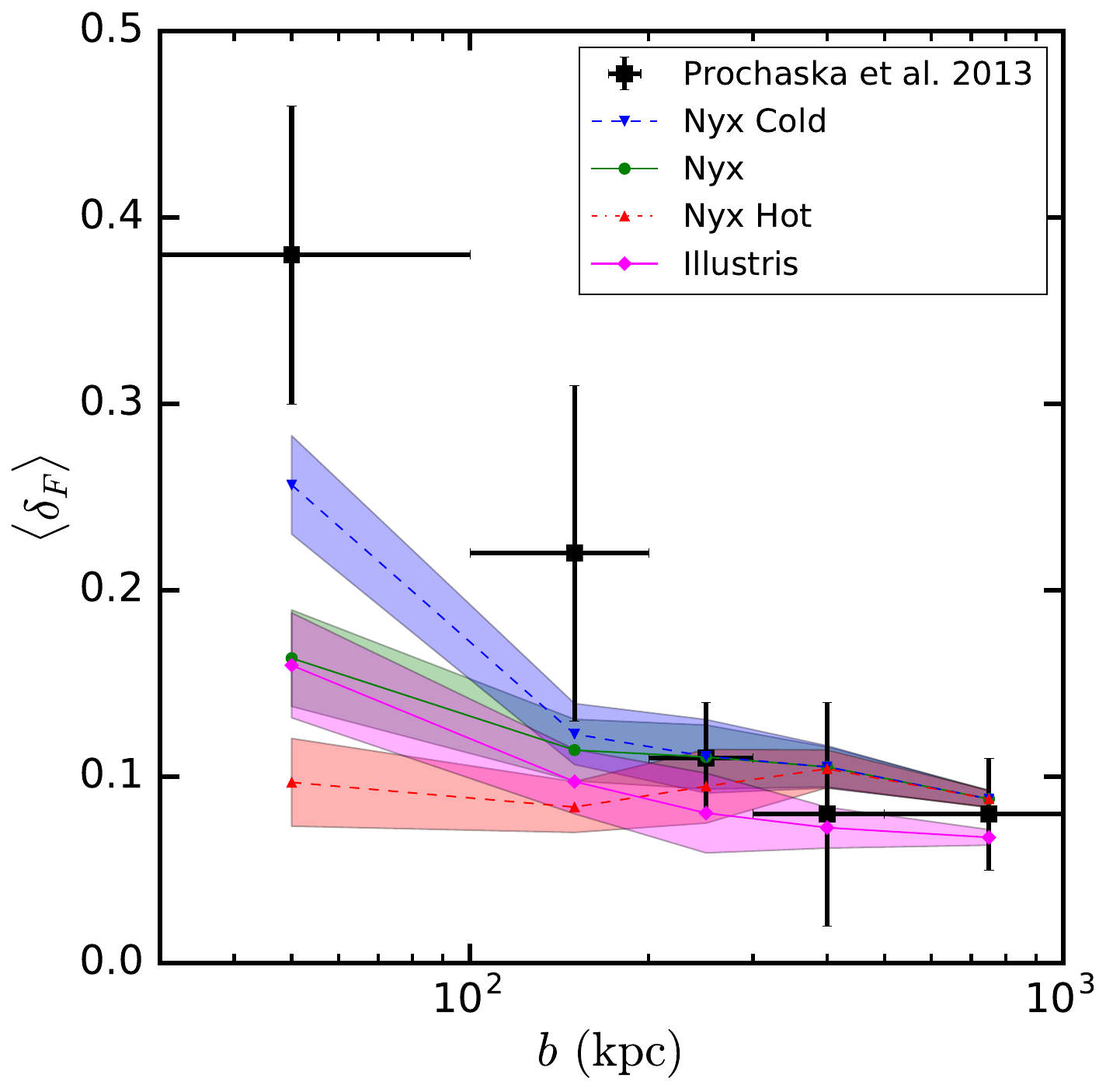} \hfill
		\includegraphics[width=0.49\textwidth]{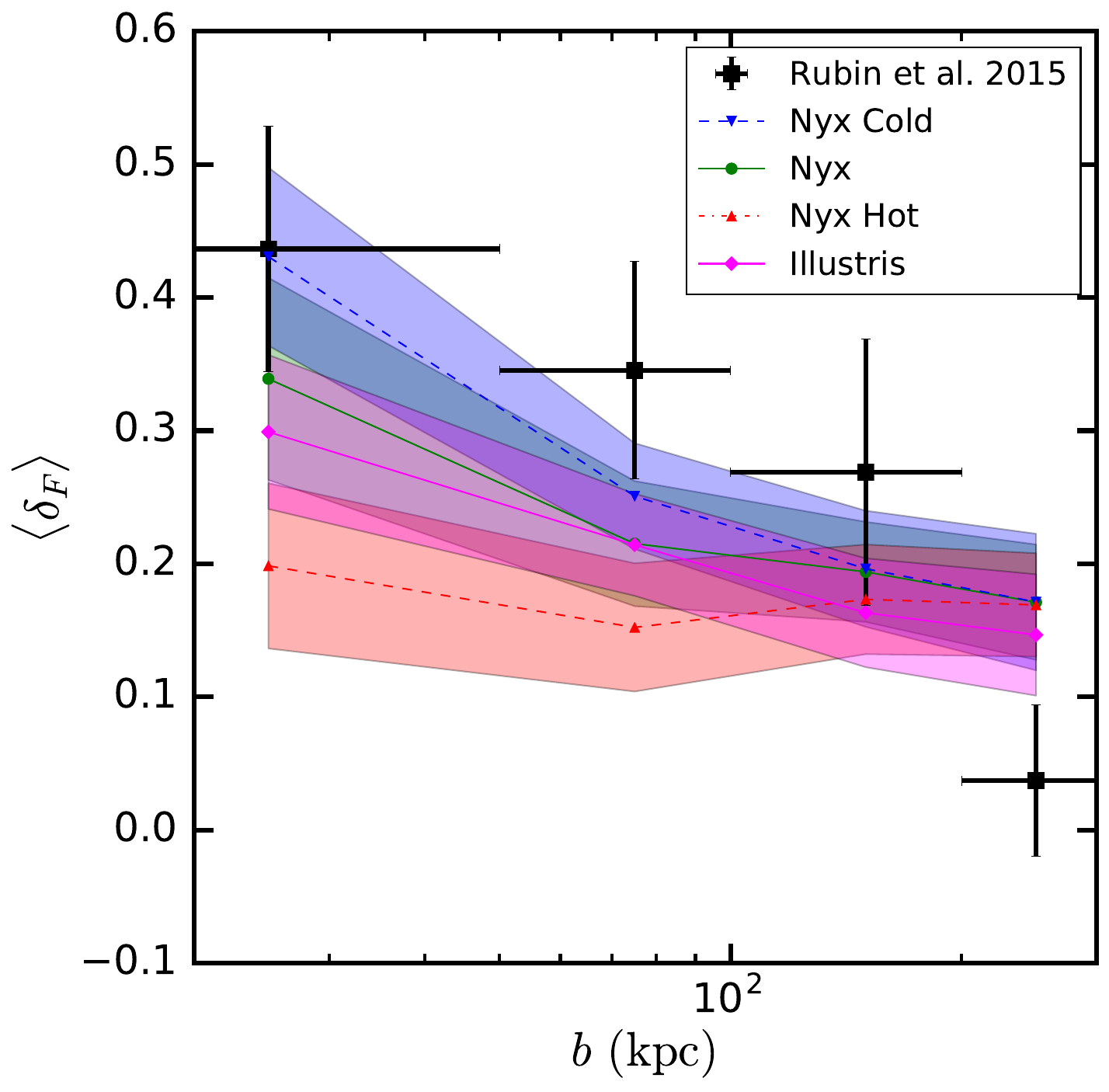}
	\caption{\textit{Left panel}: Mean \lya flux contrast around QSOs, at different transverse separations. The black squares represent the observations by \cite{Prochaska_2013}; the vertical bars are the $1\sigma$ errors of the measurements, while the horizontal bars show the extension of the impact parameter bins. The results of the simulations are represented with the same color coding, markers and line styles as in Figures \ref{fig:mean_flux} and \ref{fig:mean_flux_DLA}. The shaded areas delimit the $1\sigma$ scatter around the estimate of $\langle \delta_F \rangle$ due to the limited sample size of the observed spectra (see text for details). \textit{Right panel}: Same as the left panel, but for the mean \lya flux contrast around DLAs. The black squares represent the observations by \cite{Rubin_2015}; the vertical bars are the $1\sigma$ errors of the measurements, while the horizontal bars show the extension of the impact parameter bins.}
	\label{fig:scatter_mean}
\end{figure*}

Whereas the \lya absorption at large separation from foreground galaxies or quasars can be measured from tens of thousands of QSO spectra thanks to large-scale surveys like BOSS \citep{Font-Ribera_2012b, Font-Ribera_2013}, the number of the background quasar - foreground galaxy pairs with small transverse separations is about two orders of magnitude smaller \citep{Prochaska_2013, Turner_2014, Rubin_2015}. We want to understand to what extent the poor statistics of spectra in observations can affect the error on the estimation of the mean \lya flux contrast. This uncertainty is already accounted for by the error bars in the data, but we can use simulations to estimate its contribution to the total error in the measurements.

From our sample of simulated spectra in each impact parameter bin of the observations by \cite{Prochaska_2013}, \cite{Turner_2014} and \cite{Rubin_2015}, we draw 50 subsamples with as many skewers as the observed spectra in the bin considered. We plot our results for the observations by \cite{Prochaska_2013} and \cite{Rubin_2015} in the left and right panels of Figure \ref{fig:scatter_mean}, respectively. The black squares represent the observations; the vertical bars are the $1\sigma$ errors in the measurements, while the horizontal bars mark the bin widths. The results of the simulations follow the same color coding, marker and line styles as in Figure \ref{fig:mean_flux}. The shaded magenta, green, blue and red regions delimit the $16^{\rm th}$ and $84^{\rm th}$ percentiles of the distribution of the estimate of $\langle \delta_F \rangle$ given by the 50 subsamples around the value obtained from the entire sample. As such, the bands represent the contribution to the error of the measured $\langle \delta_F \rangle$ due to the number of observed spectra.

The left panel of Figure \ref{fig:scatter_mean} shows that the limited size of the sample of observed spectra contributes up to $\sim 60 - 70 \%$ to the total error of \cite{Prochaska_2013} measurements in the range $(200,\,500) \kpc$, and appears to be the dominant source of error in the bin $(200,\,300) \kpc$. The contribution is even bigger in the observations by \citealt{Rubin_2015} (right panel of Figure \ref{fig:scatter_mean}) in the range $(50,\,300)\kpc$, and dominates the error bar in the innermost bin. 

\begin{figure}
\begin{center}
\includegraphics[width=\columnwidth]{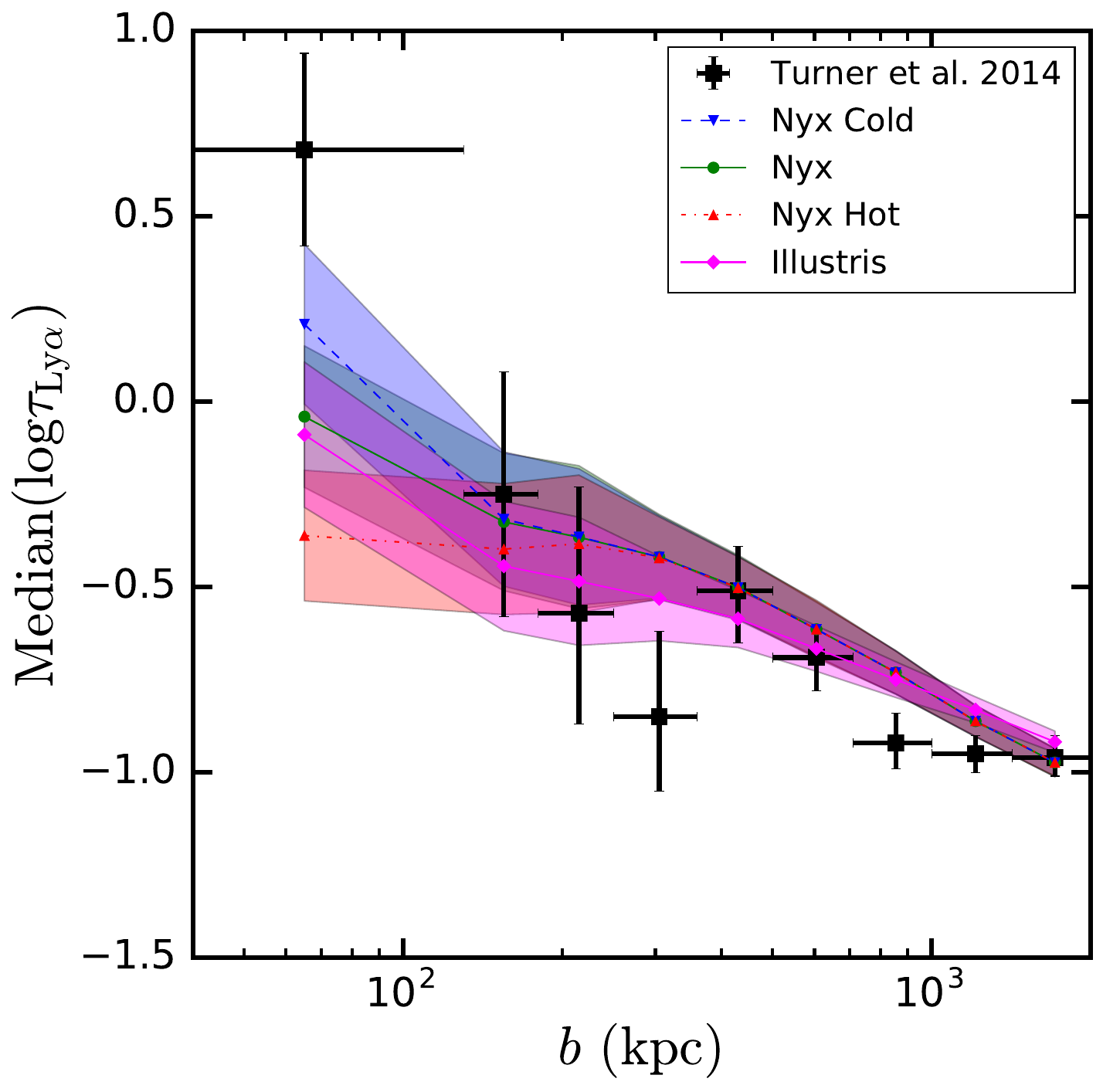}
\caption{Logarithm of the median \lya pixel optical depth around LBGs, at different transverse separations. The black squares represent the observations by \cite{Turner_2014}; the vertical bars are the $1\sigma$ errors of the measurements, while the horizontal bars show the extension of the impact parameter bins. The results of the simulations are represented with the same color coding, markers and line styles as in the right panel of Figure \ref{fig:tau_median}. The shaded areas delimit the $1\sigma$ scatter around the estimate of the median optical depth due to the limited sample size of the observed spectra (see text for details). The sample size contributes significantly to the errors of the measurements.}\label{fig:scatter_median}
\end{center}
\end{figure}

Figure \ref{fig:scatter_median} shows that the scatter in the median \lya optical depth due to the sample size of the observations by \cite{Turner_2014}. The black squares represent the data, whereas the simulations follow the color-coding, lines and marker style of the lines as in Figure \ref{fig:scatter_mean}. Also the shaded areas have the same meaning as in Figure \ref{fig:scatter_mean}. The contribution to the error in the measurements due to the size of the sample of spectra is significant also in this case, accounting for $\sim50-70\%$ of the error bar.

We could not compute the scatter in $\langle \delta_F \rangle$ due to the sample size of spectra in BOSS \citep{Font-Ribera_2012b, Font-Ribera_2013}. We verified that, assuming that the spectra uniformly populate the transverse separation bins, the scatter in $\langle \delta_F \rangle$ due to the sample size is negligible. From the size of the LBG sample in \citep{Adelberger_2003}, \cite{Adelberger_2005} and \cite{Crighton_2011}, we expect the scatter in $\langle \delta_F \rangle$ to be of the same order of magnitude as in \cite{Rubin_2015} (right panel of Figure \ref{fig:scatter_mean}).

To summarize, the simulations considered in this work predict that the relatively poor statistics of observed QSO spectra at small separation ($\lesssim 1 \Mpc)$ from foreground objects should contribute up to $\sim50\%-70\%$ to the errors on the measurements. Given that the error bars are dominated by the statistical error, and not by systematics \citep{Font-Ribera_2012b, Font-Ribera_2013, Prochaska_2013, Turner_2014, Rubin_2015}, this means that the simulations considered in this work underestimate the variance of the observations.

\subsection{Sub-resolution Physics}
\label{sec:resolution}

The limited resolution of simulations can be one of the reasons for the discrepancies between the simulations and observations shown in this work. Observations of \lya absorption around foreground $z\approx2.5$ galaxies imply the presence of large-column density, metal-enriched, $< 500 \, \rm pc$ clouds within an otherwise diffuse CGM (\citealt{Crighton_2015}; see also \citealt{Simcoe_2006, Crighton_2013}). Also, the presence of $\lesssim 20 \, \rm pc$ dense clouds has been invoked to explain the high surface brightness of extended giant \lya nebulae around quasars \citep[e.g.][]{Arrigoni-Battaia_2015}. These clouds can be resolved neither by state-of-the art cosmological hydrodynamic simulations nor zoom-in numerical simulations. In fact, \cite{Crighton_2015} showed that, to resolve the clumps inferred from their observations, AMR (Adaptive Mesh Refinement) simulations should reach a cell size of $\lesssim 140 \, \rm pc$ in the CGM, and SPH simulations should have a mass resolution better than $4 \, M_{\odot}$ \citep[see also][]{Agertz_2007, McCourt_2016, Stern_2016}. These requirements are several orders of magnitude beyond the achievements of any cosmological numerical simulation at present.

We compared the $\langle \delta_F \rangle$ profiles predicted by the high-resolution Illustris run adopted in this work with the profiles given by the two publicly available runs at lower resolutions. Specifically, their mass resolutions are a factor $8$ and $64$ worse than in the high-resolution run, respectively. We verified that the predictions of the $\langle \delta_F \rangle$ profile given by Illustris at the three available resolutions are well converged for $b>500\kpc$. The intermediate and high resolution runs agree within $5\%$ for $50\kpc < b < 100 \kpc$, and within $19\%$  for $b<50 \kpc$. Conservatively assuming that the $\langle \delta_F \rangle$ estimate in the range  $b<50 \kpc$ would increase by another 19\% if one could improve by another factor of 8 the resolution of the best Illustris run, that would still not be enough to match the data.

Although zoom-in simulations seem to capture the small-scale physics of the CGM, the resolution necessary to resolve the clumpy structure of the CGM is beyond current and near-future cosmological simulations. For such simulations, as suggested by \cite{Crighton_2015}, it may be wiser to simulate the CGM implementing sub-resolution prescriptions, as it is already the case for the modeling of star formation and galactic-scale outflows. 

\section{From Cross-Correlation to Mean Flux Contrast Profile}
\label{app:BOSS_conversion}

The cross-correlation of the \lya forest with QSOs (DLAs), measured by \citealt{Font-Ribera_2013} \citep{Font-Ribera_2012b}, is equivalent to the stacked mean \lya flux profile around QSOs (DLAs), measured by \citealt{Prochaska_2013} \citep{Rubin_2015}. As such, we can average the cross-correlation measurements within appropriate velocity windows along the line-of-sight, to translate such measurements into $\langle \delta_F \rangle$ profiles. In this section, we explain the details of the conversion done for \cite{Font-Ribera_2013} QSO-\lya cross-correlation measurements. The formalism is exactly the same also as far as the \cite{Font-Ribera_2012b} DLA-\lya cross-correlation observations are concerned.

\cite{Font-Ribera_2013} selected a sample of 61342 QSO spectra in the redshift range $2<z<3.5$ from the BOSS Data Release 9 \citep{Ahn_2012} 
and measured the \lya flux fluctuation at every pixel in each QSO spectrum. For the pixel $i$, this quantity is defined as
\begin{equation}
	\tilde{\delta}_{F_i} = \frac{f_i}{C_i \bar{F}(z_i)}-1 \, ,
\end{equation}
where $f_i$ is the measured flux, $C_i$ is the QSO continuum, and $F(z_i)$ is the mean transmitted flux obtained in the redshift bin containing the pixel redshift $z_i$. For each QSO, the pixels of all spectra are divided into bins of transverse and line-of-sight separation from the QSO ($b$ and $x$, respectively). The estimator of the cross-correlation in the bin $(b,\, x)$ is defined as
\begin{equation}
\label{eq:xi_estimator}
	\hat{\xi}_{b, \, x} = \frac{\sum _{n = 1}^N \sum_{k \in (b_n,\, x_n)} w_{nk} \tilde{\delta}_{F_{nk}}}{\sum_{n =1}^N \sum_{k \in (b_n,\, x_n)}  w_{nk}} \, .
\end{equation}
The index $n$ identifies the QSO, and $N$ is the total number of QSOs. The index $k$ identifies the pixels within the distance bin ($b_n,\,x_n)$ from the QSO $n$. The weights $w_{nk}$ are defined as
\begin{equation}
\label{eq:weight}
	w_{nk} = \left[ \sigma ^2 _F (z_{nk}) + \frac{\langle N_{nk}^2\rangle}{C_{nk}^2 \bar{F}_e^2(z_{nk})} \right]^{-1} \, .
\end{equation}
In the equation above, $\sigma ^2 _F (z_{nk})$ and $\bar{F}_e^2(z_{nk})$ are the intrinsic variance of the \lya forest flux contrast and the mean flux in the \lya forest at the pixel redshift $z_{nk}$, respectively. For the former, \cite{Font-Ribera_2013} adopt an analytic expression based on the redshift evolution of the power spectrum measured in \cite{McDonald_2006}, while for the latter the observations by \cite{Faucher-Giguere_2008b}. Finally, the term $\langle N_{nk} ^2\rangle$ is the noise at pixel $nk$, approximated as a Gaussian variance.

In \cite{Font-Ribera_2013}, the $b$-bins are delimited by (1, 4, 7, 10, 15, 20, 30, 40, 60, 80) $h^{-1} \cMpc$, while the $x$-bins are bounded by (-80, -60, -40, -30, -20, -15, -10, -6, -3, 0) $h^{-1} \cMpc$ and the same positive values. We are interested in writing the expression of the cross-correlation $\hat{\xi} _{b,\, \Delta v}$ for a transverse distance bin $b$ and a velocity window $\Delta v$, centered around the QSOs. If the velocity window contains $M$ line-of-sight distance bins as chosen by \cite{Font-Ribera_2013}, equation \eqref{eq:xi_estimator} can be re-written as
\begin{equation}
	\label{eq:xi_midstep}
	\hat{\xi}_{b, \, \Delta v} = \frac{  \sum_{m=1}^M \hat{\xi}_{b, \, x_m} \sum _{n = 1}^N  \sum_{k \in (b_n,\, x_{mn})}  w_{nk} }{ \sum_{n =1}^N  \sum_{m=1}^M  \sum_{k \in (b_n,\, x_{mn})} w_{nk}}
	 \, .
\end{equation}
To compute $\hat{\xi} _{b,\, \Delta v}$, one would need to have access to all spectra, in order to properly compute the weight function at each pixel. Since we do not have access to such data, we assume that the weight function is a constant. Physically, this is equivalent to assuming that the noise term in \eqref{eq:weight} is the same for all pixels and that the intrinsic variance of the \lya forest is approximately constant in the redshift range considered. Within such approximation, we can write
\begin{equation}
	\label{eq:appr_estimator}
	\hat{\xi}_{b, \, \Delta v} \approx \frac{\sum_{m=1}^M K_m \hat{\xi}_{b, \, x_m}}{\sum_{m=1}^M K_m }  \, ,
\end{equation}
where $K_m$ is the number of pixels in each bin $(b_n, \, x_{mn})$. On the other hand, within the same approximation that led to \eqref{eq:xi_midstep}, \eqref{eq:xi_estimator} becomes:
\begin{equation}
\label{eq:xi_last}
	\hat{\xi}_{b, \, \Delta v} \approx -\langle \delta_F(b,\,\Delta v) \rangle \, ,
\end{equation}
where $\delta_F(b,\,\Delta v) $ is the mean \lya flux contrast at impact parameter $b$ within a velocity window $\Delta v$ around all foreground QSOs (see the definition in equation\eqref{eq:delta_F} and the relative discussion). Therefore, comparing \eqref{eq:appr_estimator} with \eqref{eq:xi_last}, we obtain
\begin{equation}
\label{eq:QSO_conversion}
	\langle \delta _F (b, \, \Delta v) \rangle \approx - \hat{\xi}_{b, \, \Delta v} \approx  - \frac{\sum _{m=1} ^M K_m \, \hat{\xi}_{b, \, x_m} }{\sum _{m=1} ^M K_m } \, ,
\end{equation}
within the aforementioned assumption that the weights in \eqref{eq:xi_estimator} are constant.

We used \eqref{eq:QSO_conversion} to convert \cite{Font-Ribera_2013} observations to the same quantity measured by \cite{Prochaska_2013}. The velocity window is $\Delta v = 2000 \, \rm km \, s^{-1}$ (i.e. $\pm 1000 \, \rm km\, s^{-1}$ around the foreground object), which corresponds to $\sim 20 \, h^{-1} \, \cMpc$ assuming the same cosmology as \cite{Font-Ribera_2013}. 
The results of our analysis are listed in Table \ref{tab:Font-Ribera13}, where the errors on $\langle \delta_F \rangle$ are determined propagating the errors in \cite{Font-Ribera_2013} measurements. As a caveat, we point out that our estimate \eqref{eq:QSO_conversion} would be exact if \cite{Font-Ribera_2013} data were re-analyzed computing the weights defined in \eqref{eq:weight} pixel by pixel. For convenience, we report the analogous measurements by \cite{Prochaska_2013} at smaller impact parameter in Table \ref{tab:Prochaska13}. 

The comparison between \cite{Rubin_2015} and \cite{Font-Ribera_2012b} can be done following the same argument explained in this appendix, using a velocity window of $1000 \, \rm km \, s^{-1}$. In this case, the line-of-sight bins chosen by \cite{Font-Ribera_2012b} do not cover exactly the desired velocity window, so we linearly interpolate between their data points in order to average them within \cite{Rubin_2015} velocity window. The results are tabulated in table \ref{tab:Font-Ribera12}. It is the first time that large-scale measurements of the \lya cross-correlation function \citep{Font-Ribera_2012b, Font-Ribera_2013} are used together with observations of \lya absorption in the CGM \citep{Prochaska_2013, Rubin_2015} to jointly constrain the physics of the IGM and the CGM.

\begin{table}[h!]
	\begin{center}
		\caption{\lya absorption at large impact parameter from QSOs, inferred from \cite{Font-Ribera_2013}}
		\label{tab:Font-Ribera13}
		\begin{threeparttable}
			\begin{tabular}{rrc}
				\hline
				$b_{\rm min}$ \tnote{a} & $b_{\rm max}$ \tnote{b} & $\langle \delta_F \rangle$ \tnote{c}\\
				$(h^{-1} \, \cMpc)$ & $(h^{-1} \, \cMpc)$ & \\
				\hline
				1 & 4 & $0.0675\pm0.0031$ \\
				4 & 7 & $0.0514\pm0.0020$ \\
				7 & 10 & $0.0382\pm0.0017$ \\
				10 & 15 & $0.0261\pm0.0011$ \\
				15 & 10 & $0.01942\pm0.00092$ \\
				20 & 30 & $0.01061\pm0.00056$ \\
				30 & 40 & $0.00410\pm0.00048$ \\
				40 & 60 & $0.00271\pm0.00030$ \\
				60 & 80 & $0.00126\pm0.00026$\\
				\hline
			\end{tabular}
			\begin{tablenotes}
				\item[a] Inner edge of the impact parameter bin.
				\item[b] Outer edge of the impact parameter bin.
				\item[c] Mean \lya flux contrast.
		\end{tablenotes}
		\end{threeparttable}
	\end{center}
\end{table}

\begin{table}[h!]
	\begin{center}
		\caption{\lya absorption at small impact parameter from QSOs, quoted from \cite{Prochaska_2013}}
		\label{tab:Prochaska13}
		\begin{threeparttable}
			\begin{tabular}{rrc}
				\hline
				$b_{\rm min}$ \tnote{a} & $b_{\rm max}$ \tnote{b} & $\langle \delta_F \rangle$ \tnote{c}\\
				$(\rm kpc)$ & $(\rm Mpc)$ & \\
				\hline
				
				0 & 100 & $0.38\pm0.08$ \\
				100 & 200 & $0.22\pm0.09$ \\
				200 & 300 & $0.11\pm0.03$ \\
				300 & 500 & $0.08\pm0.06$ \\
				500 & 1000 & $0.08\pm0.03$ \\
				\hline
			\end{tabular}
			\begin{tablenotes}
				\item[a] Inner edge of the impact parameter bin.
				\item[b] Outer edge of the impact parameter bin.
				\item[c] Mean \lya flux contrast.
		\end{tablenotes}
		\end{threeparttable}
	\end{center}
\end{table}

\begin{table}[h!]
	\begin{center}
		\caption{\lya absorption at large impact parameter from DLAs, inferred from \cite{Font-Ribera_2012b}}
		\label{tab:Font-Ribera12}
		\begin{threeparttable}
			\begin{tabular}{rrc}
				\hline
				$b_{\rm min}$ \tnote{a} & $b_{\rm max}$ \tnote{b} & $\langle \delta_F \rangle$ \tnote{c} \\
				$(h^{-1} \, \cMpc)$ & $(h^{-1} \, \cMpc)$ & \\
				\hline
				1 & 4	 & $0.083\pm0.012$ \\
				4 & 7	 & $0.0513\pm0.0079$ \\
				7 & 10 & $0.0523\pm0.0070$ \\
				10 & 15 & $0.0271\pm0.0044$ \\
				15 & 20 & $0.0182\pm0.0038$ \\
				20 & 30 & $0.0105\pm0.0023$ \\
				30 & 40 & $0.0056\pm0.0019$ \\
				40 & 60 & $0.0027\pm0.0012$\\
				\hline
			\end{tabular}
			\begin{tablenotes}
				\item[a] Inner edge of the impact parameter bin.
				\item[b] Outer edge of the impact parameter bin.
				\item[c] Mean \lya flux contrast.
		\end{tablenotes}
		\end{threeparttable}
	\end{center}
\end{table}

\section{Mean Flux Contrast at Small Separations from DLA\lowercase{s}}
\label{app:Rubin}

\cite{Rubin_2015} stacked the absorption spectra of background QSOs passing at different transverse separation from foreground DLAs in four bins of impact parameter. After re-normalizing the spectra to the pseudo-continuum 
measured in the velocity intervals $(-4000,\,-3500)\, \rm km \, s^{-1}$ and $(3500,\, 4000), \rm km \, s^{-1}$, they determined the average equivalent width of the \lya absorption $\langle W_{\mathrm{Ly}\alpha} \rangle$ in the velocity window $\Delta v = 1000 \, \rm km \, s^{-1}$ around the DLAs. From the definition of equivalent width \citep{Draine_book}, the mean 
\lya flux $\langle F \rangle_{\Delta v}$ within a velocity window $\Delta v$ centered in the foreground DLA along a single absorption spectrum 
can be inferred as
\begin{equation}
	\langle F \rangle_{\Delta v} =F_0 \left(1- \frac{c \, W_{\mathrm{Ly}\alpha}}{\Delta v \,  \lambda_{\mathrm{Ly}\alpha}} \right)\, ,
\end{equation}
where $c$ is the speed of light in vacuum, $\lambda_{\mathrm{Ly}\alpha}$ the rest-frame wavelength of the \lya transition, and $F_0$ the pseudo-continuum. Since the pseudo-continuum measured by \cite{Rubin_2015} is meant to represent the mean flux of the IGM at the redshift of their observations, we can infer the mean \lya flux contrast at a certain impact parameter from a sample of foreground DLAs simply as
\begin{equation}
	\langle \delta_F \rangle =  \frac{c \, \langle W_{\mathrm{Ly}\alpha}\rangle}{\Delta v \, \lambda_{\mathrm{Ly}\alpha}} \, .
\end{equation}
The results are listed in Table \ref{tab:Rubin15}.

\begin{table}[h!]
	\begin{center}
		\caption{\lya absorption around DLAs obtained from \cite{Rubin_2015}}
		\label{tab:Rubin15}
		\begin{threeparttable}
			\begin{tabular}{rrc}
				\hline
				$b_{\rm min}$ \tnote{a} & $b_{\rm max}$ \tnote{b} & $\langle \delta_F \rangle$ \tnote{c}\\
				$(\rm kpc)$ & $(\rm kpc)$ & \\
				\hline
				0 & 50 & $0.436\pm0.092$ \\
				50 & 100 & $0.345\pm0.082$ \\
				100 & 200 & $0.269\pm0.010$ \\
				200 & 300 & $0.037\pm0.057$ \\
				\hline
			\end{tabular}
			\begin{tablenotes}
				\item[a] Inner edge of the impact parameter bin.
				\item[b] Outer edge of the impact parameter bin.
				\item[c] Mean \lya flux contrast.
			\end{tablenotes}
		\end{threeparttable}
	\end{center}
\end{table}

\section{Gas Velocity and \lya Absorption}
\label{app:velocity}

\begin{figure*}

\begin{adjustbox}{addcode={\begin{minipage}{\textheight}}{\caption{%
      Radial velocity - density relationship of hydrogen at different radial bins from a foreground QSO. The volume-weighted 2D histograms are plotted upon stacking 100 randomly drawn QSO-hosting halos in Nyx ($> 10^{12.5}\,M_{\odot}$, top panels) and all 72 QSO-hosting halos in Illustris ($> 10^{12.4}\,M_{\odot}$, bottom panels), in different bins of distance from the center of the halos. From left to right, the histograms refer to the radial bins $(0,\, r_{\rm vir})$, $(r_{\rm vir},\, 2r_{\rm vir})$, $(2r_{\rm vir},\, 3r_{\rm vir})$, $(3r_{\rm vir},\, 5r_{\rm vir})$ and $(5r_{\rm vir},\, 1 \Mpc)$. Negative velocities denote inflowing gas. Overall, the two simulations yield qualitatively similar diagrams at all virial radii. Illustris presents an overall offset of $\sim + 50 \, \rm km \, s^{-1}$ with respect to Nyx in the three innermost bins, and generally more outflowing gas than Nyx (see text for details).
      } \label{fig:vrad-nH_CGM}\end{minipage}},rotate=90,left}
     \hspace{0.15\textheight}
     \includegraphics[width=1.01\textwidth]{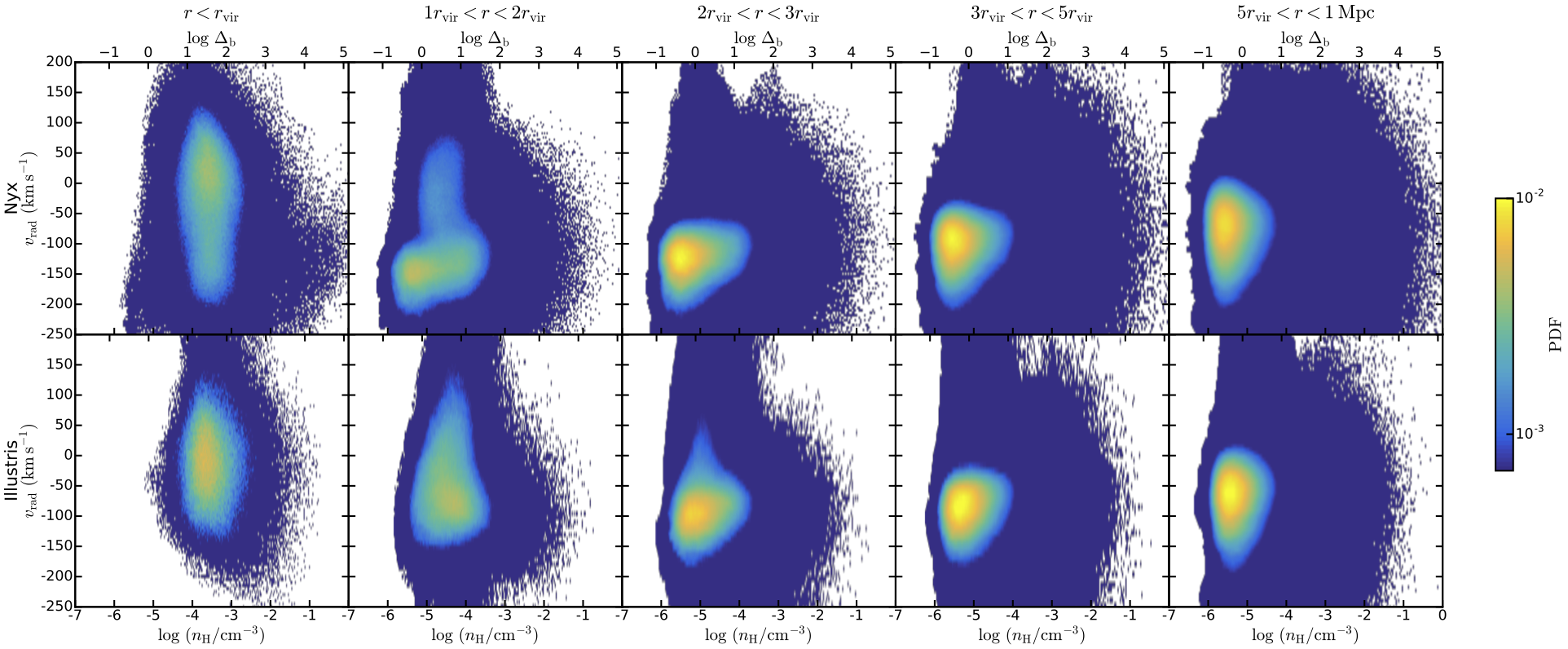}%

\end{adjustbox}

\vspace{-\textheight}
  \begin{adjustbox}{addcode={\begin{minipage}{\textheight}}{\caption{%
      Same as in Figure \ref{fig:vrad-nH_CGM}, but for a sample of 100 randomly chosen DLA/LBG-hosting halos in Nyx ($> 10^{11.7}\,M_{\odot}$, top panels) and Illustris ($> 10^{11.6}\,M_{\odot}$, bottom panels). Also in this case, the two simulations yield overall qualitatively similar diagrams at all virial radii. Illustris presents an overall offset of $\sim + 20 \, \rm km \, s^{-1}$ with respect to Nyx in the three innermost bins, and generally more outflowing gas than Nyx (see text for details).
      }\label{fig:vrad-nH_CGM_LBG}\end{minipage}},rotate=90,right}
      \hspace{0.17\textheight}
     \includegraphics[width=1.01\textwidth]{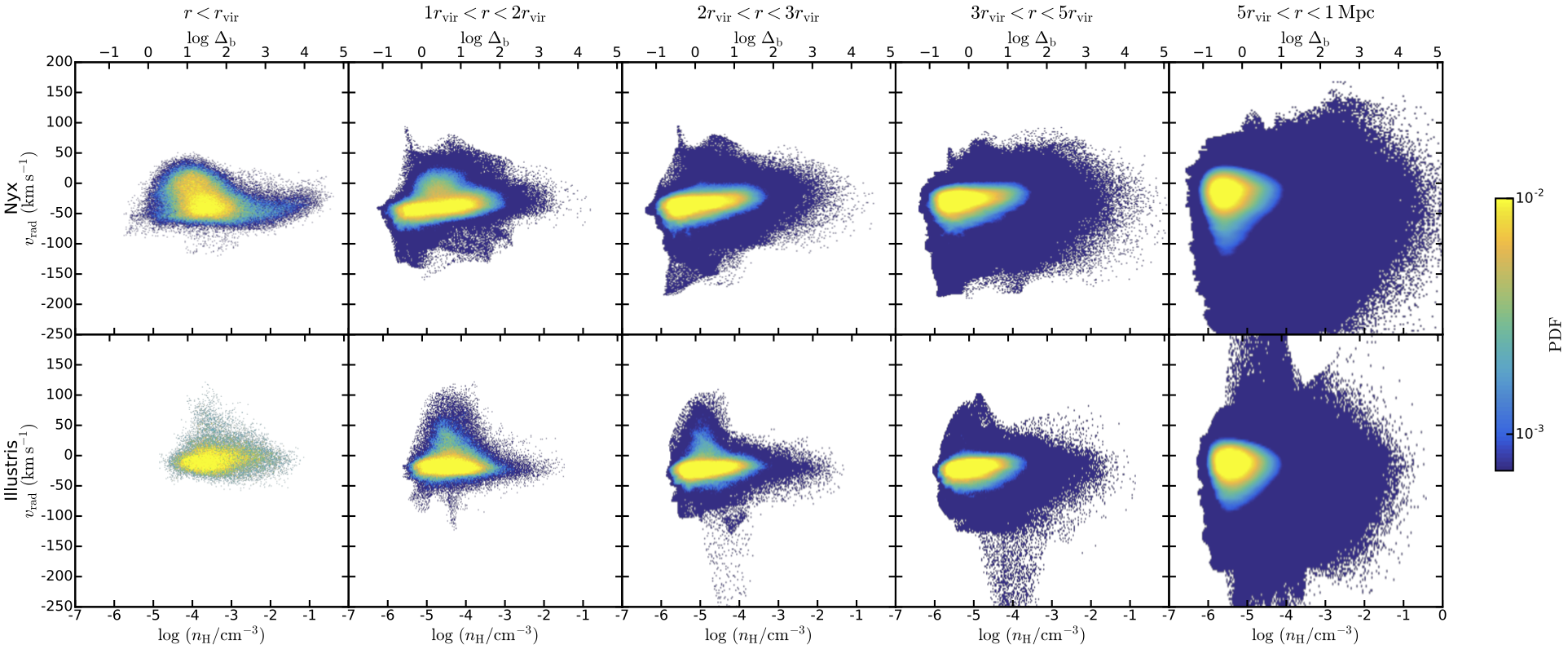}%
\end{adjustbox}
\end{figure*}

In \S~\ref{sec:discussion} we focused on the impact of density and
temperature of the gas on \lya absorption. Nonetheless, the \lya
optical depth depends also on the peculiar velocity of the gas. In
fact, the smoothness of the velocity field can have a significant
impact on the statistics of the \lya absorption lines
\citep{Sorini_2016}. Moreover, various hydrodynamic simulations
(including Illustris) underpredict 
the line width distribution of the \lya forest \citep{Viel_2017} or the line width - \hi column density relationship \citep{Gaikwad_2017} obtained from HST-COS QSO absorption spectra. The agreement with data of the latter statistics can be improved adding a turbulent broadening contribution to the line width in the simulations (\citealt{Gaikwad_2017}; see also \citealt{Oppenheimer_2009} and \citealt{Viel_2017}). This term is not a thermal broadening, but velocity broadening coming from motions not captured by the simulations.

In this section, we focus on the connection between the radial component of the velocity of the gas around galaxies and \lya absorption in the CGM. Following what we did in \S~\ref{sec:discussion}, we investigate the radial velocity - hydrogen density relationship of the gas in the CGM in Nyx and Illustris. In Figure \ref{fig:vrad-nH_CGM} we plot this relationship around $M\gtrsim 10^{12.5}\,M_{\odot}$ ($M\gtrsim 10^{12.4}\,M_{\odot}$) halos from Nyx (Illustris), within the same radial bins as in \S~\ref{sec:discussion}. In Figure \ref{fig:vrad-nH_CGM_LBG} we show an analogous plot for the $M\gtrsim 10^{11.7}\,M_{\odot}$ ($M\gtrsim 10^{11.6}\,M_{\odot}$) halos from Nyx (Illustris).

Figure \ref{fig:vrad-nH_CGM} shows that, at any given radial bin, the shape of the radial velocity - hydrogen density relationship is qualitatively similar in Nyx and Illustris. However, the gas in Illustris presents an overall offset of $\sim+50 \, \rm km \, s^{-1}$ in the radial velocity with respect to Nyx. Furthermore, in the innermost bin, Nyx presents a larger spread in radial velocity: inflowing gas rarely reaches a radial velocity of $200 \, \rm km \, s^{-1}$ in Nyx, while in Illustris the bulk of the inflowing gas is slower than $150 \, \rm km \, s^{-1}$. In the interval $(r_{\rm vir},\,2 r_{\rm vir})$, the majority of the gas in Nyx lies in the radial velocity range $(-200,\,-100)$. On the contrary, there is a larger amount of gas with positive radial velocity (i.e. outflowing) in Illustris. Finally, in the bin $(2r_{\rm vir},\,3 r_{\rm vir})$, Illustris presents a plume toward more positive velocities, 
corresponding to gas with radial velocity $\gtrsim -50 \, \rm km \, s^{-1}$ and density in the range $(10^{-5.5} ,\, 10^{-4.5})\, \rm cm^{-3}$. Such feature is absent in Nyx, instead. For $r>3 r_{\rm vir}$, Nyx and Illustris present very similar radial velocity - density diagrams.

In Figure \ref{fig:vrad-nH_CGM_LBG}, the radial velocity - hydrogen density relationships in Nyx and Illustris look qualitatively even more similar than in Figure \ref{fig:vrad-nH_CGM}. The most different bin is the innermost one, where Nyx presents a larger spread in the diagram. In all radial bins, the gas in Illustris appears to have an overall offset of $\sim + 20 \, \rm km \, s^{-1}$ with respect to the gas in Nyx. In the bin $(2r_{\rm vir},\,3 r_{\rm vir})$, Illustris presents an excess of outflowing gas in the density range $(10^{-5.5} ,\,10^{-4.5})\, \rm cm^{-3}$ with respect to Nyx. 

Analyzing the radial velocity - density relationships of hydrogen in the CGM in the two simulations, we can conclude that, in general, Illustris presents more outflowing gas than Nyx. However, the velocity offsets are small compared to the velocity window within which $\delta_F$ is computed ($1000\,\rm km\,s^{-1}$ and $2000\,\rm km\,s^{-1}$ for DLAs/LBGs and QSOs, respectively; $340\,\rm km \, s^{-1}$ for \citealt{Turner_2014} measurements). For this reason, and considering that the radial velocity - density relationships in the two simulations look much more alike than the respective temperature - density relationships, we think that the higher CGM temperature in Illustris has a greater impact on the \lya absorption profiles than the larger amount of outflowing gas.

\bibliographystyle{apj}
\bibliography{Sorini18_accepted}

\end{document}